\documentclass[aps,10pt,twocolumn,showpacs,amsmath,amssymb,nofootinbib]{revtex4-1}

\usepackage{graphicx}
\usepackage{color}
\usepackage{soul}
\usepackage{float}
\usepackage{widetable}

\usepackage{longtable}

\usepackage{graphicx}
\usepackage{booktabs}
\usepackage{placeins}
\usepackage{multirow}
\usepackage{url}

\usepackage{amsmath}
\usepackage{amssymb}

\newcommand{\C}[1]{{\cal C}_{#1}}
\newcommand{\B}{{\cal B}}

\newcommand{\Cc}[1]{{\cal C}_{#1}}

\begin{document}

\title{Emerging patterns of New Physics with and without\\ Lepton Flavour Universal contributions}

\author{Marcel Alguer\'o$^{a,b}$, Bernat Capdevila$^{a,b,c}$, Andreas Crivellin$^{d,e}$, S\'ebastien Descotes-Genon$^{f}$, Pere Masjuan$^{a,b}$, Joaquim Matias$^{a,b}$, Mart\'{i}n Novoa-Brunet$^{f}$ and Javier Virto$^{g}$.
\vspace{0.3cm}}

\affiliation{
$^{a}$Grup de F\'isica Te\`orica (Departament de F\'isica), Universitat Aut\`onoma de Barcelona, E-08193 Bellaterra (Barcelona), Catalunya. \\
$^b$ Institut de F\'isica d'Altes Energies (IFAE), The Barcelona Institute of Science and Technology, Campus UAB, E-08193 Bellaterra (Barcelona), Catalunya.\\
$^{c}$Universit\`a di Torino and INFN Sezione di Torino, Via P. Giuria 1, Torino I-10125, Italy.\\
$^{d}$ Paul Scherrer Institut, CH--5232 Villigen PSI, Switzerland.\\
$^{e}$Physik-Institut, Universit\"at Z\"urich, Winterthurerstrasse 190, CH-8057 Z\"urich, Switzerland.\\
$^{f}$ Universit\'e Paris-Saclay, CNRS/IN2P3, IJCLab, 91405 Orsay, France.\\
$^g$Departament de F\'{i}sica Qu\`antica i Astrof\'{i}sica, Institut de Ci\`encies del Cosmos, Universitat de Barcelona, Mart\'{i} Franqu\`{e}s 1, E08028 Barcelona, Catalunya.
}

\begin{abstract}
We perform a model-independent global fit to $b\to s\ell^+\ell^-$ observables to confirm existing New Physics  (NP) patterns (or scenarios) and to identify new ones emerging from the inclusion of the updated LHCb and Belle measurements of $R_K$ and $R_{K^*}$, respectively. Our analysis,  updating Refs.~\cite{Capdevila:2017bsm,Alguero:2018nvb} and including these new data,
suggests the presence of right-handed couplings encoded in the Wilson coefficients $\Cc{9'\mu}$ and $\Cc{10'\mu}$. It also strengthens our earlier observation that a lepton flavour universality violating (LFUV) left-handed lepton coupling ($\Cc{9\mu}^{\rm V}=-\Cc{10\mu}^{\rm V}$), often preferred from the model building point of view, accommodates the data better if lepton-flavour universal (LFU) NP is allowed,  in particular in $\Cc{9}^{\rm U}$. Furthermore, this  scenario with LFU NP provides a simple and model-independent connection
to the $b\to c\tau \nu$ anomalies, showing a preference of $\approx 7\,\sigma$ with respect to the SM. It
may also explain why fits to the whole set of $b\to s\ell^+\ell^-$ data or to the subset of LFUV data exhibit stronger preferences for different NP scenarios. Finally, motivated by $Z^\prime$ models with vector-like quarks, we propose four new scenarios with LFU and LFUV NP contributions that give a very good fit to data. We provide also an addendum collecting our updated results after including the data for the $B\to K^*\mu\mu$ angular distribution  released in 2020 by the LHCb collaboration.
\end{abstract}


\pacs{13.25.Hw, 11.30.Hv}

\maketitle

\section{Introduction}

The flavour anomalies in $b\to s\ell^+\ell^-$ processes are at present among the most promising signals of new physics (NP). Their analyses can be efficiently and consistently performed in a model-independent effective field theory (EFT) framework~(see, for instance, \cite{Capdevila:2017bsm,Alguero:2018nvb,global}), where all short-distance physics (including NP) is encoded in Wilson coefficients, i.e. the coefficients of higher-dimension operators. A central open question is then which pattern(s) in the space of the Wilson coefficients is (are) preferred by $b \to s \ell^+\ell^-$ observables. More precise measurements, in particular for the observables showing deviations from the Standard Model (SM) expectations ($P_5^\prime$~\cite{implications}, $R_{K,K^*,\phi}$, $Q_5$\cite{assessing} \ldots), help us to improve the results of this EFT analysis, which can then be used as a guideline for the construction of phenomenologically accurate NP models.

In this context we present here an update and extension of our recent works in Refs.~\cite{Capdevila:2017bsm,Alguero:2018nvb},
 in the light of new measurements of key observables involved in $b\to s\ell^+\ell^-$ anomalies. We update the experimental value of the ratio probing lepton flavour universality (LFU) defined as $R_{K}=\frac{\B(B\to K\mu^+\mu^-)}{\B(B\to Ke^+e^-)}$:
\begin{align}\nonumber
R^{[1.1,6]}_{K_{\rm{LHCb}}}&=0.846_{-0.054 \, -0.014}^{+0.060 \, +0.016} \, ,  \\
R^{[1,6]}_{K_{\rm{Belle}}}&=0.98^{+0.27}_{-0.23}\pm 0.06 \, ,  \\ \nonumber
R^{[q^2>14.18]}_{K_{\rm{Belle}}}&=1.11^{+0.29}_{-0.26}\pm 0.07  \, ,
\end{align}
as announced recently by the LHCb collaboration~\cite{Aaij:2019wad}, corresponding to the average of Run-1 and part of Run-2 (2015-2016) measurements, and the Belle collaboration~\cite{BelleRK}, combining the data from charged and neutral modes.
The correlations with the (finely binned) measurements of $\B(B\to K\mu^+\mu^-)$~\cite{Aaij:2014pli} are tiny and therefore neglected here. In addition the Belle collaboration has also presented new results for $R_{K^*}$, the equivalent LFU-violating (LFUV) ratio for $B\to K^*\ell\ell$, in three bins~\cite{Abdesselam:2019wac}, again considering both charged and neutral channels:

\begin{align}\nonumber
R_{K^*}^{[0.045,1.1]}&=0.52_{-0.26}^{+0.36}\pm 0.05 \, , \\
R_{K^*}^{[1.1,6]}&=0.96_{-0.29}^{+0.45}\pm 0.11 \, , \\
R_{K^*}^{[15,19]}&=1.18_{-0.32}^{+0.52}\pm 0.10\, . \nonumber
\end{align}

\begin{table*}
\begin{tabular}{c||c|c|c|c||c|c|c|c} 
 & \multicolumn{4}{c||}{All} &  \multicolumn{4}{c}{LFUV}\\
\hline
1D Hyp.   & Best fit& 1 $\sigma$/2 $\sigma$   & Pull$_{\rm SM}$ & p-value & Best fit & 1 $\sigma$/ 2 $\sigma$  & Pull$_{\rm SM}$ & p-value\\
\hline\hline
\multirow{2}{*}{$\Cc{9\mu}^{\rm NP}$}    & \multirow{2}{*}{-0.98} &    $[-1.15,-0.81]$ &    \multirow{2}{*}{5.6}   & \multirow{2}{*}{65.4\,\%}
&   \multirow{2}{*}{-0.89}   &$[-1.23,-0.59]$&   \multirow{2}{*}{3.3}  & \multirow{2}{*}{52.2\,\%}  \\
 &  & $[-1.31,-0.64]$ &  & &  &  $[-1.60,-0.32]$ & \\
 \multirow{2}{*}{$\Cc{9\mu}^{\rm NP}=-\Cc{10\mu}^{\rm NP}$}    &   \multirow{2}{*}{-0.46} &    $[-0.56,-0.37]$ &   \multirow{2}{*}{5.2}  & \multirow{2}{*}{55.6\,\%}
 &  \multirow{2}{*}{-0.40}   &   $[-0.53,-0.29]$ & \multirow{2}{*}{4.0}   & \multirow{2}{*}{74.0\,\%}  \\
 &  & $[-0.66,-0.28]$ &  & & & $[-0.63,-0.18]$  &    \\
 \multirow{2}{*}{$\Cc{9\mu}^{\rm NP}=-\Cc{9'\mu}$}     & \multirow{2}{*}{-0.99} &    $[-1.15,-0.82]$   &  \multirow{2}{*}{5.5}  & \multirow{2}{*}{62.9\,\%}
 &  \multirow{2}{*}{-1.61}   &    $[-2.13,-0.96]$  & \multirow{2}{*}{3.0} & \multirow{2}{*}{42.5\,\%} \\
 &  & $[-1.31,-0.64]$ &  & & & $[-2.54,-0.41]$ &    \\
\hline
 \multirow{2}{*}{$\Cc{9\mu}^{\rm NP}=-3 \Cc{9e}^{\rm NP}$} & \multirow{2}{*}{-0.87}  & $[-1.03,-0.71]$ & \multirow{2}{*}{5.5}  & \multirow{2}{*}{61.9\,\%}
  &   \multirow{2}{*}{-0.66} &    $[-0.90,-0.44]$ & \multirow{2}{*}{3.3}  & \multirow{2}{*}{52.2\,\%}
\\
 & & $[-1.19,-0.55]$ &  & & & $[-1.17,-0.24]$     & \\
\end{tabular} \scriptsize
\caption{Most prominent 1D patterns of NP in $b\to s\mu^+\mu^-$. Pull$_{\rm SM}$ is quoted in units of standard deviation. The $p$-value of the SM hypothesis is 11.0\% for the fit ``All" and 8.0\% for the fit LFUV.} \scriptsize
\label{tab:results1D}
\end{table*}

\begin{table*} 
\begin{tabular}{c||c|c|c||c|c|c} 
 & \multicolumn{3}{c||}{All} &  \multicolumn{3}{c}{LFUV}\\
\hline
 2D Hyp.  & Best fit  & Pull$_{\rm SM}$ & p-value & Best fit & Pull$_{\rm SM}$ & p-value\\
\hline\hline
$(\Cc{9\mu}^{\rm NP},\Cc{10\mu}^{\rm NP})$ & (-0.91,0.18) & 5.4 & 68.7\,\% & (-0.16,0.56) & 3.4 & 76.9\,\% \\
$(\Cc{9\mu}^{\rm NP},\Cc{7^{\prime}})$  & (-1.00,0.02) & 5.4 & 67.9\,\% & (-0.90,-0.04) & 2.9 & 55.1\,\% \\
$(\Cc{9\mu}^{\rm NP},\Cc{9^\prime\mu})$  & (-1.10,0.55) & 5.7 & 75.1\,\% & (-1.79,1.14) & 3.4 & 76.1\,\% \\
$(\Cc{9\mu}^{\rm NP},\Cc{10^\prime\mu})$  & (-1.14,-0.35) & 5.9 & 78.6\,\% & (-1.88,-0.62) & 3.8 & 91.3\,\% \\
\hline
$(\Cc{9\mu}^{\rm NP}, \Cc{9e}^{\rm NP})$ & (-1.05,-0.23) & 5.3 & 66.2\,\% & (-0.73,0.16) & 2.8 & 52.3\,\%  \\
\hline
Hyp. 1 & (-1.06,0.26) & 5.7 & 75.7\,\% & (-1.62,0.29) & 3.4 & 77.6\,\% \\
Hyp. 2 & (-0.97,0.09) & 5.3 & 65.2\,\% & (-1.95,0.25) & 3.2 & 66.6\,\% \\
Hyp. 3 & (-0.47,0.06) & 4.8 & 55.7\,\% & (-0.39,-0.13) & 3.4 & 76.2\,\% \\
Hyp. 4  & (-0.49,0.12) & 5.0 & 59.3\,\% & (-0.48,0.17) & 3.6 & 84.3\,\% \\
Hyp. 5  & (-1.14,0.24) & 5.9 & 78.7\,\% & (-2.07,0.52) & 3.9 & 92.5\,\% \\
\end{tabular}
\caption{Most prominent 2D patterns of NP in $b\to s\mu^+\mu^-$. The last five rows correspond to Hypothesis 1: $(\Cc{9\mu}^{\rm NP}=-\Cc{9^\prime\mu} , \Cc{10\mu}^{\rm NP}=\Cc{10^\prime\mu})$,  2: $(\Cc{9\mu}^{\rm NP}=-\Cc{9^\prime\mu} , \Cc{10\mu}^{\rm NP}=-\Cc{10^\prime\mu})$, 3: $(\Cc{9\mu}^{\rm NP}=-\Cc{10\mu}^{\rm NP} , \Cc{9^\prime\mu}=\Cc{10^\prime\mu}$), 4: $(\Cc{9\mu}^{\rm NP}=-\Cc{10\mu}^{\rm NP} , \Cc{9^\prime\mu}=-\Cc{10^\prime\mu})$ and 5: $(\Cc{9\mu}^{\rm NP} , \Cc{9^\prime\mu}=-\Cc{10^\prime\mu})$.}
\label{tab:results2D}
\end{table*}

Our treatment for the Belle observables within the global fit follows the same strategy as described in Ref.~\cite{Capdevila:2017bsm} for $Q_{4,5}$ where we introduced a nuisance parameter accounting for the relative weight of each isospin component.


\begin{table*} 
\begin{tabular}{c||c|c|c|c|c|c}
 & $\Cc7^{\rm NP}$ & $\Cc{9\mu}^{\rm NP}$ & $\Cc{10\mu}^{\rm NP}$ & $\Cc{7^\prime}$ & $\Cc{9^\prime \mu}$ & $\Cc{10^\prime \mu}$  \\
\hline\hline
Best fit & +0.01 & -1.10 & +0.15 & +0.02 & +0.36 & -0.16 \\ \hline
1 $\sigma$ & $[-0.01,+0.05]$ & $[-1.28,-0.90]$ & $[-0.00,+0.36]$ & $[-0.00,+0.05]$ & $[-0.14,+0.87]$ &$[-0.39,+0.13]$ \\
2 $\sigma$ & $[-0.03,+0.06]$ & $[-1.44,-0.68]$ & $[-0.12,+0.56]$ & $[-0.02,+0.06]$ & $[-0.49,+1.23]$ &$[-0.58,+0.33]$
\end{tabular}
\caption{1 and 2~$\sigma$ confidence intervals for the NP contributions to Wilson coefficients in
the 6D hypothesis allowing for NP in $b\to s\mu^+\mu^-$ operators dominant in the SM and their chirally-flipped counterparts, for the fit ``All''. The Pull$_{\rm SM}$ is 5.1~$\sigma$ and the \textit{p}-value is $81.6 \%$.}
\label{tab:Fit6D}
\end{table*}

\begin{table*}  \begin{center}
\begin{tabular}{c|c|c|c|c|c|c|c|c|c|c}
$\alpha_{0\mu}$ & $\alpha_{1\mu}$ & $\alpha_{2\mu}$ & $\alpha_{3\mu}$ & $\alpha_{4\mu}$ & $\alpha_{5\mu}$ & $\alpha_{6\mu}$ &  $\alpha_{7\mu}$ & $\alpha_{8\mu}$ & $\alpha_{9\mu}$ & $\alpha_{10\mu}$\\
\hline
4.00 & 0.92 & 0.12 & 0.92 & 0.12 & 0.24 & -1.06 & 0.12 & -1.06 & 0.12 & 0.25\\
\hline\hline
$\alpha_{0e}$ & $\alpha_{1e}$ & $\alpha_{2e}$ & $\alpha_{3e}$ & $\alpha_{4e}$ & $\alpha_{5e}$ & $\alpha_{6e}$ &  $\alpha_{7e}$ & $\alpha_{8e}$ & $\alpha_{9e}$ & $\alpha_{10e}$\\
\hline
3.99 & 0.92 & 0.12 & 0.92 & 0.12 & 0.24 & -1.05 & 0.12 & -1.05 & 0.12 & 0.24\\
\end{tabular} \end{center}
\caption{Coefficients for the polynomial parameterisation of the numerator and denominator of $R_K^{[1.1,6]}$ in the vicinity of the SM point. \label{tab:polpar}}
\end{table*}

\begin{figure*}
\begin{center}
\includegraphics[width=0.315\textwidth]{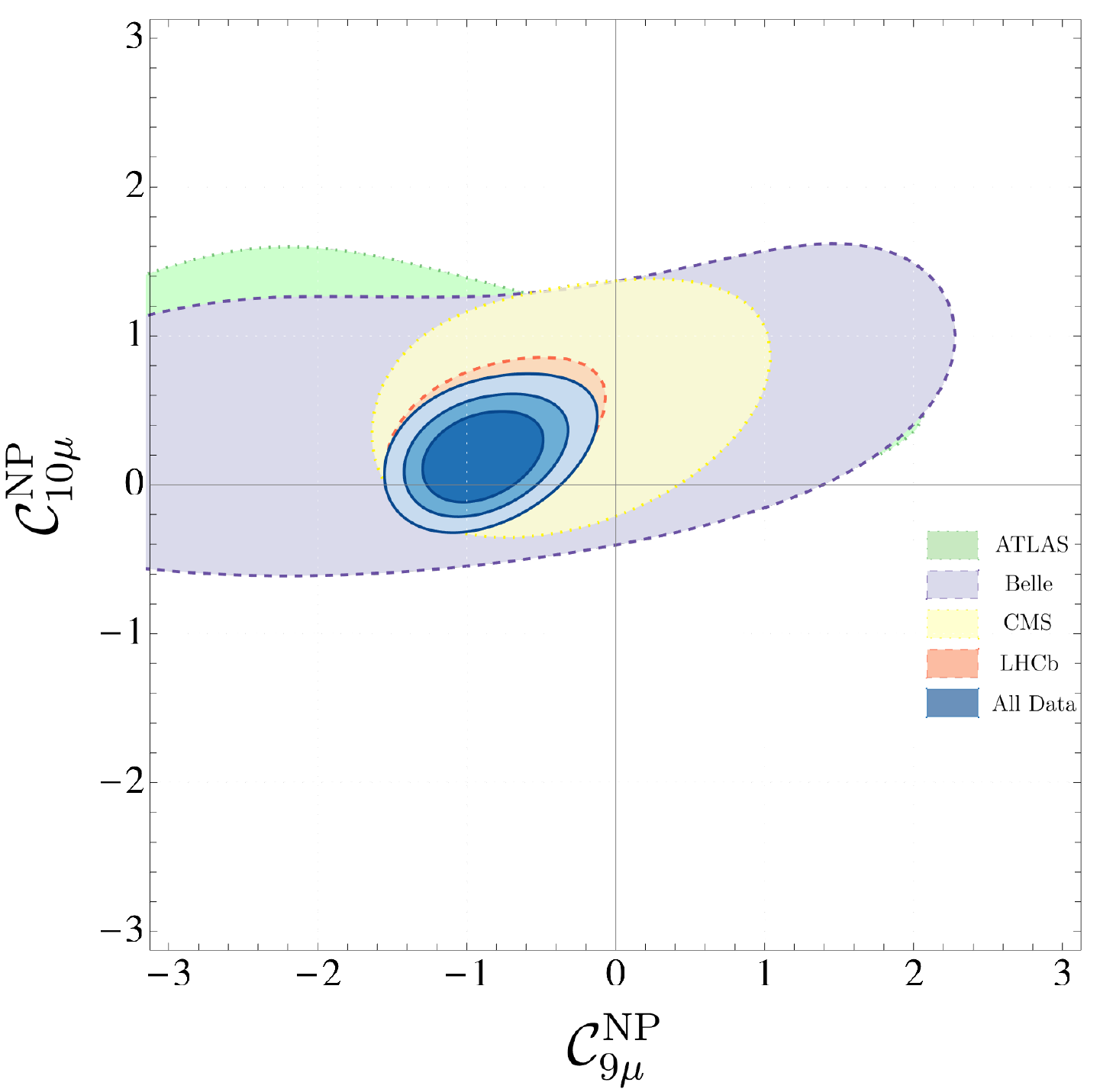}\hspace{2mm}
\includegraphics[width=0.315\textwidth]{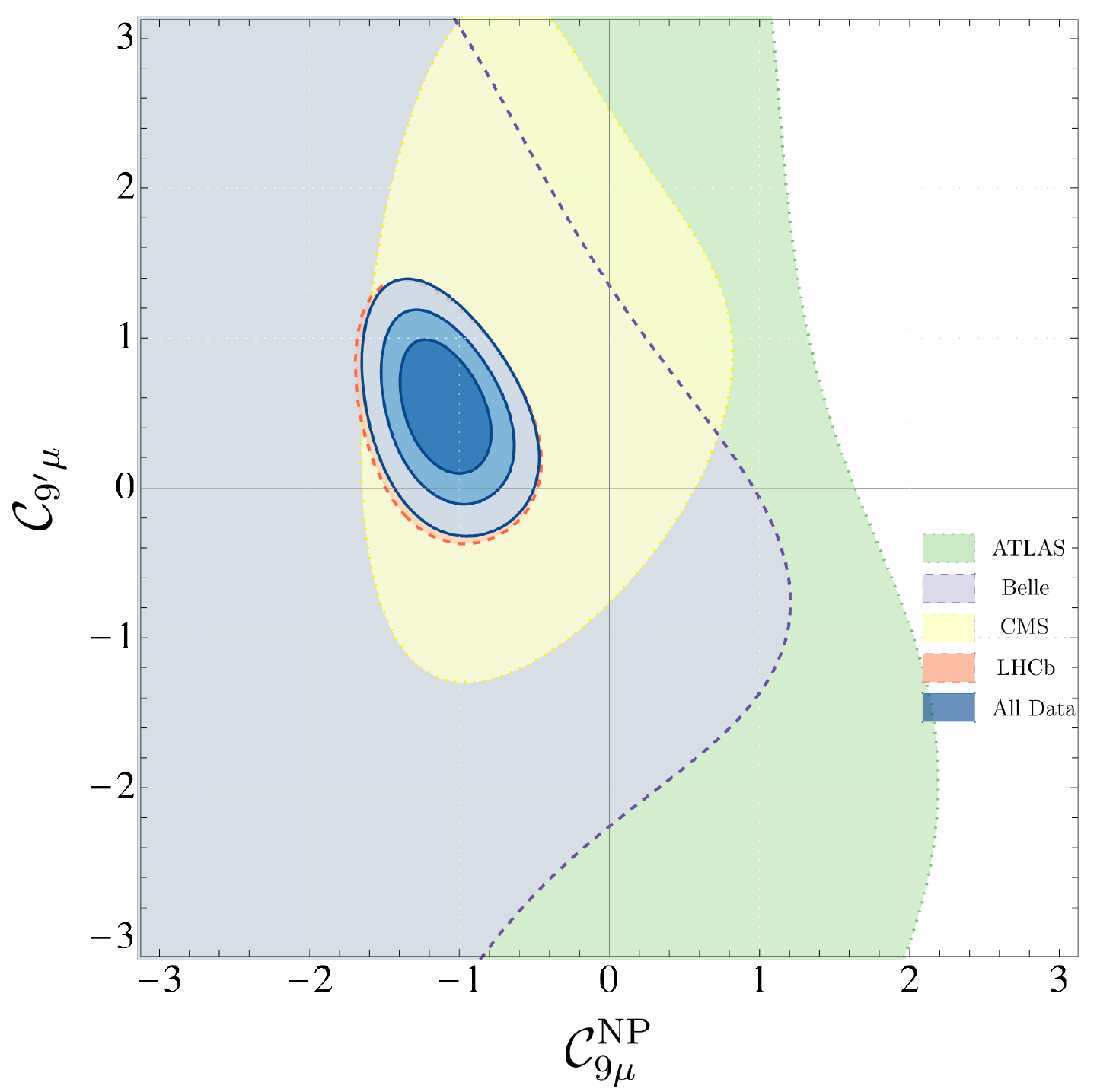}\hspace{2mm}
\includegraphics[width=0.315\textwidth]{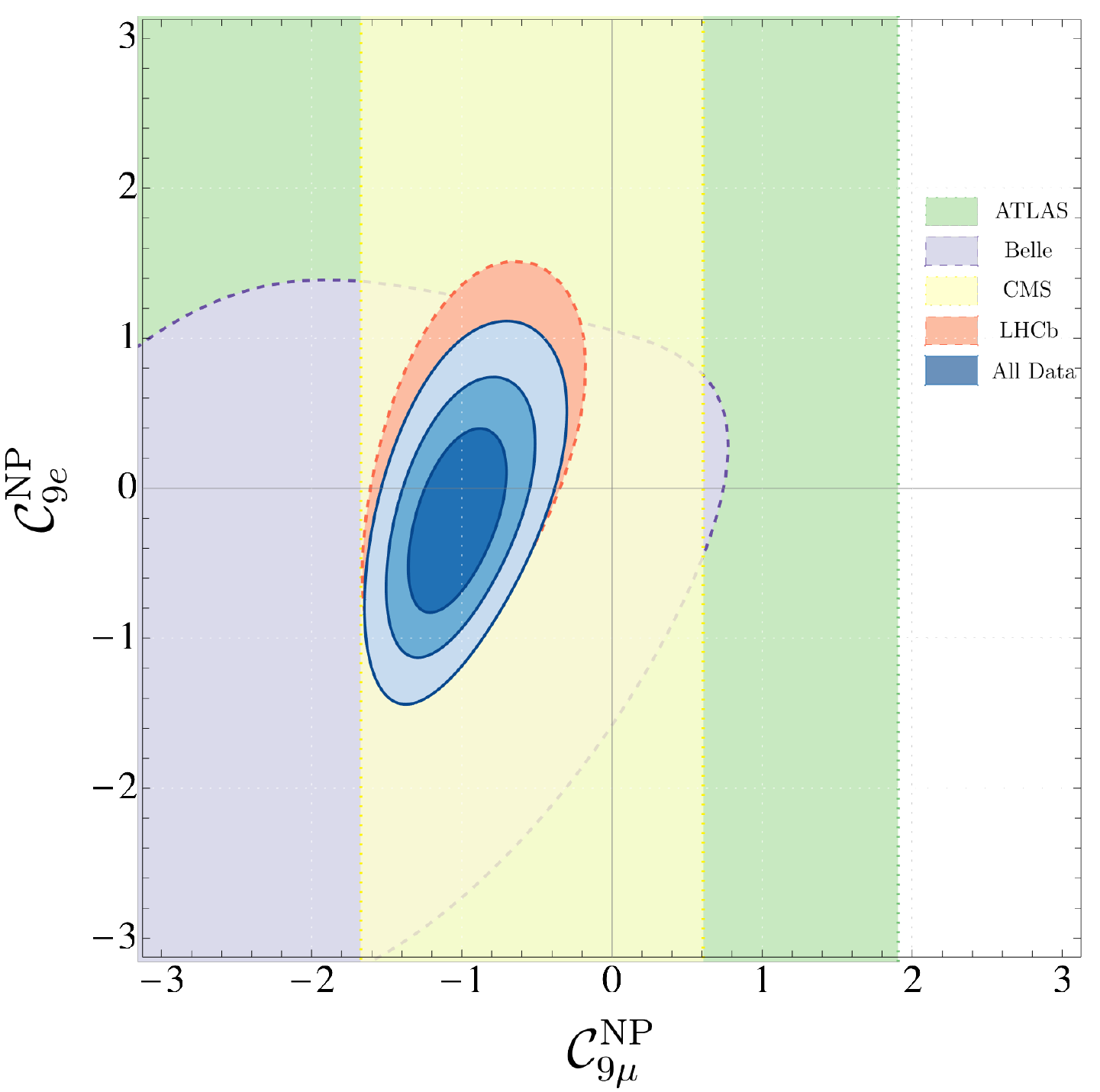}
\includegraphics[width=0.315\textwidth]{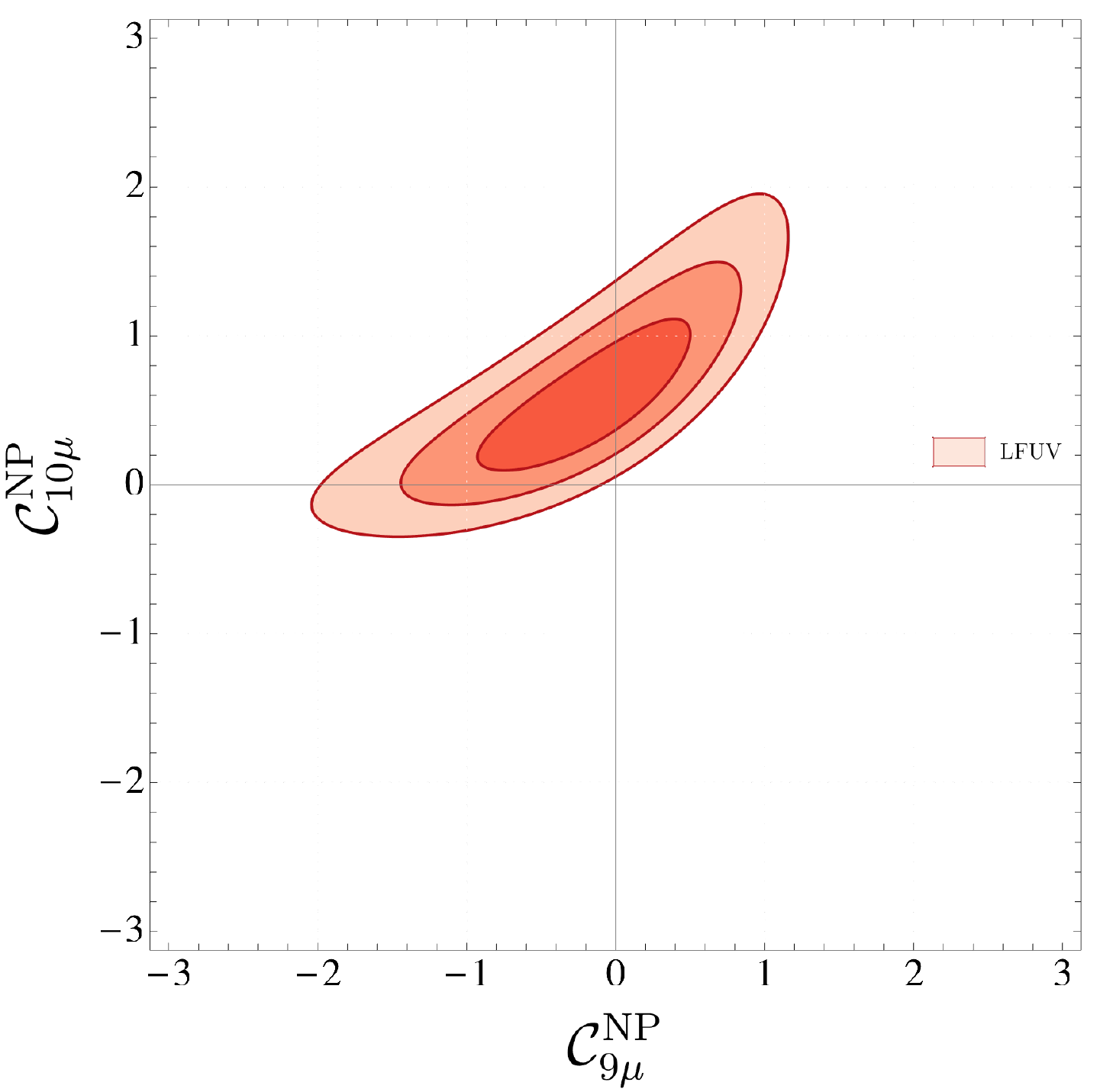}\hspace{2mm}
\includegraphics[width=0.315\textwidth]{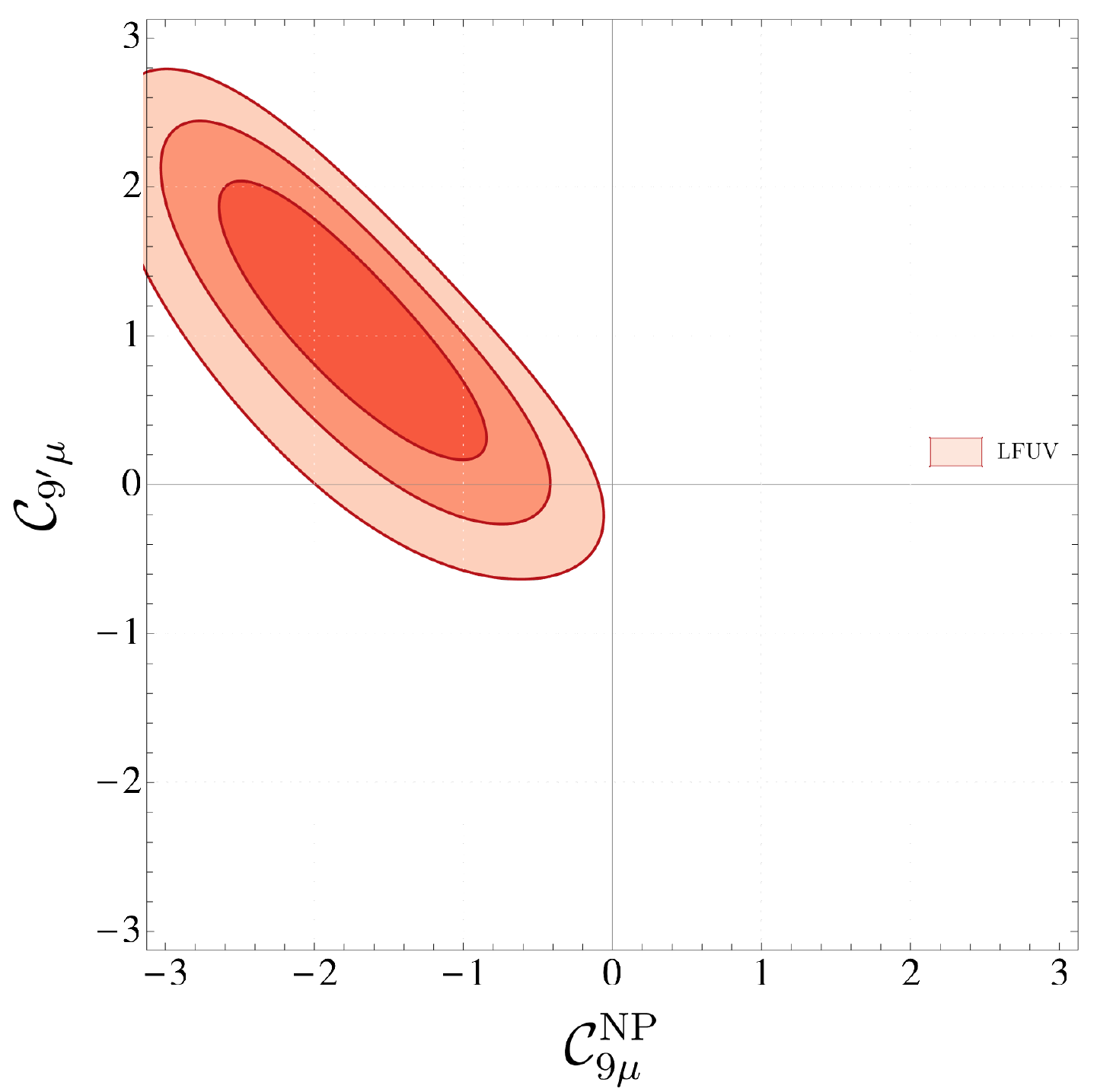}\hspace{2mm}
\includegraphics[width=0.315\textwidth]{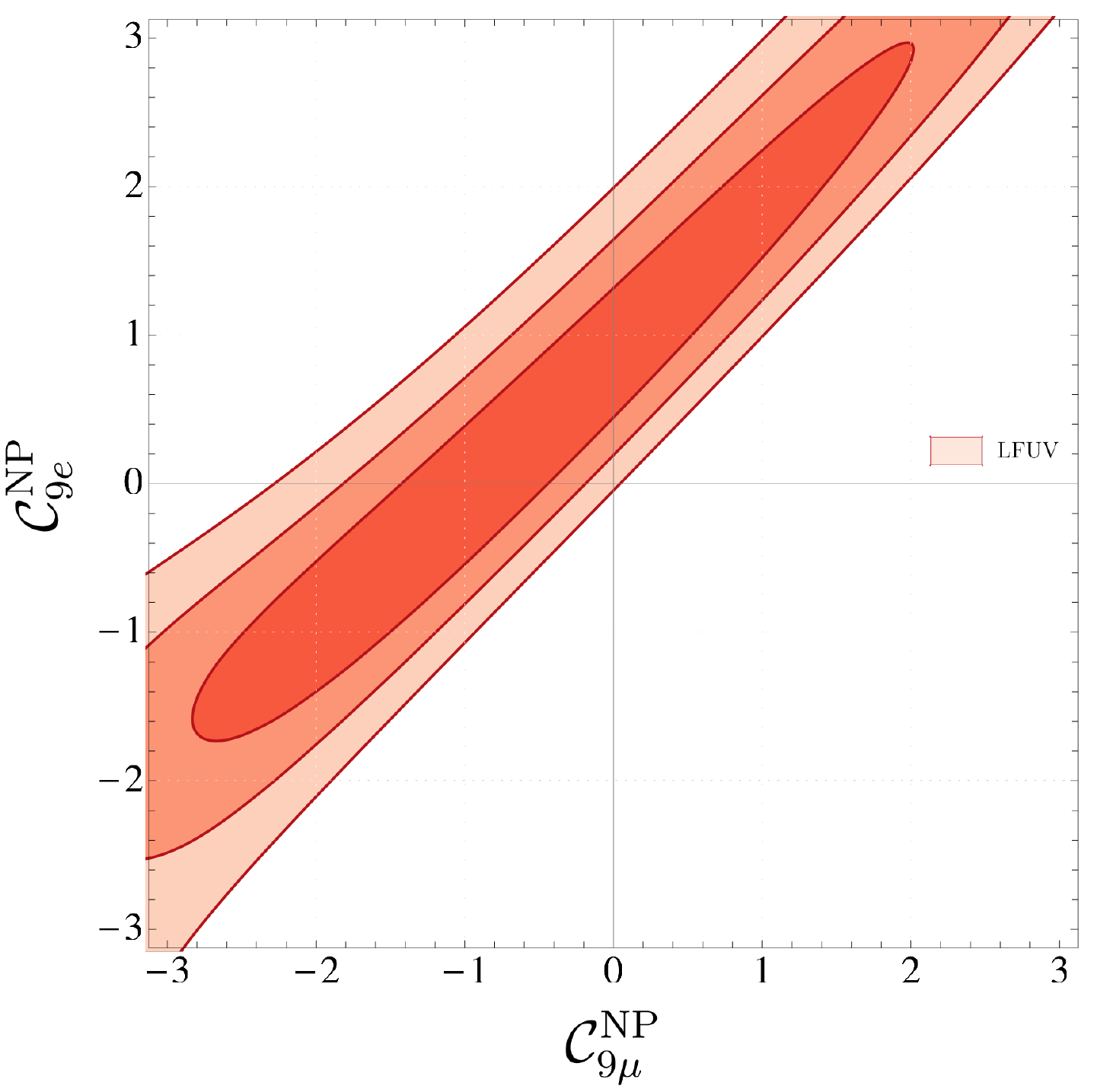}
\end{center}
\caption{From left to right: Allowed regions in the $(\Cc{9\mu}^{\rm NP},\Cc{10\mu}^{\rm NP})$, $(\Cc{9\mu}^{\rm NP},\Cc{9^\prime\mu})$ and $(\Cc{9\mu}^{\rm NP},\Cc{9e}^{\rm NP})$ planes for the corresponding 2D hypotheses, using all available data (fit ``All'') upper row or LFUV fit  lower row.}
\label{fig:FitResultAll}
\end{figure*}


\begin{table*} \small \begin{center}
\begin{tabular}{lc||c|c|c|c|c}
\multicolumn{2}{c||}{Scenario} & Best-fit point & 1 $\sigma$ & 2 $\sigma$ & Pull$_{\rm SM}$ & p-value \\
\hline\hline
\multirow{ 3}{*}{Scenario 5} &$\Cc{9\mu}^{\rm V}$ & $-0.36$ & $[-0.86,+0.10]$ & $[-1.41,+0.52]$ &
\multirow{ 3}{*}{5.2} & \multirow{ 3}{*}{71.2\,\%} \\
&$\Cc{10\mu}^{\rm V}$ & $+0.67$ & $[+0.24,+1.03]$ & $[-1.73,+1.36]$ & \\
&$\Cc{9}^{\rm U}=\Cc{10}^{\rm U}$ & $-0.59$ & $[-0.90,-0.12]$ & $[-1.13,+0.68]$ &\\
\hline
\multirow{ 2}{*}{Scenario 6}&$\Cc{9\mu}^{\rm V}=-\Cc{10\mu}^{\rm V}$ & $-0.50$ & $[-0.61,-0.38]$ & $[-0.72,-0.28]$ &
\multirow{ 2}{*}{5.5} & \multirow{ 2}{*}{71.0\,\%} \\
&$\Cc{9}^{\rm U}=\Cc{10}^{\rm U}$ & $-0.38$ & $[-0.52,-0.22]$ & $[-0.64,-0.06]$ &\\
\hline
\multirow{ 2}{*}{Scenario 7}&$\Cc{9\mu}^{\rm V}$ & $-0.78$ & $[-1.11,-0.47]$ & $[-1.45,-0.18]$ &
\multirow{ 2}{*}{5.3} & \multirow{ 2}{*}{66.2\,\% }  \\
&$\Cc{9}^{\rm U}$ & $-0.20$ & $[-0.57,+0.18]$ & $[-0.92,+0.55]$  &\\
\hline
\multirow{ 2}{*}{Scenario 8}&$\Cc{9\mu}^{\rm V}=-\Cc{10\mu}^{\rm V}$ & $-0.30$ & $[-0.42,-0.20]$ & $[-0.53,-0.10]$ &
\multirow{ 2}{*}{5.7} & \multirow{ 2}{*}{75.2\,\%} \\
&$\Cc{9}^{\rm U}$ & $-0.74$ & $[-0.96,-0.51]$ & $[-1.15,-0.25]$ &\\
\hline\hline
\multirow{ 2}{*}{Scenario 9}&$\Cc{9\mu}^{\rm V}=-\Cc{10\mu}^{\rm V}$ & $-0.57$ & $[-0.73,-0.41]$ & $[-0.87,-0.28]$ &
\multirow{ 2}{*}{5.0} & \multirow{ 2}{*}{60.2\,\%} \\
&$\Cc{10}^{\rm U}$ & $-0.34$ & $[-0.60,-0.07]$ & $[-0.84,+0.18]$ &\\
\hline
\multirow{ 2}{*}{Scenario 10}&$\Cc{9\mu}^{\rm V}$ & $-0.95$ & $[-1.13,-0.76]$ & $[-1.30,-0.57]$ &
\multirow{ 2}{*}{5.5} & \multirow{ 2}{*}{69.5\,\%} \\
&$\Cc{10}^{\rm U}$ & $+0.27$ & $[0.08,0.47]$ & $[-0.09,0.66]$ &\\
\hline
\multirow{ 2}{*}{Scenario 11}&$\Cc{9\mu}^{\rm V}$ & $-1.03$ & $[-1.22,-0.84]$ & $[-1.38,-0.65]$ &
\multirow{ 2}{*}{5.6} & \multirow{ 2}{*}{73.6\,\%} \\
&$\Cc{10'}^{\rm U}$ & $-0.29$ & $[-0.47,-0.12]$ & $[-0.63,0.05]$ &\\
\hline
\multirow{ 2}{*}{Scenario 12}&$\Cc{9'\mu}^{\rm V}$ & $-0.03$ & $[-0.22,0.15]$ & $[-0.40,0.32]$ &
\multirow{ 2}{*}{1.6} & \multirow{ 2}{*}{15.7\,\%} \\
&$\Cc{10}^{\rm U}$ & $+0.41$ & $[0.21,0.63]$ & $[0.02,0.83]$ &\\
\hline
\multirow{ 4}{*}{Scenario 13}&$\Cc{9\mu}^{\rm V}$ & $-1.11$ & $[-1.28,-0.91]$ & $[-1.41,-0.71]$ &
\multirow{ 4}{*}{5.4} & \multirow{ 4}{*}{78.7\,\%} \\
&$\Cc{9'\mu}^{\rm V}$ & $+0.53$ & $[0.24,0.83]$ & $[-0.10,1.11]$ &\\
&$\Cc{10}^{\rm U}$ & $+0.24$ & $[0.01,0.48]$ & $[-0.21,0.69]$ &\\
&$\Cc{10'}^{\rm U}$ & $-0.04$ & $[-0.28,0.20]$ & $[-0.48,0.42]$ &\\
\end{tabular}
\caption{Most prominent patterns for LFU and LFUV NP contributions from Fit ``All''.
Scenarios 5 to 8 were introduced in Ref.~\cite{Alguero:2018nvb}.  Scenarios 9 (motivated by 2HDMs~\cite{Crivellin:2019dun}) and 10 to 13  (motivated by $Z^\prime$ models with vector-like quarks~\cite{Bobeth:2016llm}) are new.}\label{Fit3Dbis} \end{center}
\end{table*}
We have also updated our average for ${\cal B}(B_s \to \mu^+\mu^-)$ including the latest measurement from the ATLAS collaboration~\cite{Aaboud:2018mst} and taking into account the most recent lattice update of $f_{B_s}$ for $N_f=2+1+1$ simulations collected in Ref.~\cite{FLAG}.

A relatively small numerical impact of such updates has been found. As  in Ref.~\cite{Capdevila:2017bsm}, our analysis also includes  the latest update of $P_{4,5}^\prime$ from the Belle collaboration~\cite{Wehle:2016yoi} where the muon and electron modes are considered separately (averaging charged and neutral modes), superseding the previous measurement in Ref.~\cite{Abdesselam:2016llu} where an average over both leptonic modes is presented. This allows us to include an additional measurement $P_{5\mu}^\prime$ (exhibiting a 2.6 $\sigma$ discrepancy with respect to the SM) as well as the LFUV observable $Q_5$ in our analysis (see Ref.~\cite{Ciuchini:2019usw} for another recent analysis including this update).

In addition to updating the experimental inputs, our analysis explores new emerging directions in the parameter space spanned by the effective operators driven by data within two different frameworks. First, following Ref.~\cite{Capdevila:2017bsm} we assume in Sec.~\ref{sec:LFUVNP} that NP affects only muons and is thus purely Lepton-Flavour Universality Violating (LFUV).
In Sec.~\ref{sec:LFUV-LFU-NP} we follow the complementary approach discussed in Ref.~\cite{Alguero:2018nvb}, where we consider  the consequences of removing the frequently made hypothesis that NP is purely LFUV. We then explore the implications of allowing both LFU and LFUV NP contributions to the Wilson coefficients $\Cc{9^{(\prime)}}$ and $\Cc{10^{(\prime)}}$.

Motivated by the new emerging directions in the LFUV case we also extend our analysis of NP scenarios to allow for the presence of LFU NP right handed-currents (RHC). In Sec.~\ref{sec:btoc}, we focus on a particular scenario (scenario 8) which can, within an EFT framework, link the flavour anomalies in $b\to s\ell^+\ell^-$ and $b\to c\ell\nu$ processes.  Furthermore, we consider new patterns, motivated by $Z^\prime$ models with vector-like quarks, which naturally predict LFU effects in $\Cc{10^{(\prime)}}$ complemented by LFUV ones. Finally, we sum up our results in Sec.~\ref{sec:conc}. An appendix is devoted to the description of the correlations obtained for the various Wilson coefficients in the most relevant scenarios considered in this article.

\section{Global fits in presence of LFUV NP}\label{sec:LFUVNP}

We start by considering the fits for NP scenarios which affect muon modes only.
Tabs.~\ref{tab:results1D}-\ref{tab:Fit6D} and Fig.~\ref{fig:FitResultAll} update the corresponding tables and figures of Ref.~\cite{Capdevila:2017bsm} based on fits to the full set of data (``All") or restricted to quantities assessing LFUV. While we do not observe any significant difference in the 1D scenarios with ``All" data compared to Ref.~\cite{Capdevila:2017bsm}, some of the Pulls (with respect to the SM) for the LFUV 1D fits get reduced by half a standard deviation. A few other comments are in order:
\begin{enumerate}
\item The scenario $\Cc{9\mu}^{\rm NP}=-\Cc{9'\mu}$, which favours a SM-like value of $R_K^{[1.1,6]}$~\cite{Alguero:2018nvb,Alguero:2019pjc}, has an increased significance in the ``All" fit compared to our earlier analysis.
\item  The scenario $\Cc{9\mu}^{\rm NP}$ has the largest $p$-value in the ``All" fit while $\Cc{9\mu}^{\rm NP}=-\Cc{10\mu}^{\rm NP}$ has the largest $p$-value in the LFUV fit, a difference which can be solved through the introduction of LFU NP (see Ref.~\cite{Alguero:2018nvb} and next section).
\item The best-fit point for the scenario $\Cc{9\mu}^{\rm NP}$ coincides now in the ``All" and LFUV fits.
\item The scenario with only $\Cc{10\mu}^{\rm NP}$ has a significance in the ``All" fit of only  4.0$\sigma$ level and 3.9$\sigma$ for the LFUV fit, which explains its absence from Tab.~\ref{tab:results1D} as happens in Ref.~\cite{Capdevila:2017bsm}.
\end{enumerate}

\begin{figure*}
	\begin{center}
		\includegraphics[width=7.3cm]{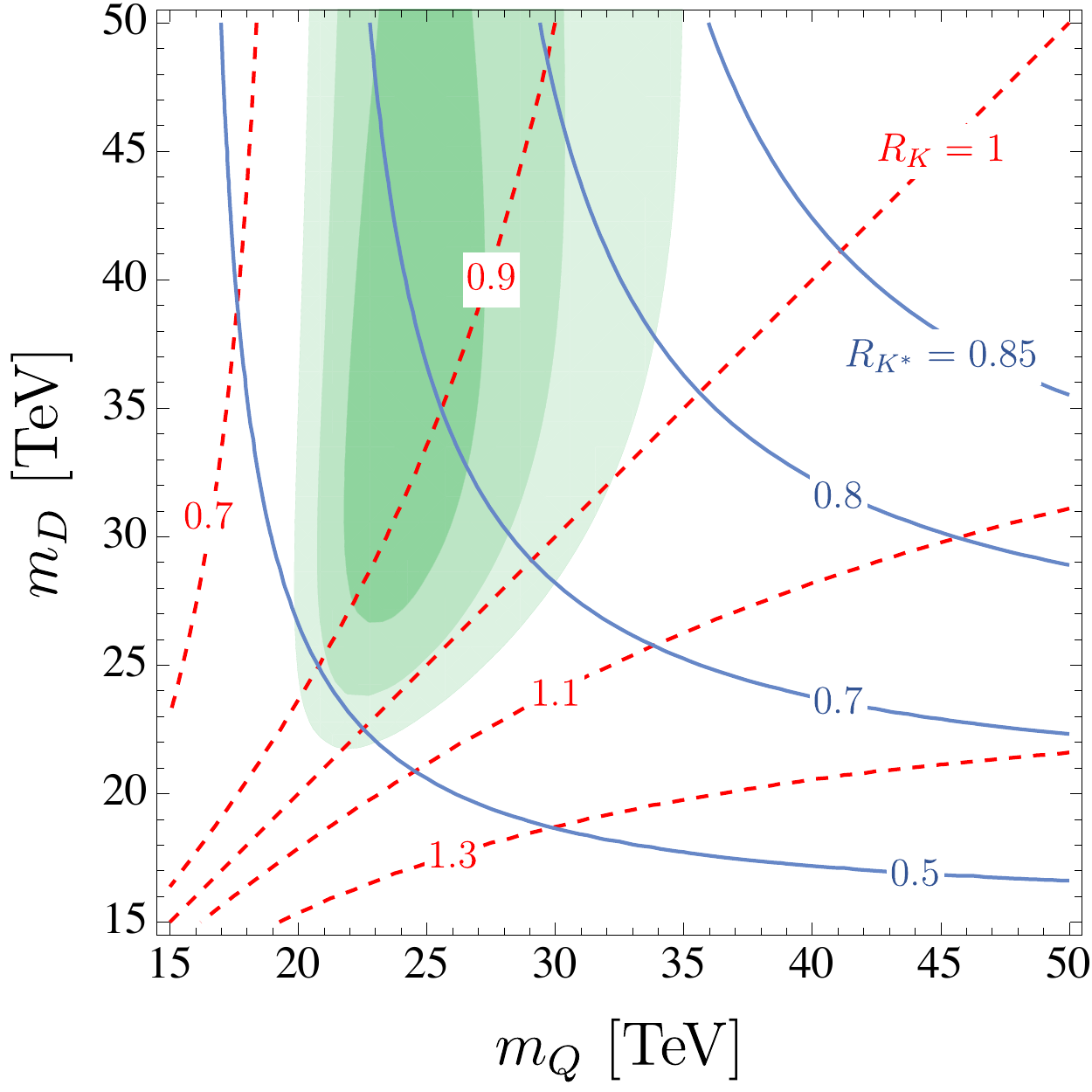}
	\end{center}
	\caption{Preferred regions (at the 1, 2 and 3$\,\sigma$ level) for the $L_\mu-L_\tau$ model of Ref.~\cite{Altmannshofer:2014cfa} from $b\to s\ell^+\ell^-$ data (green) in the $(m_Q,\, m_D)$ plane with $Y^{D,Q}=1$. The contour lines denote the predicted values for $R_K^{[1.1,6]}$ (red, dashed) and $R_{K^*}^{[1.1,6]}$ (blue, solid).}
	\label{fig:modelfits2}
\end{figure*}

\begin{figure*}
\begin{center}
\includegraphics[width=0.315\textwidth]{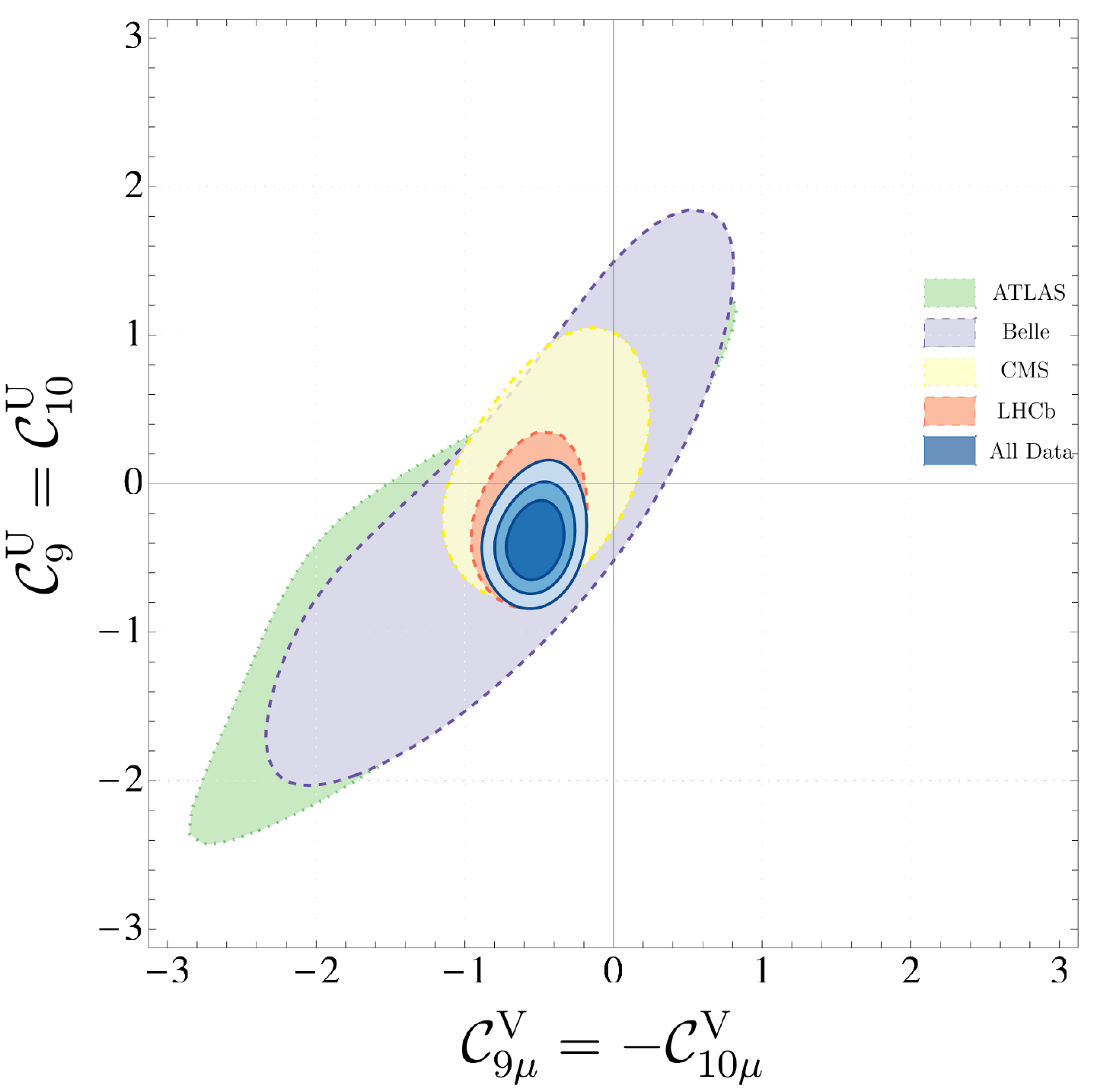}\hspace{5mm}
\includegraphics[width=0.315\textwidth]{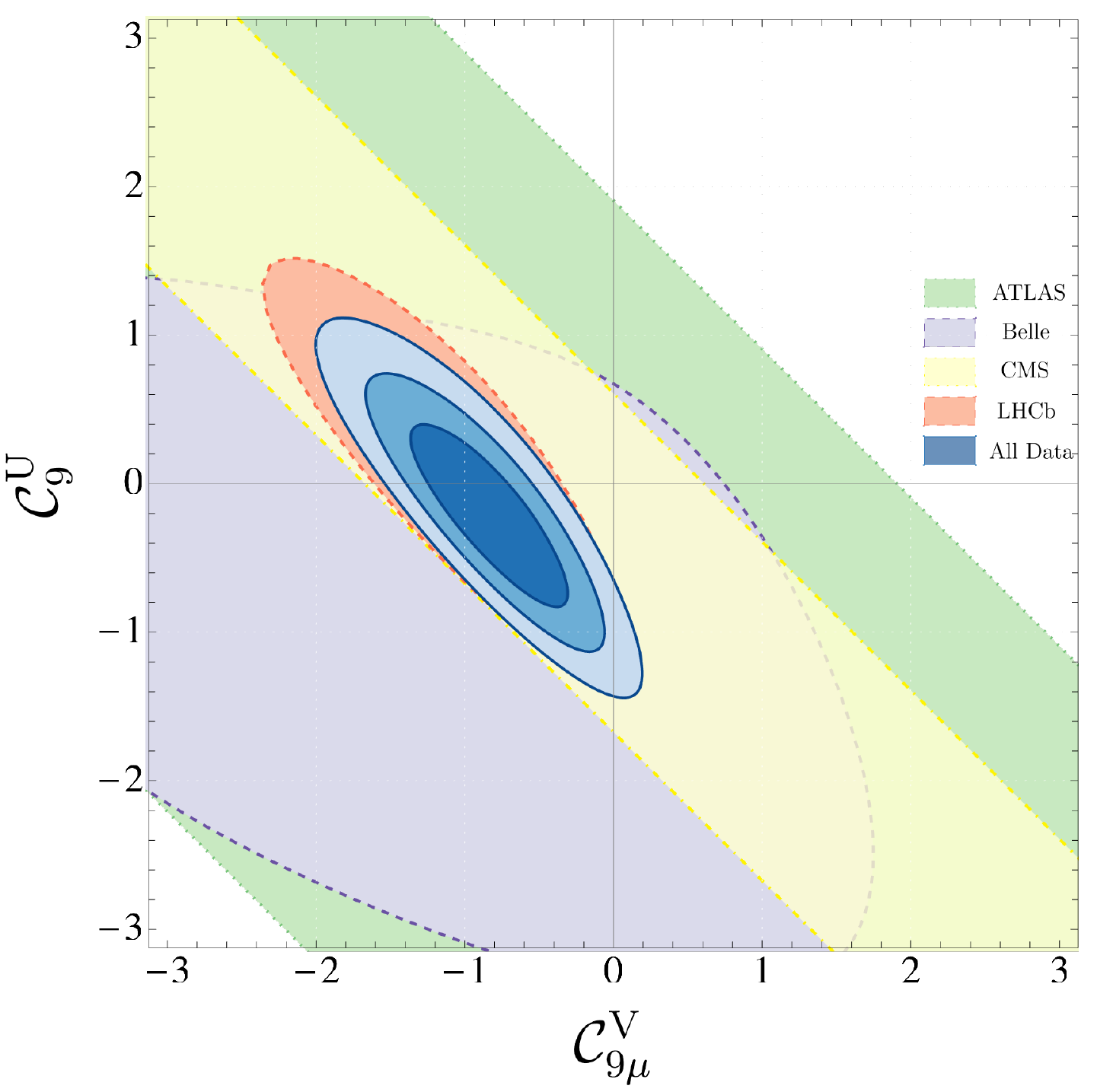}
\\
\includegraphics[width=0.315\textwidth]{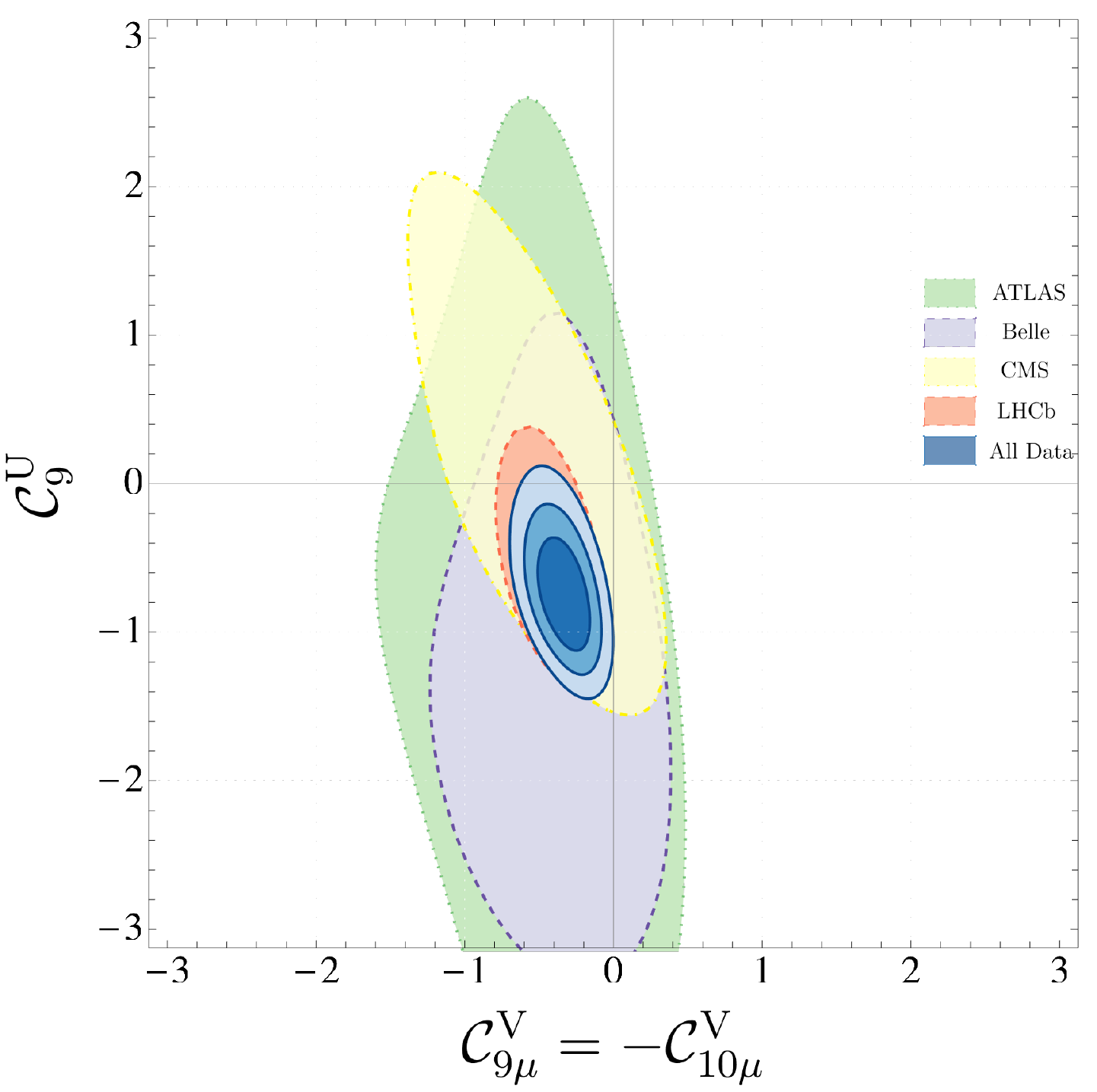}\hspace{5mm}
\includegraphics[width=0.315\textwidth]{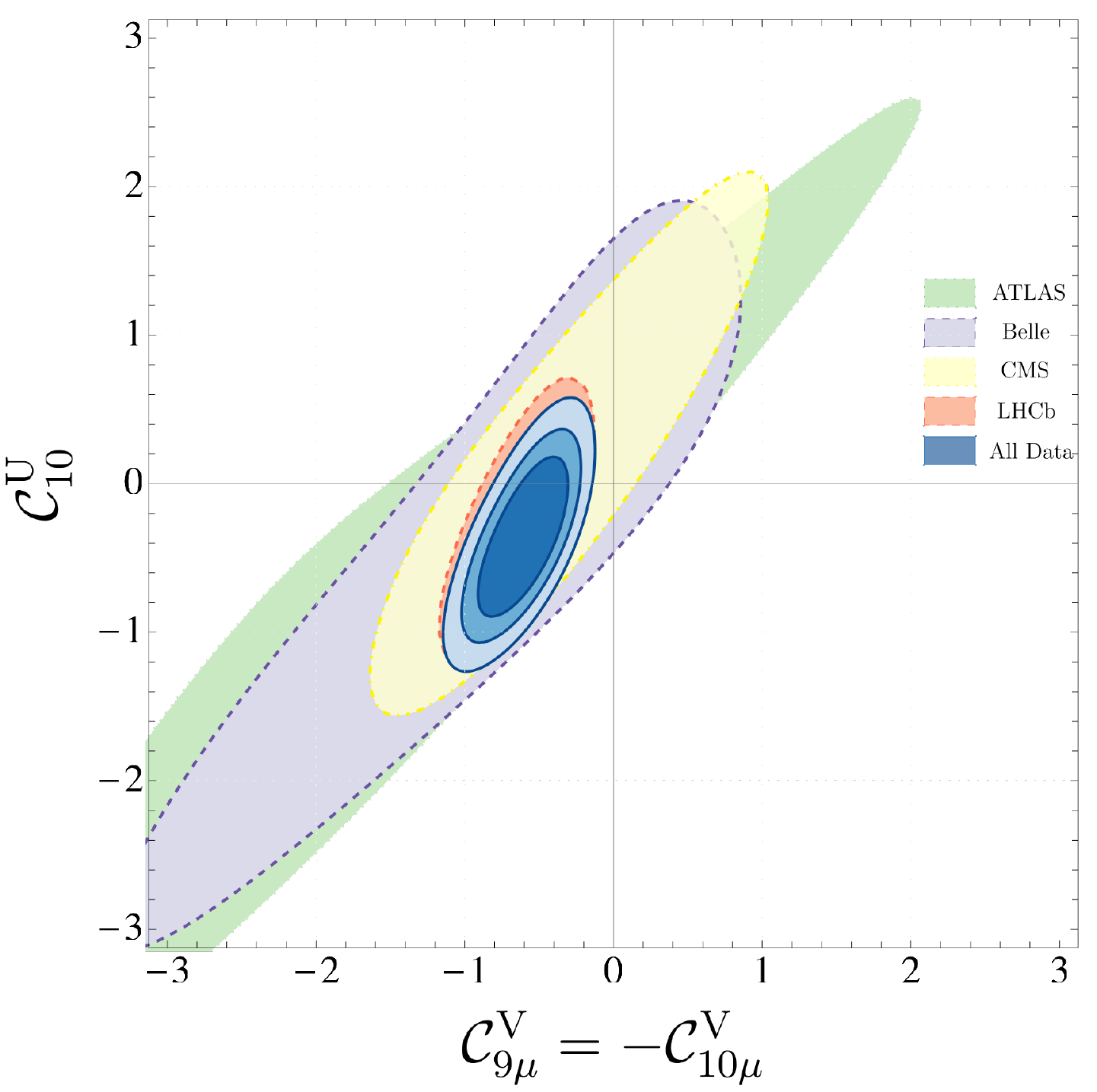}
\end{center}
\caption{Updated plots of Ref.~\cite{Alguero:2018nvb} corresponding to Scenarios 6,7,8 and the new Scenario 9.} \label{LFU1}
\end{figure*}

\begin{figure*}
\begin{center}
\includegraphics[width=0.315\textwidth]{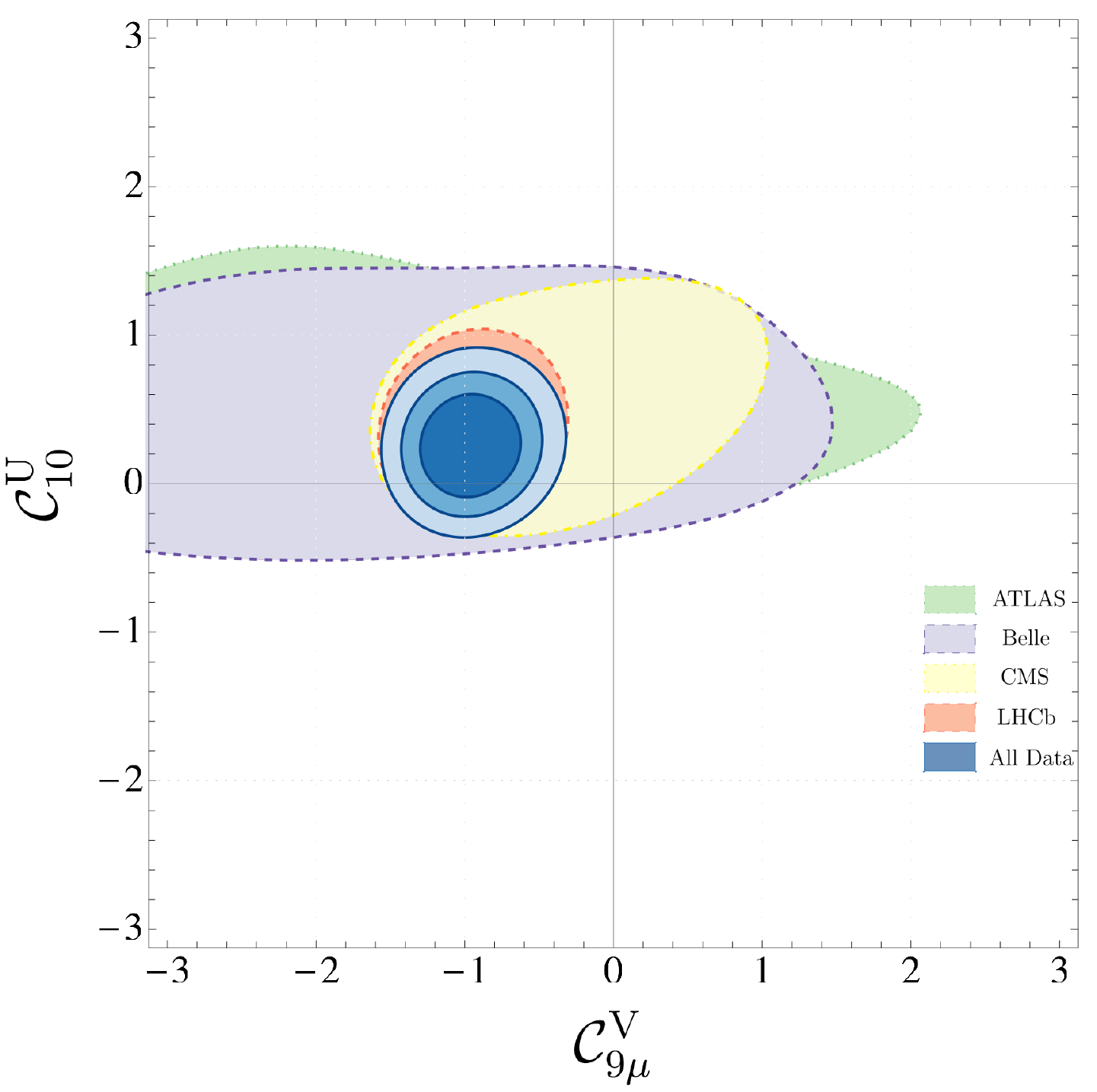} \hspace{2mm}
\includegraphics[width=0.315\textwidth]{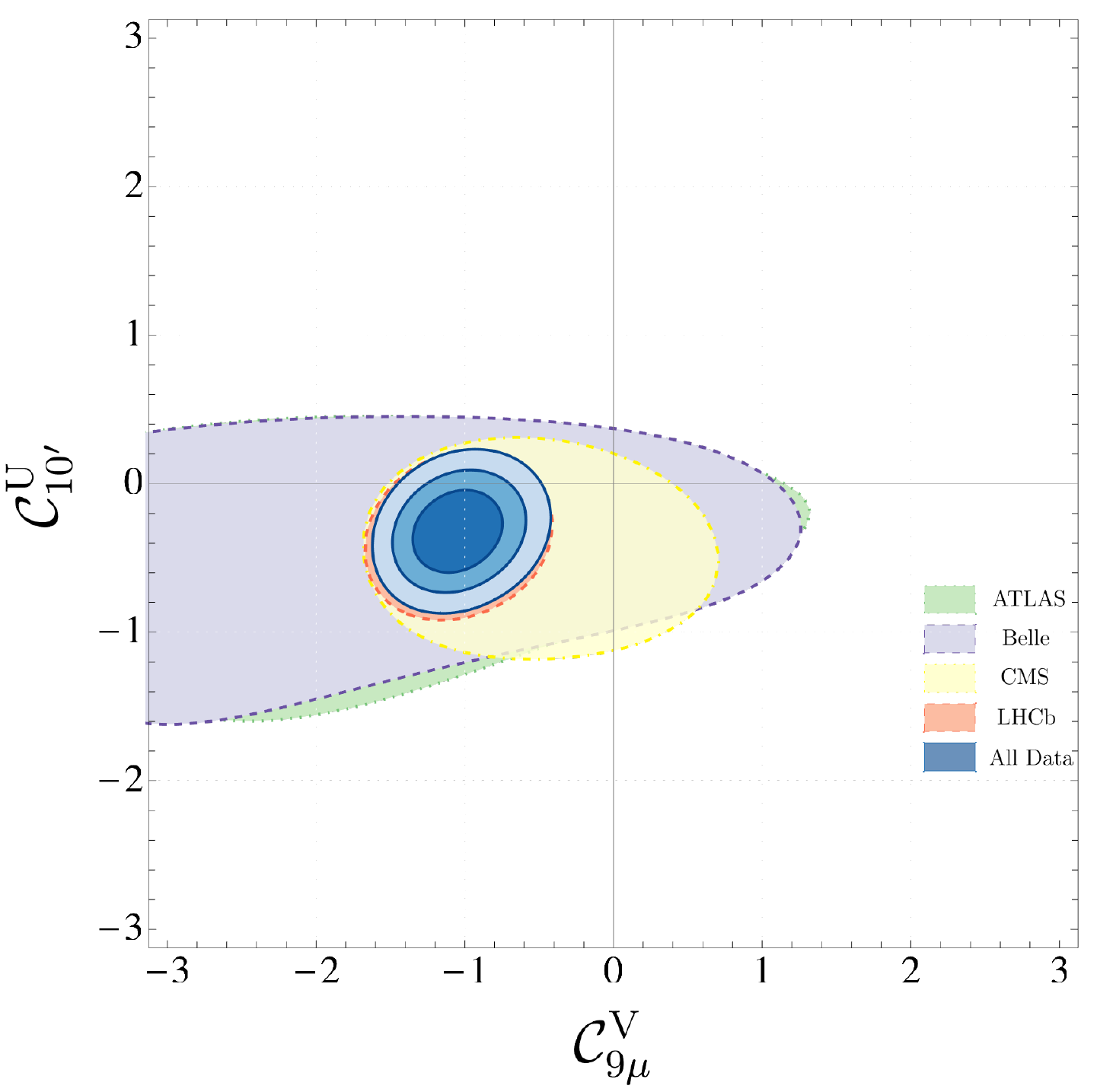}\hspace{2mm}
\includegraphics[width=0.315\textwidth]{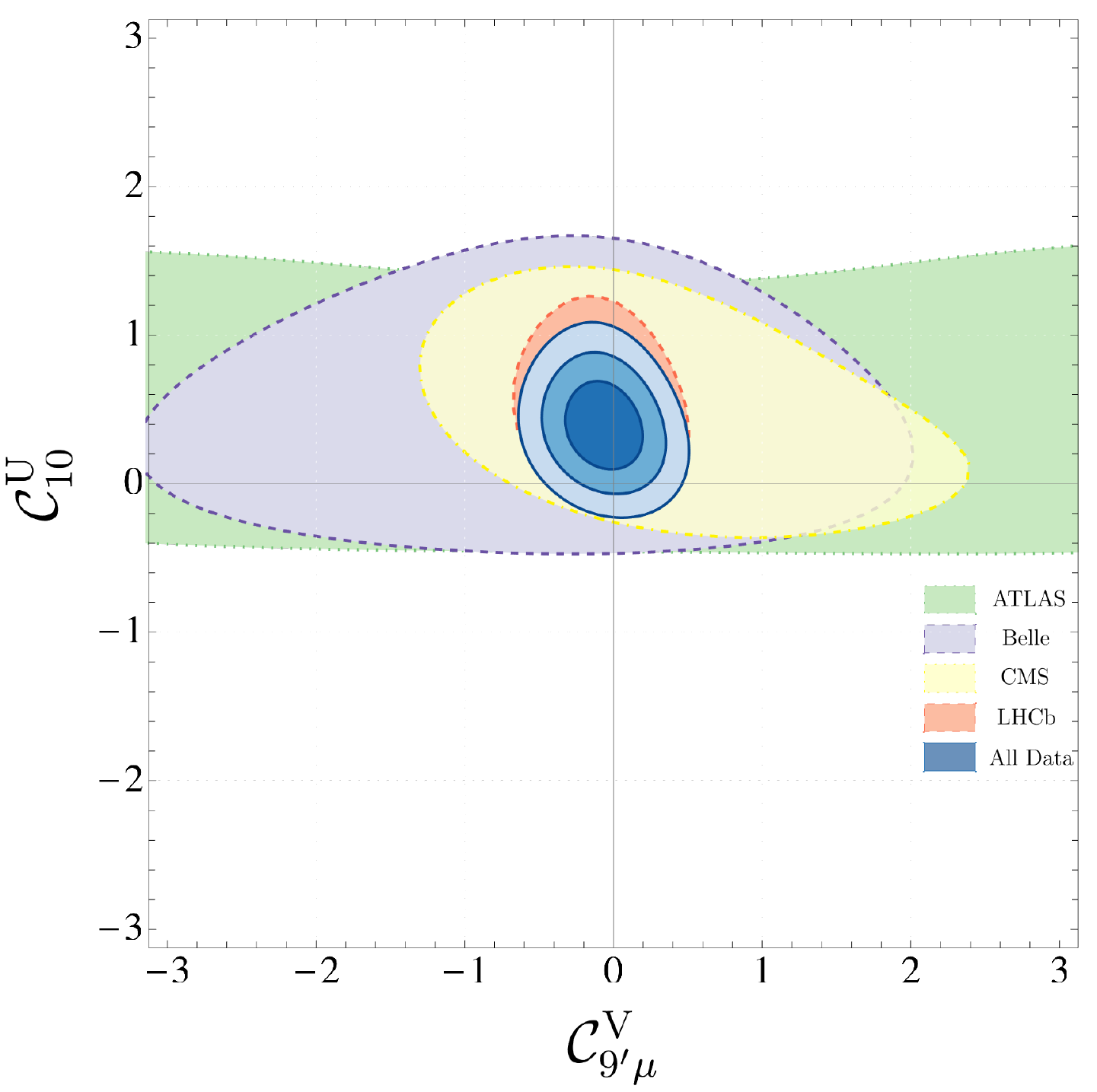}
\end{center}
\caption{Updated plots of Ref.~\cite{Alguero:2018nvb} corresponding to the new Scenarios 10,11,12. } \label{LFU2}
\end{figure*}

Concerning the 2D scenarios collected in Tab.~\ref{tab:results2D}, the same picture arises as in Ref.~\cite{Capdevila:2017bsm}, except that $\Cc{9e}^{\rm NP}$ is now basically zero and small contributions to RHC seem slightly favoured ($\Cc{9^\prime\mu}>0, \Cc{10'\mu}<0$)~\footnote{Interestingly, these small contributions also reduce slightly the mild tension between $P_5^\prime$ at large and low recoils pointed out in Ref.~\cite{Alguero:2019pjc} compared to the scenario with only $\Cc{9\mu}^{\rm NP}$.}. Indeed, these RHC contributions tend to increase the value
of $R_K^{[1.1,6]}$ while $\Cc{9\mu}^{\rm NP}<0$ tend to decrease it as can be seen from the explicit expression of $R_K^{[1.1,6]}=A_{\mu}/A_e$ where the numerator and the denominator can be given by an approximate polynomial parameterisation near the SM point
\begin{eqnarray}
A_\ell =&&\ \alpha_{0\ell} + \alpha_{1\ell}\,\Cc{9\ell}^{\rm NP} + \alpha_{2\ell}\,\left(\Cc{9\ell}^{\rm NP}\right)^2 + \alpha_{3\ell}\,\Cc{9'\ell}^{} \nonumber \\[2mm] &&\qquad+ \alpha_{4\ell}\,\left(\Cc{9'\ell}^{}\right)^2 + \alpha_{5\ell}\,\Cc{9\ell}^{\rm NP}\Cc{9'\ell}^{} \nonumber \\[2mm]
&&\qquad + \alpha_{6\ell}\,\Cc{10\ell}^{\rm NP} + \alpha_{7\ell}\,\left(\Cc{10\ell}^{\rm NP}\right)^2 \\[2mm]
&& + \alpha_{8\ell}\,\Cc{10'\ell}^{} + \alpha_{9\ell}\,\left(\Cc{10'\ell}^{}\right)^2 + \alpha_{10\ell}\,\Cc{10\ell}^{\rm NP}\Cc{10'\ell}^{}\nonumber
\end{eqnarray}
 with the coefficients provided in Tab.~\ref{tab:polpar} (for linearised expressions, see Refs.~\cite{Celis:2017doq,Alguero:2018nvb}). We introduce a new Hyp. 5 in Tab.~\ref{tab:results2D}. The comparison between Hyps.~4 and 5 shows that the scenario $\Cc{9'\mu}=-\Cc{10'\mu}$ (left-handed lepton coupling for right-handed quarks) prefers to be associated with $\Cc{9\mu}^{\rm NP}$ (vector lepton coupling for left-handed quarks) rather than $\Cc{9\mu}^{\rm NP}=-\Cc{10\mu}^{\rm NP}$ (left-handed lepton coupling for left-handed quarks).
Finally, no significant changes are observed in the 6D fit, except for the slight increase in the Pull$_{\rm SM}$, see Tab.~\ref{tab:Fit6D}.

With the updated data, little change is observed among the preferred 2D NP models. Nevertheless, with an $R_K^{[1.1,6]}$ value closer to one, scenarios with right-handed currents (RHC), namely $(\Cc{9\mu}^{\rm NP}, \Cc{9'\mu})$ and $(\Cc{9\mu}^{\rm NP}, \Cc{10'\mu})$, seem to emerge.
The first scenario is naturally generated in a $Z^\prime$ model with opposite couplings to right-handed and left-handed quarks and  was proposed in Ref.~\cite{Altmannshofer:2014cfa} within the context of a gauged $L_\mu-L_\tau$ symmetry with vector-like quarks. The latter (of masses $m_D$ and $m_Q$) are charged under $L_\mu-L_\tau$ and have the same SM quantum numbers as right-handed down quarks and left-handed quark doublets, respectively. The vector-like quarks couple to the SM ones and to a scalar $\phi$ which breaks the $L_\mu-L_\tau$ symmetry with couplings $Y^{D,Q}$. We show the update of Fig.~2 of Ref.~\cite{Altmannshofer:2014cfa} assuming $Y^{D,Q}=1$ in Fig.~\ref{fig:modelfits2}. Since the current fit allows for $\Cc{9'\mu}=0$ at the two sigma level, the  $SU(2)$ singlet vector-like quark can still be decoupled~\cite{Crivellin:2015mga}.

\section{Global fits in presence of LFUV and LFU NP}\label{sec:LFUV-LFU-NP}

We turn to scenarios that allow also for the presence of LFU NP~\cite{Alguero:2018nvb,Alguero:2019pjc} (in addition to LFUV contributions to muons only), leading to the value of the Wilson coefficients
\begin{equation}
\Cc{ie}=\Cc{i}^{\rm U}\,,\qquad \Cc{i\mu}=\Cc{i}^{\rm U}+\Cc{i}^{\rm V}\,.
\end{equation}
(with $i=9,10$) for $b\to se^+e^-$ and $b\to s\mu^+\mu^-$ transitions respectively.

We update some of the scenarios considered in Ref.~\cite{Alguero:2018nvb} in Tab.~\ref{Fit3Dbis}. Concerning new directions in parameter space we allow for RHC, motivated by the results of the previous section, and focus on scenarios that could be fairly easily obtained in simple NP models.

With the updated experimental inputs, we confirm our earlier result~\cite{Alguero:2018nvb} that a LFUV left-handed lepton coupling structure (corresponding to $\Cc{9}^{\rm V}=-\Cc{10}^{\rm V}$ and preferred from a model-building point of view)  yields a better description of data with the addition of LFU-NP in the coefficients $\Cc{9,10}$, as shown by the scenarios 6,8 in Tab.~\ref{Fit3Dbis} with $p$-values larger than $70\%$.

We observe a very slight decrease in significance for the scenarios 5--7, with the exception of scenario 8 which exhibits one of the most significant pulls with respect to the SM.

Scenario 8 of Ref.~\cite{Alguero:2018nvb} can actually be realized via off-shell photon penguins~\cite{Crivellin:2018yvo} in a leptoquark model explaining also $b\to c\tau\nu$ data (we will return to this point in the following section).

Updated plots of the 2D LFU-LFUV scenarios discussed in Ref. \cite{Alguero:2018nvb} are shown in Fig.~\ref{LFU1}.

The new scenarios 9--13 are characterized by a $\Cc{10(')}^{\rm U}$ contribution. This arises naturally in models with modified $Z$ couplings (to a good approximation $\Cc{9(')}^{\rm U}$ can be neglected).
The pattern of scenario 9 occurs in Two-Higgs-Doublet models where this flavour universal effect can be supplemented by a $\Cc{9}^{\rm V}=-\Cc{10}^{\rm V}$ effect~\cite{Crivellin:2019dun}.

In case of scenarios 11 to 13, one can invoke models with vector-like quarks where modified $Z$ couplings are even induced at tree level. The LFU effect in $\Cc{10(')}^{\rm U}$ can be accompanied by a $\Cc{9,10(')}^{\rm V}$ effect from $Z^\prime$ exchange~\cite{Bobeth:2016llm}.
Vector-like quarks with the quantum numbers of right-handed down quarks (left-handed quarks doublets) generate effect in $\Cc{10}^{\rm U}$ and $\Cc{9'}^{\rm V}$ ($\Cc{10(')}^{\rm U}$ and $\Cc{9}^{\rm V}$) for a $Z^\prime$ boson with vector couplings to muons~\cite{Bobeth:2016llm}.

The comparison of scenarios 10 and 12 illustrates that $\Cc{9\mu}^{\rm V}$ plays an important role in LFU NP scenarios and cannot be swapped for its chirally-flipped counterpart without consequences.
Finally, the allowed regions for the new LFU scenarios are displayed in  Fig.~\ref{LFU2}.

\begin{figure*}[t]
	\begin{center}
		\includegraphics[height=8.0cm]{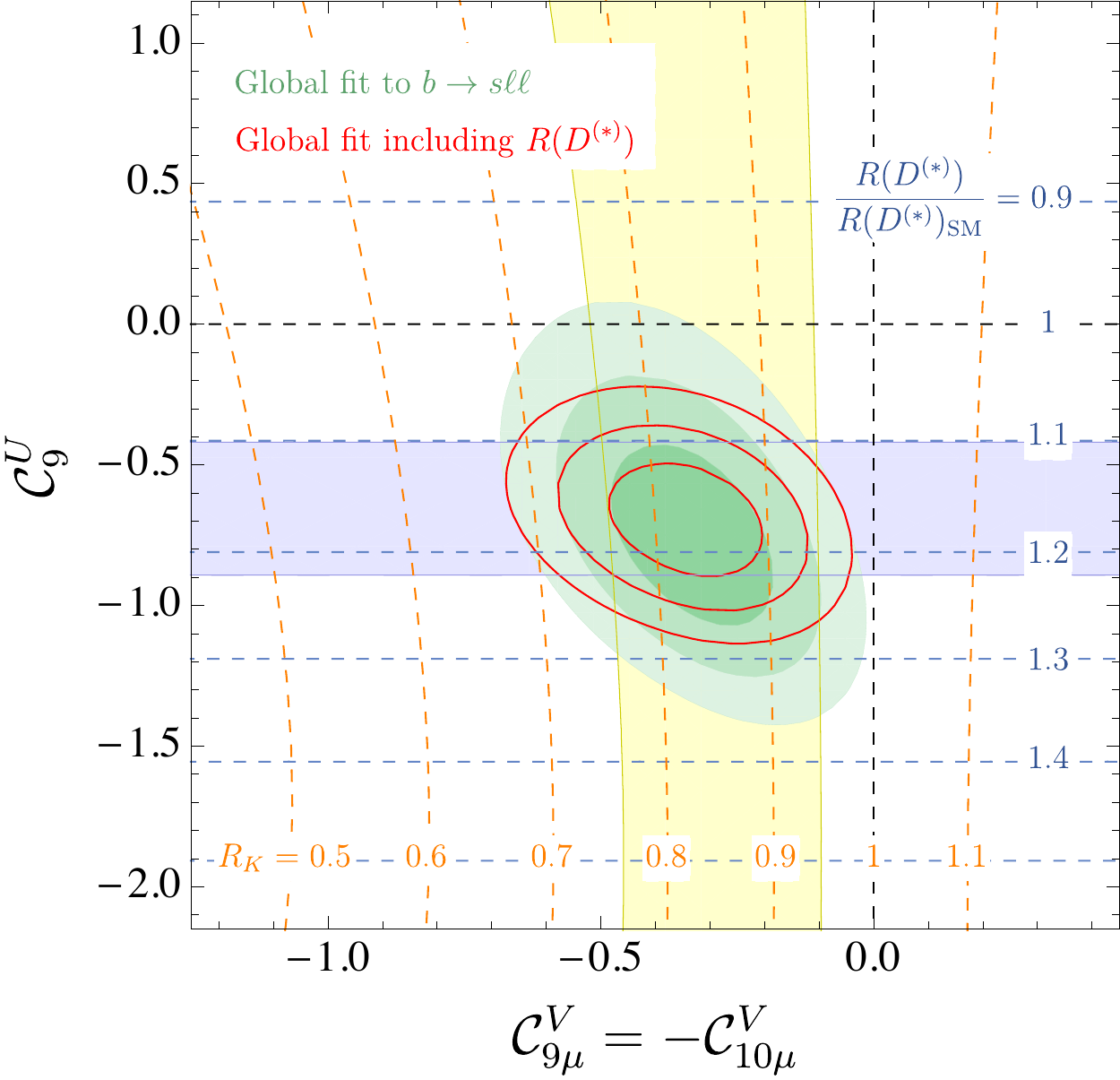}
		\hspace{5mm}
		\raisebox{2.6mm}{\includegraphics[height=7.86cm]{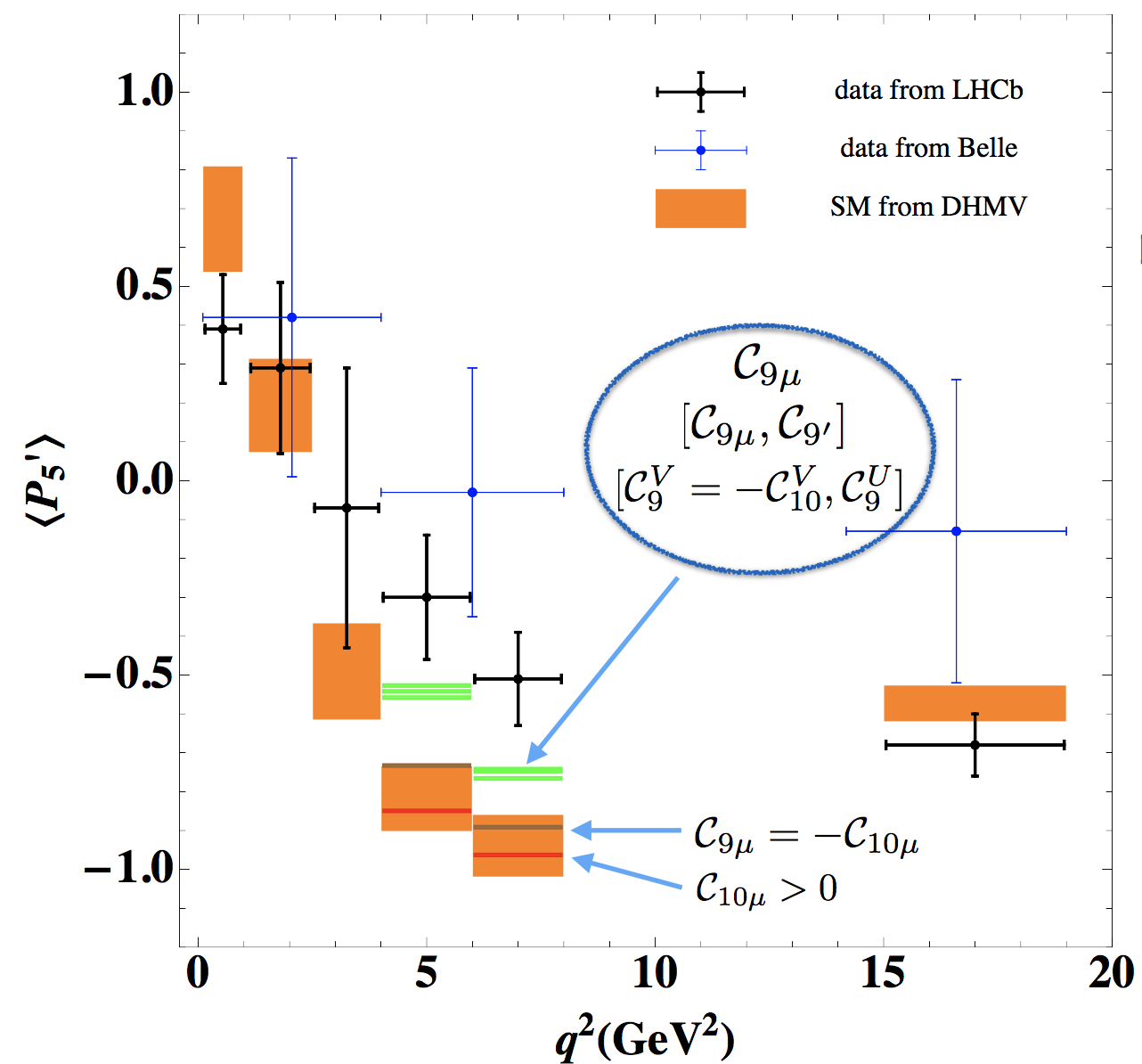}}
	\end{center}
	\caption{Left: Preferred regions at the 1, 2 and 3$\,\sigma$ level (green) in the $(\Cc{9\mu}^{\rm V}=-\Cc{10\mu}^{\rm V},\,\Cc{9}^{\rm U})$ plane from $b\to s\ell^+\ell^-$ data. The red contour lines show the corresponding regions once $R_{D^{(*)}}$ is included in the fit (for $\Lambda=2$~TeV). The horizontal blue (vertical yellow) band is consistent with $R_{D^{(*)}}$ ($R_{K}$) at the $2\,\sigma$ level and the contour lines show the predicted values for these ratios. Right:  Impact of favoured NP scenarios on the observable $P_5^\prime$. Only central values for the NP scenarios are displayed. The most interesting scenarios cluster together while traditional scenarios like $\Cc{9\mu}^{\rm NP}=-\Cc{10\mu}^{\rm NP}$ or the scenario $\Cc{10\mu}^{\rm NP}$ considered in Ref.~\cite{Aebischer:2019mlg} fail to explain this anomaly.}
	\label{fig:modelfits}
\end{figure*}


\section{Model-independent connection to $b\to c\ell\nu$}\label{sec:btoc}

In complement with the above EFT analysis, we focus now on the NP interpretation of scenario 8. Indeed, this scenario allows for a model-independent connection between the anomalies in $b\to s\ell^+\ell^-$ and  those in $b\to c\tau\nu$, which are now at the $3.1\sigma$ level~\cite{Amhis:2016xyh}.

Such a correlation arises in the SMEFT scenario where $\Cc{}^{(1)}=\Cc{}^{(3)}$ expressed in terms of gauge-invariant dimension-6 operators~\cite{Grzadkowski:2010es,Capdevila:2017iqn}. This scenario stems naturally from models with an  $SU(2)$ singlet vector leptoquark~\cite{Alonso:2015sja,Greljo:2015mma,Calibbi:2015kma}.
The operator involving-third generation leptons explains $R_{D^{(*)}}$ and the one involving the second generation gives a LFUV effect in $b\to s\mu^+\mu^-$ processes. The constraint from $b\to c\tau\nu$ and $SU(2)_L$ invariance leads generally to large contributions to the operator $\bar s \gamma^\mu P_Lb \bar \tau \gamma_\mu P_L \tau$, which enhances $b\to s\tau^+\tau^-$ processes~\cite{Capdevila:2017iqn}, but also mixes into ${\cal O}_9$ and generates $\Cc{9}^{\rm U}$ at $\mu=m_b$~\cite{Crivellin:2018yvo}. Note that not all models addressing the charged and neutral current anomalies simultaneously have an anarchic flavour structure. In fact, in the case of alignment in the down-sector~\cite{Blanke:2018sro,DiLuzio:2017vat} one does not find large effects in $b\to s\tau^+\tau^-$ or ${\cal C}_9^{\rm U}$.

Therefore, scenario 8 is reproduced in this setup with an additional correlation between $\Cc{9}^{\rm U}$ and $R_{D^{(*)}}$. Assuming a generic flavour structure so that small CKM elements can be neglected~\cite{Capdevila:2017iqn,Crivellin:2018yvo}, we get
\begin{equation}
\Cc{9}^{\rm U}\! \approx \! 7.5\left(1-\sqrt{\frac{R_{D^{(*)}}}{R_{D^{(*)}{\rm SM}}}}\right)\!\! \left(1+\frac{\log(\Lambda^2/(1{\rm TeV}^2))}{10.5}\right). \\  \\
\end{equation}

Realizations of this scenario in specific NP models yield also an effect in $\Cc{7}$ generally~\cite{Crivellin:2018yvo}. However, since this effect is model dependent (and in fact small in some UV complete models~\cite{Calibbi:2017qbu,Bordone:2017bld}), we neglect it here, leading to the left plot in Fig.~\ref{fig:modelfits}, where we include the recent update of Ref.~\cite{Abdesselam:2019dgh} to draw the band for $R_{D(*)}$. Note that this scenario has a pull of {7.0$\,\sigma$} due to the inclusion of $R_{D^{(*)}}$, which increases $\Delta\chi^2$ by~$\sim20$.

\section{Conclusions}\label{sec:conc}

In summary, including recent updates ($R_K$, $R_{K^*}$ and ${\cal B}(B_s \to\mu^+\mu^-)$) our global model-independent analysis yields a very similar picture to the one previously found in Refs.~\cite{Capdevila:2017bsm,Alguero:2018nvb} for the various NP scenarios of interest with some important peculiarities. In presence of LFUV NP contributions only, the 1D fits to ``All'' observables remain basically unchanged showing the preference for $\Cc{9\mu}^{\rm NP}$ scenario over $\Cc{9\mu}^{\rm NP}=-\Cc{10\mu}^{\rm NP}$. If only LFUV observables are considered the situation is reversed, as already found in Ref.~\cite{Capdevila:2017bsm}, but now with an increased gap between the significances. This difference between the preferred hypotheses, depending on the data set used, can be solved introducing LFU NP contributions~\cite{Alguero:2018nvb}.

The main differences arise for the 2D scenarios: the cases including RHC, ($\Cc{9\mu}^{\rm NP}, \Cc{10^\prime\mu}$), ($\Cc{9\mu}^{\rm NP}, \Cc{9^\prime\mu}$) or ($\Cc{9\mu}^{\rm NP}, \Cc{9^\prime\mu}=-\Cc{10^\prime\mu}$), can accommodate better the recent updates, which enhances the significance of these scenarios compared to Ref.~\cite{Capdevila:2017bsm}, pointing to new patterns including RHC. A more precise experimental measurement of the observable $P_1$\cite{kruger, complete}  would be very useful to confirm or not the presence of RHC NP encoded in $\Cc{9^\prime\mu}$ and $\Cc{10^\prime\mu}$.

We also observe interesting changes in the 2D fits in the presence of LFU NP, where new scenarios (not considered in Ref.~\cite{Alguero:2018nvb}) give a good fit to data with $\Cc{10^{(\prime)}}^{\rm U}$ and additional LFUV contributions. For example scenario 11 ($\Cc{9\mu}^{\rm V}, \Cc{10^\prime\mu}$) can accommodate $b\to s\ell^+\ell^-$ data very well, at the same level as
scenario 8. Scenarios including LFU NP in left-handed currents (discussed in Ref.~\cite{Alguero:2018nvb}) stay practically unchanged but with some preference for scenarios 6 and 8, which have a $(V-A)$ structure for the LFUV-NP and a $V$ or $(V+A)$ structure for the LFU-NP. Furthermore, we have included additional scenarios 9 and 10 that exhibit a significance of 5.0$\sigma$ and 5.5$\sigma$ respectively.

We note that the amount of LFU NP is sensitive to the structure of the LFUV component. For instance, in scenario 7 ($\Cc{9\mu}^{\rm V}$ and $\Cc{9}^{\rm U}$) the LFU component is negligible at its best fit point. On the contrary, if the LFUV-NP has a $(V-A)$ structure, the LFU-NP component  ($\Cc{9}^{\rm U}$) is  large, as illustrated by scenarios 6, 8 and 9. Scenarios with NP in RHC (either LFU or LFUV) prefer such contributions at the $2\sigma$ level (see scenarios 11 and 13) with the exception of scenario 12 with negligible $\Cc{9'\mu}^{\rm V}$.  The new values of $R_K$ and $R_{K^*}$ seem thus to open a window for RHC contributions while the new ${\cal B}(B_s \to \mu\mu)$ update (theory and experiment) helps only marginally scenarios with $\Cc{10\mu}^{\rm NP}$.

Finally, we showed that scenario 8, which allows for a model-independent connection between the $b\to c\tau \nu$ anomalies and the ones in $b\to s\ell^+\ell^-$, can explain all data consistently and is preferred over the SM by $7\,\sigma$.

Fig.~\ref{fig:modelfits} illustrates the impact on the largest anomaly ($P_5^\prime$) of some of the most significant scenarios. Interestingly, several of the scenarios currently favoured cluster around the same values for the bins showing deviations with respect to the SM.

We have thus identified a number of NP scenarios with similarly good $p$-values and pulls with respect to the SM, which are able to reproduce the $b\to s\ell^+\ell^-$ data very well. Hierarchies among these scenarios can be identified, but additional data and reduced uncertainties are required to come to a final conclusion. The full exploitation of LHC run-2 data by the LHCb experiment (as well as by ATLAS and CMS) and the forthcoming results from the Belle and Belle II collaborations are expected to improve the situation very significantly in the forthcoming years, helping us to pin down the actual NP pattern hinted at by the $b\to s\ell^+\ell^-$ anomalies currently observed and to build accurate phenomenological models to be confirmed through other experimental probes such as direct production experiments.

\vspace{0.3cm}

\emph{Note added:} After the completion of this work, several global analyses have been performed to assess NP scenarios affecting $b\to s\ell^+\ell^-$ processes~\cite{Aebischer:2019mlg,Ciuchini:2019usw,Alok:2019ufo,Arbey:2019duh}. They agree well with our findings, with small differences stemming mainly from slightly different theoretical approaches as well as theoretical and experimental inputs. The improvement brought
by RHC has been observed in Refs.~\cite{Alok:2019ufo,Ciuchini:2019usw}, whereas the interest of LFU NP contributions is also identified in Refs.~\cite{Aebischer:2019mlg,Ciuchini:2019usw,London}. Most of the analyses observe that the slight deviation from ${\cal B}(B_s\to\mu^+\mu^-)$ plays no specific role in the global fit~\cite{Arbey:2019duh,Alok:2019ufo}, apart from Ref.~\cite{Aebischer:2019mlg}. In the latter analysis, the significance of a scenario with only $\Cc{10\mu}^{\rm NP}$ is much more important than in our case, and the hierarchies between the significances of 2D scenarios is different. After discussion with the authors of Ref.~\cite{Aebischer:2019mlg}, this difference comes from their inclusion of
$B_s$-$\bar{B}_s$ mixing and the assumption that $\Delta F=2$ observables are purely governed by the SM, which helps them sharpening the prediction for ${\cal B}(B_s\to\mu^+\mu^-)$ and increase the weight of this observable in the fit. Our present analysis does not rely on this strong hypothesis, which should be contrasted with the fact that most models invoked to explain $b\to s\ell^+\ell^-$ anomalies typically affect also $\Delta F=2$ observables.

\begin{acknowledgments}

We warmly thank M. Misiak for useful discussions on the theoretical update of ${\cal B}_{B_s\to \mu\mu}$
discussed in App.~\ref{app:addendum}.
This work received financial support from European Regional Development Funds under the Spanish Ministry of Science, Innovation and Universities (projects FPA2014-55613-P and FPA2017-86989-P) and from the Agency for Management of University and Research Grants of the Government of Catalonia (project SGR 1069) [MA, BC, PM, JM] and from European Commission (Grant Agreements 690575, 674896 and 69219) [SDG]. The work of PM is supported by the Beatriu de Pinos postdoctoral program co-funded by the Agency for Management of University and Research Grants of the Government of Catalonia and by the COFUND program of the Marie Sklodowska-Curie actions under the framework program Horizon 2020 of the European Commission.  JM gratefully acknowledges the financial support by ICREA under the ICREA Academia programme. JV acknowledges funding from the European Union's Horizon 2020 research and innovation programme under the Marie Sklodowska-Curie grant agreement No 700525, `NIOBE' and support from SEJI/2018/033 (Generalitat Valenciana). The work of AC is supported by a Professorship Grant (PP00P2\_176884) of the Swiss National Science Foundation.

\end{acknowledgments}

\appendix

\section{Correlations among fit parameters}

In addition to the confidence regions provided for the various scenarios in this article, we display here the correlation matrices for the most interesting NP scenarios.

\subsection{Correlation Matrices of Fits to LFUV NP}

First, we present the correlations between fit parameters of the NP scenarios defined in Tab. II and Tab. III. These are all NP solutions whose parameters assess LFUV NP.

By order of appearance in Tab. II, the correlations between the coefficients of all 2D scenarios with $\text{Pull}_\text{SM}\gtrsim 5.3\sigma$ are,

\[%
\text{Corr}(\mathcal{C}_{9\mu}^\text{NP},\mathcal{C}_{10\mu}^\text{NP})= \begin{pmatrix}%
1.00&0.30\\%
0.30&1.00%
\end{pmatrix}%
\]%

\[%
\text{Corr}(\mathcal{C}_{9\mu}^\text{NP},\mathcal{C}_{9'\mu})= \begin{pmatrix}%
1.00&-0.39\\%
-0.39&1.00%
\end{pmatrix}%
\]%

\[%
\text{Corr}(\mathcal{C}_{9\mu}^\text{NP},\mathcal{C}_{10'\mu})= \begin{pmatrix}%
1.00&0.33\\%
0.33&1.00%
\end{pmatrix}%
\]%

\[%
\text{Corr}(\mathcal{C}_{9\mu}^\text{NP},\mathcal{C}_{9e}^\text{NP})= \begin{pmatrix}%
1.00&0.51\\%
0.51&1.00%
\end{pmatrix}%
\]%

\[%
\text{Corr}(\mathcal{C}_{9\mu}^\text{NP}=-\mathcal{C}_{9'\mu},\mathcal{C}_{10\mu}^\text{NP}=\mathcal{C}_{10'\mu})= \begin{pmatrix}%
1.00&-0.17\\%
-0.17&1.00%
\end{pmatrix}%
\]%

\[%
\text{Corr}(\mathcal{C}_{9\mu}^\text{NP},\mathcal{C}_{9'\mu}=-\mathcal{C}_{10'\mu})= \begin{pmatrix}%
1.00&-0.34\\%
-0.34&1.00%
\end{pmatrix}%
\]%

The last two matrices correspond to Hyp. 1 and Hyp. 5 as defined in Tab. II. Despite the high $\text{Pull}_\text{SM}$ of the 2D scenario $\{\mathcal{C}_{9\mu}^\text{NP},\mathcal{C}_{7'}\}$ ($5.4\sigma$), its correlation matrix is not collected here due to the value of $\mathcal{C}_{7'}$ being negligible, with tiny errors.

Regarding the 6D fit of Tab. III,

\[%
\text{Corr}_\text{6D}= \begin{pmatrix}%
1.00&-0.34&-0.07&0.06&0.02&-0.03\\%
-0.34&1.00&0.24&-0.06&0.04&0.24\\%
-0.07&0.24&1.00&-0.13&0.61&0.59\\%
0.06&-0.06&-0.13&1.00&-0.13&-0.08\\%
0.02&0.04&0.61&-0.13&1.00&0.85\\%
-0.03&0.24&0.59&-0.08&0.85&1.00%
\end{pmatrix}%
\]

where the columns are ordered as $\{\mathcal{C}_{7}^\text{NP},\mathcal{C}_{9\mu}^\text{NP},\mathcal{C}_{10\mu}^\text{NP},\mathcal{C}_{7'},\mathcal{C}_{9'\mu},\mathcal{C}_{10'\mu}\}$.

Interesting information can be extracted from $\text{Corr}_\text{6D}$. Most of the coefficients do not show particularly strong correlations with the others except for the pairs $\{\mathcal{C}_{10\mu}^\text{NP},\mathcal{C}_{9'\mu}\}$, $\{\mathcal{C}_{10\mu}^\text{NP},\mathcal{C}_{10'\mu}\}$ and $\{\mathcal{C}_{9'\mu},\mathcal{C}_{10'\mu}\}$, being the latter the highest in correlation. While $\mathcal{C}_{9\mu}^\text{NP}$ and $\mathcal{C}_{9'\mu}$ show a non-negligible correlation in the fit to these coefficients only, in the 6D fit the aforementioned parameters are uncorrelated to a large extent. On the contrary, the correlation between $\mathcal{C}_{9\mu}^\text{NP}$ and $\mathcal{C}_{10\mu}^\text{NP}$ is very similar for both the global 6D and the 2D fit to these parameters alone.

\vspace{0.4cm}

\subsection{Correlation Matrices of Fits to LFUV-LFU NP}

Second, the correlations between fit parameters of scenarios with both LFUV and LFU NP have also been considered. Below one can find the correlation matrices of scenarios 5 to 11, in that order.

\[%
\text{Corr}(\mathcal{C}_{9\mu}^\text{V},\mathcal{C}_{9}^\text{U}=\mathcal{C}_{10}^\text{U},\mathcal{C}_{10\mu}^\text{V})= \begin{pmatrix}%
1.00&-0.93&0.91\\%
-0.93&1.00&-0.94\\%
0.91&-0.94&1.00%
\end{pmatrix}%
\]%

\[%
\text{Corr}(\mathcal{C}_{9\mu}^\text{V}=-\mathcal{C}_{10\mu}^\text{V},\mathcal{C}_{9}^\text{U}=\mathcal{C}_{10}^\text{U})= \begin{pmatrix}%
1.00&0.17\\%
0.17&1.00%
\end{pmatrix}%
\]%

\[%
\text{Corr}(\mathcal{C}_{9\mu}^\text{V},\mathcal{C}_{9}^\text{U})= \begin{pmatrix}%
1.00&-0.85\\%
-0.85&1.00%
\end{pmatrix}%
\]%

\[%
\text{Corr}(\mathcal{C}_{9\mu}^\text{V}=-\mathcal{C}_{10\mu}^\text{V},\mathcal{C}_{9}^\text{U})= \begin{pmatrix}%
1.00&-0.44\\%
-0.44&1.00%
\end{pmatrix}%
\]%

\[%
\text{Corr}(\mathcal{C}_{9\mu}^\text{V}=-\mathcal{C}_{10\mu}^\text{V},\mathcal{C}_{10}^\text{U})= \begin{pmatrix}%
1.00&0.69\\%
0.69&1.00%
\end{pmatrix}%
\]%

\[%
\text{Corr}(\mathcal{C}_{9\mu}^\text{V},\mathcal{C}_{10}^\text{U})= \begin{pmatrix}%
1.00&0.05\\%
0.05&1.00%
\end{pmatrix}%
\]%

\[%
\text{Corr}(\mathcal{C}_{9\mu}^\text{V},\mathcal{C}_{10'}^\text{U})= \begin{pmatrix}%
1.00&0.20\\%
0.20&1.00%
\end{pmatrix}%
\]%

No significant changes can be observed when comparing with the results in App. 2 of Ref.~\cite{Alguero:2018nvb}. As expected, $\mathcal{C}_{9\mu}^\text{V}$ and $\mathcal{C}_{9}^\text{U}$ are highly anti-correlated, with its nominal value somewhat smaller than in~\cite{Alguero:2018nvb}. Fit estimates of the parameters in scenario $\{\mathcal{C}_{9\mu}^\text{V}=-\mathcal{C}_{10\mu}^\text{V},\mathcal{C}_{9}^\text{U}=\mathcal{C}_{10}^\text{U}\}$ are now slightly correlated, while before their correlation was negligible. Interestingly, however, we find the parameters of the new scenario $\{\mathcal{C}_{9\mu}^\text{V},\mathcal{C}_{10}^\text{U}\}$ statistically independent up to a large extent.

\newpage




\section{State-of-the-art $b \to s \ell \ell$ global fits in March 2020 (Addendum)} \label{app:addendum}

\begin{figure*}
\begin{center}
\includegraphics[width=0.315\textwidth]{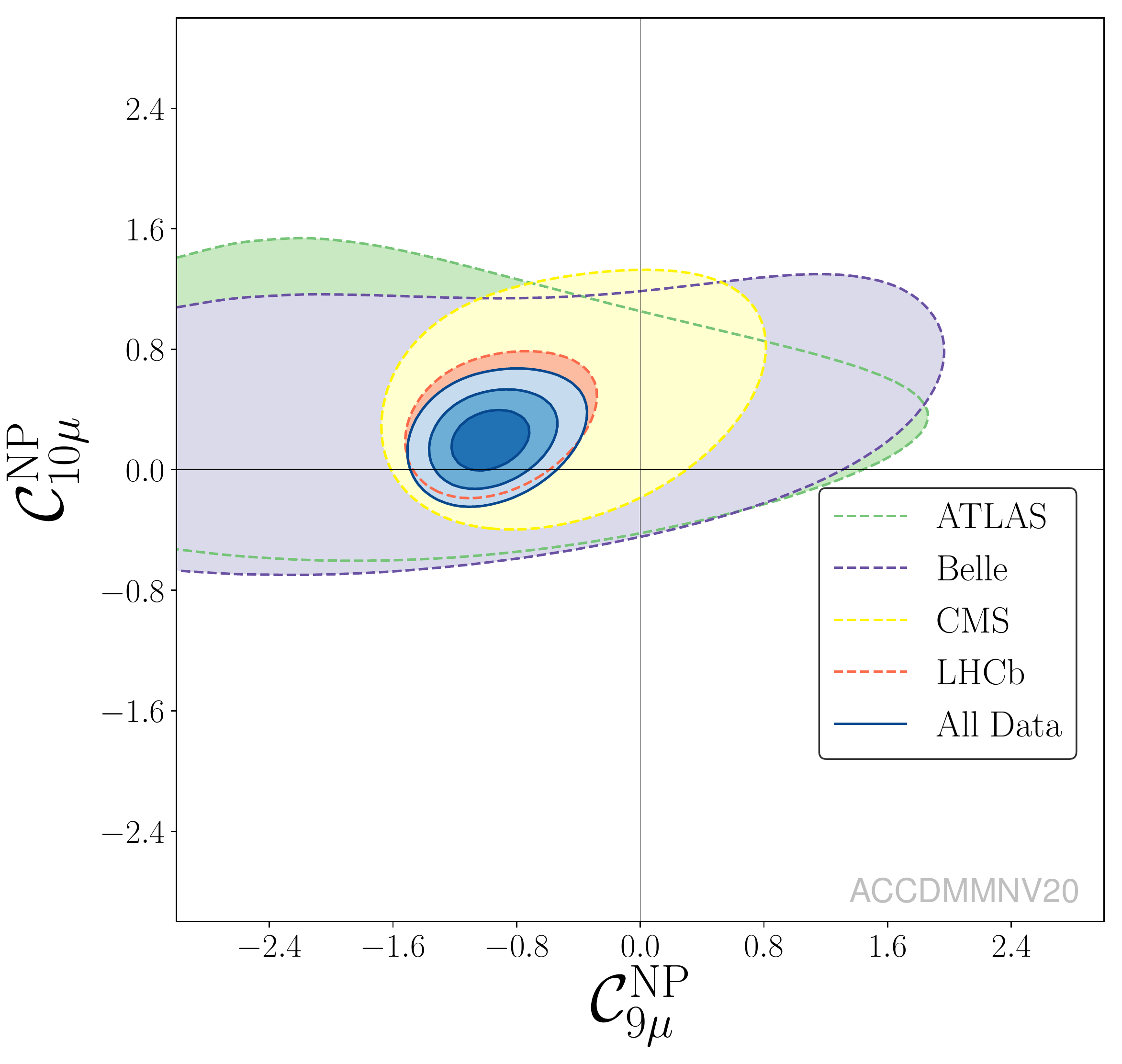}\hspace{2mm}
\includegraphics[width=0.315\textwidth]{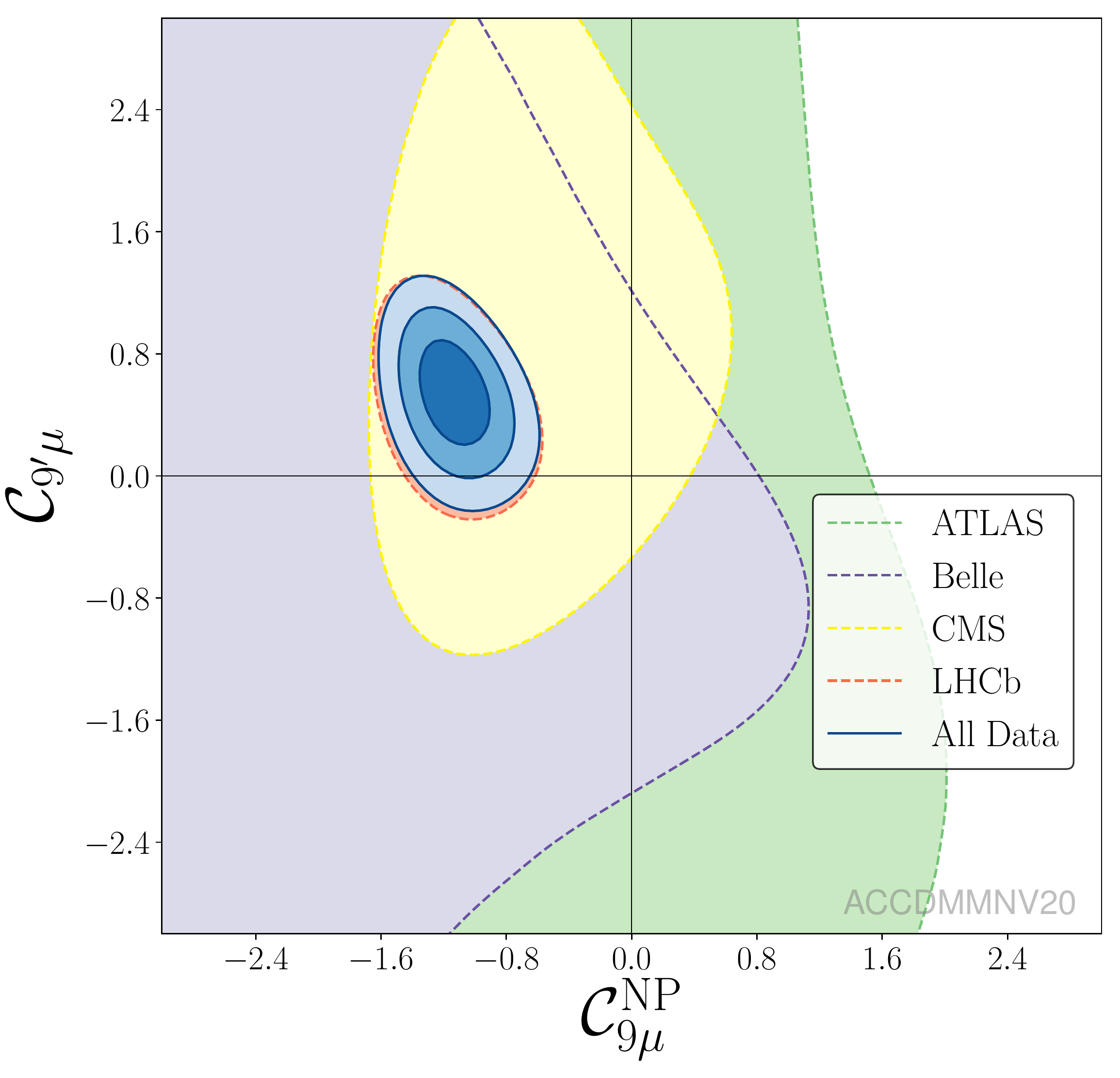}\hspace{2mm}
\includegraphics[width=0.315\textwidth]{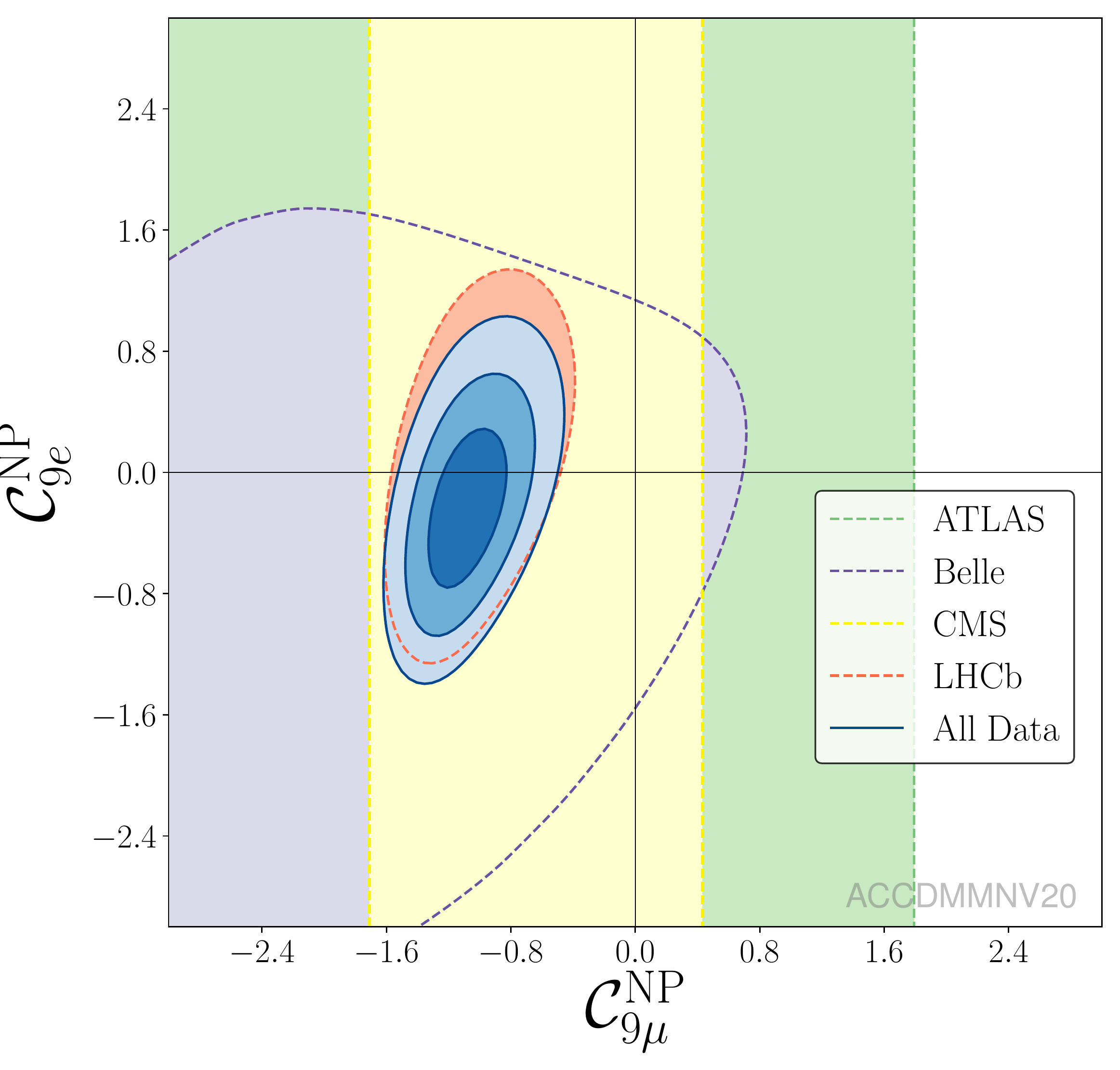}
\includegraphics[width=0.315\textwidth]{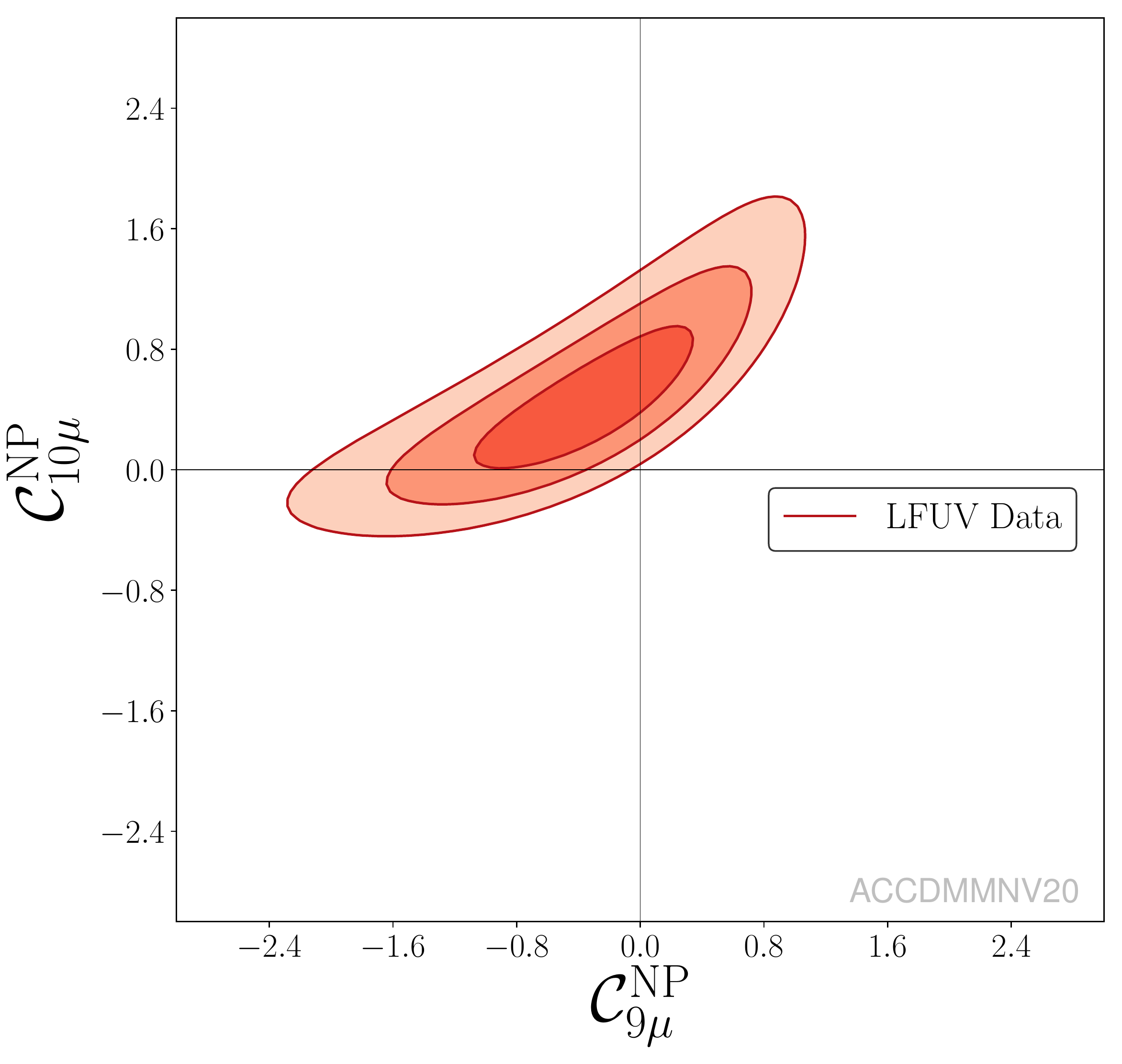}\hspace{2mm}
\includegraphics[width=0.315\textwidth]{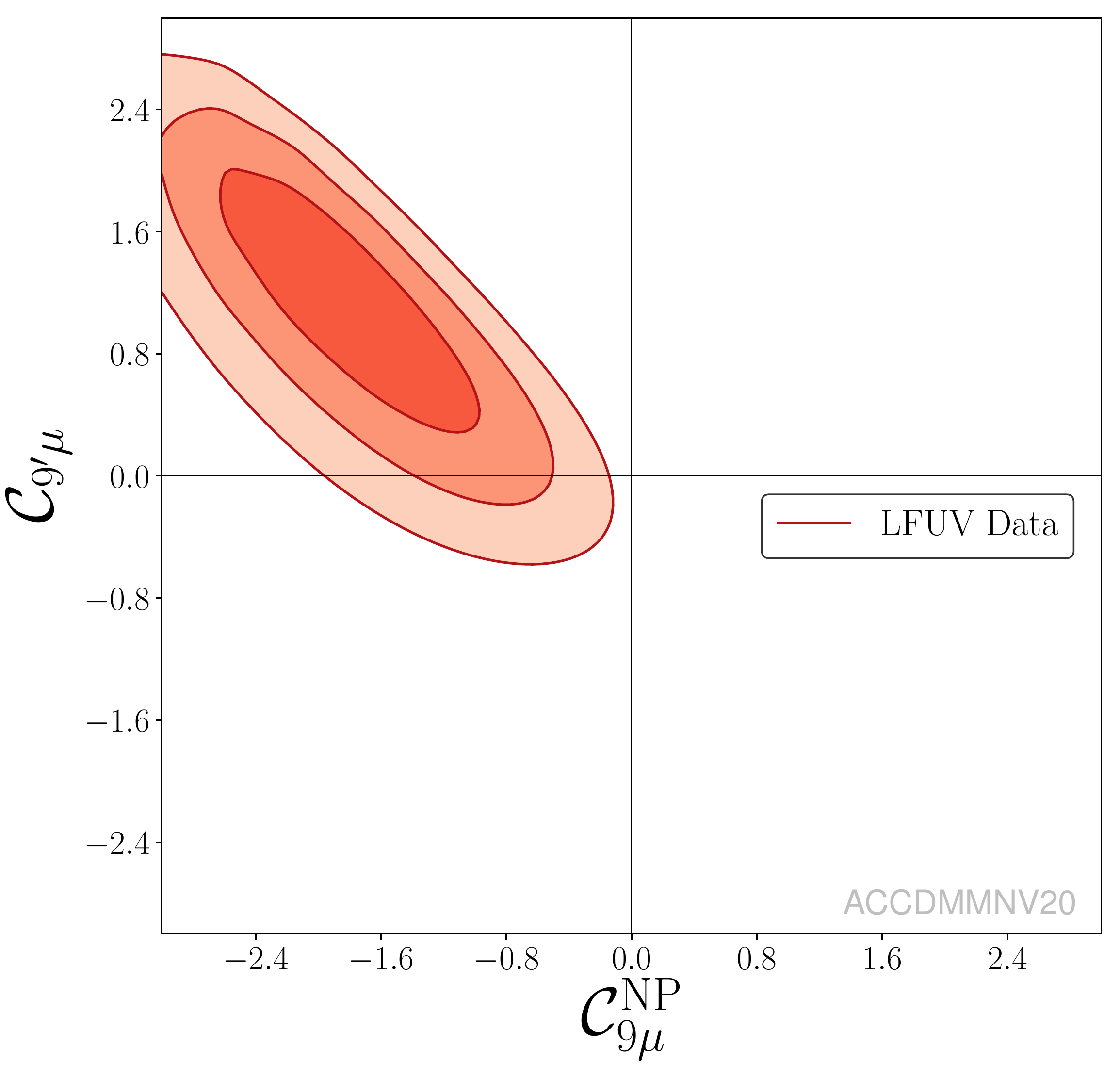}\hspace{2mm}
\includegraphics[width=0.315\textwidth]{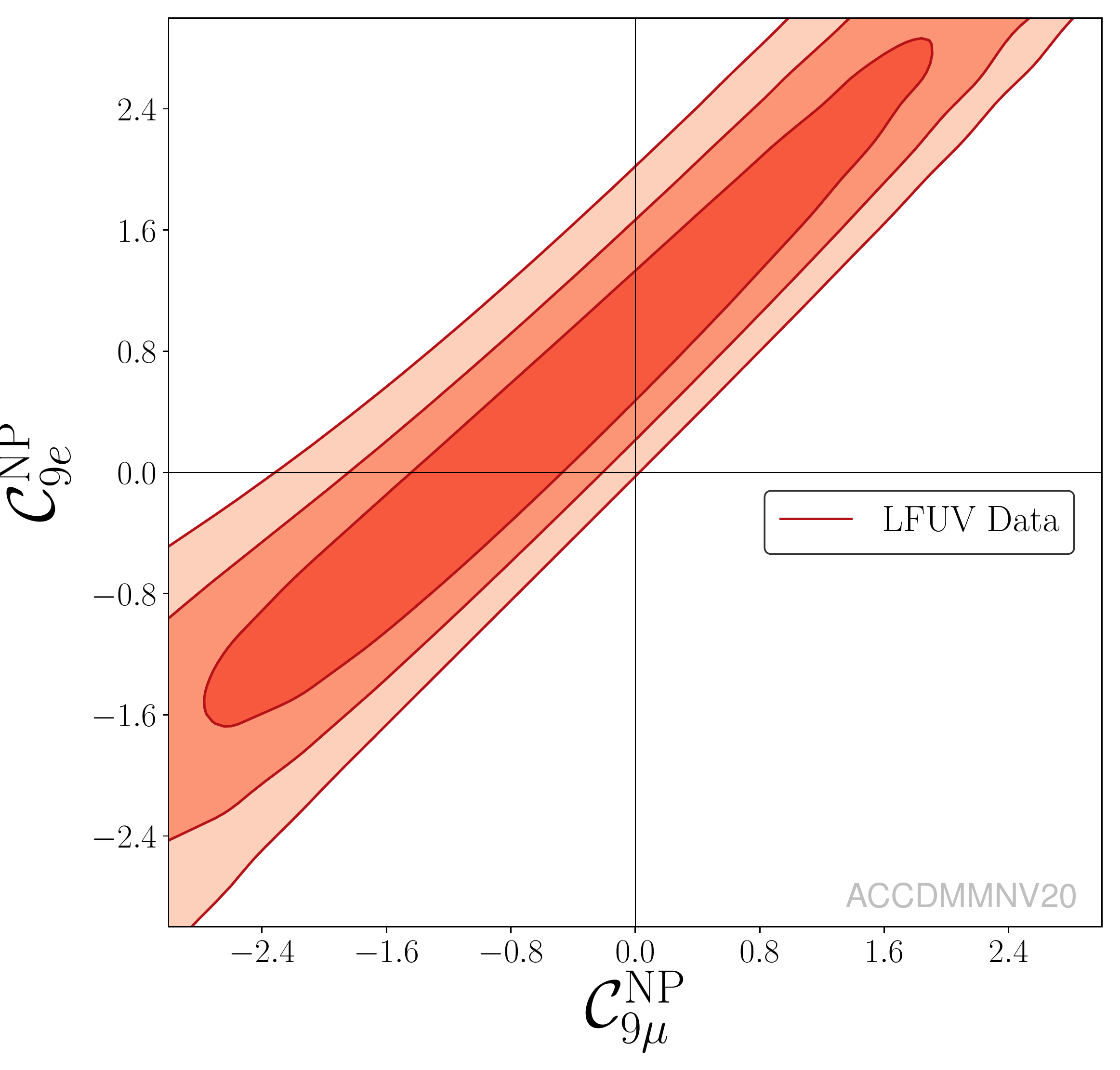}
\end{center}
\caption{From left to right: Allowed regions in the $(\Cc{9\mu}^{\rm NP},\Cc{10\mu}^{\rm NP})$, $(\Cc{9\mu}^{\rm NP},\Cc{9^\prime\mu})$ and $(\Cc{9\mu}^{\rm NP},\Cc{9e}^{\rm NP})$ planes for the corresponding 2D hypotheses, using all available data (fit ``All'') upper row or LFUV fit  lower row.}
\label{fig:FitResultAllApp}
\end{figure*}

This addendum updates the results presented in the main text and in Ref.~\cite{Alguero:2018nvb} after including the most recent $B \to K^*\mu\mu$ angular distribution data from the LHCb collaboration~\cite{lhcbupdated}, released in March 2020. As such, the tables and figures presented in the following supersede the ones in the main text of this article:
\begin{itemize}
\item Fig.~\ref{fig:FitResultAll} is superseded by Fig.~\ref{fig:FitResultAllApp}, Fig.~\ref{fig:modelfits2} by Fig.~\ref{fig:modelfits2App} (left), Fig.~\ref{fig:modelfits} (right) by Fig.~\ref{fig:12App} (left), Fig.~\ref{fig:modelfits} (left) by Fig.~\ref{fig:modelfits2App} (right)
 and Figs.~\ref{LFU1} and \ref{LFU2} by Figs.~\ref{LFU1App} and \ref{LFU2App} respectively.
\item Tables~\ref{tab:results1D}, \ref{tab:results2D}, \ref{tab:Fit6D}, and \ref{Fit3Dbis} by Tables~\ref{tab:Fits1D2020}, \ref{tab:Fits2D2020}, \ref{tab:Fit6D2020}, and \ref{tab:FitsLFU2020} respectively.
\item Fig. \ref{fig:12App} (right) and Figs. \ref{fig:c9} are new.
\end{itemize}

\begin{table}[b]
\begin{tabular}{|c|c|} 
\hline
{\rm Re} $\lambda_u$ & $3.383 \times 10^{-4}$ \\
{\rm Im} $\lambda_u$ & $7.555 \times 10^{-4}$ \\
$\lambda_t$ & \,\,\, $(4.124 \pm 0.063)\times 10^{-2}$ \\
$\tau_{B^0}$ & $1.520 \times 10^{-12}$ s \\
$\tau_{B^+}$ & $1.638 \times 10^{-12}$ s\\
$\tau_{B_s}$ & $1.509 \times 10^{-12}$ s \\
\hline
\end{tabular}
\caption{List of updated input parameters  in the present analysis.}
\label{tab:inputspar}
\end{table}

The data presented in Ref.~\cite{lhcbupdated} corresponds to an integrated luminosity of 4.7 fb$^{-1}$ collected by LHCb collaboration. Our global analysis now includes 180 observables corresponding to: i) all previous data \cite{Capdevila:2017bsm} ii) updates discussed in the main text 
and iii) the combined Run1+2016 data for optimized observables presented in Ref.~\cite{lhcbupdated}. The combined Run1+2016 data share two main features: on the one hand, the global picture is very coherent with respect to the Run-1 and part of Run-2 (2015-2016) data used in the main text. On the other hand, errors are generally reduced, specially in the bins $[1.1,2.5]$ and $[2.5,4.0]$. These two features, by themselves, will reduce the p-value of the SM as we will see below.

In the analysis presented in this addendum, besides updating the data, we have also updated some input parameters (see Table~\ref{tab:inputspar}) and
improved the theoretical prediction for $B_s \to \mu\mu$ using the results from Refs.~\cite{Beneke:2017vpq,MisiakOrsay}.
However, it turns out that this theory update has a relatively marginal impact on our results.

\begin{table*} 
\renewcommand{\arraystretch}{1.5}
\setlength{\tabcolsep}{8pt}
\begin{tabular}{@{}c||c|c|c|c||c|c|c|c@{}} 
\hline
 & \multicolumn{4}{c||}{All} &  \multicolumn{4}{c}{LFUV}\\
\hline
1D Hyp.   & Best fit& 1 $\sigma$/2 $\sigma$   & Pull$_{\rm SM}$ & p-value & Best fit & 1 $\sigma$/ 2 $\sigma$  & Pull$_{\rm SM}$ & p-value\\
\hline\hline
\multirow{2}{*}{$\Cc{9\mu}^{\rm NP}$}    & \multirow{2}{*}{-1.03} &    $[-1.19,-0.88]$ &    \multirow{2}{*}{6.3}   & \multirow{2}{*}{37.5\,\%}
&   \multirow{2}{*}{-0.91}   &$[-1.25,-0.61]$&   \multirow{2}{*}{3.3}  & \multirow{2}{*}{60.7\,\%}  \\
 &  & $[-1.33,-0.72]$ &  & &  &  $[-1.63,-0.34]$ & \\
 \multirow{2}{*}{$\Cc{9\mu}^{\rm NP}=-\Cc{10\mu}^{\rm NP}$}    &   \multirow{2}{*}{-0.50} &    $[-0.59,-0.41]$ &   \multirow{2}{*}{5.8}  & \multirow{2}{*}{25.3\,\%}
 &  \multirow{2}{*}{-0.39}   &   $[-0.50,-0.28]$ & \multirow{2}{*}{3.7}   & \multirow{2}{*}{75.3\,\%}  \\
 &  & $[-0.69,-0.32]$ &  & & & $[-0.62,-0.17]$  &    \\
 \multirow{2}{*}{$\Cc{9\mu}^{\rm NP}=-\Cc{9'\mu}$}     & \multirow{2}{*}{-1.02} &    $[-1.17,-0.87]$   &  \multirow{2}{*}{6.2}  & \multirow{2}{*}{34.0\,\%}
 &  \multirow{2}{*}{-1.67}   &    $[-2.15,-1.05]$  & \multirow{2}{*}{3.1} & \multirow{2}{*}{53.1\,\%} \\
 &  & $[-1.31,-0.70]$ &  & & & $[-2.54,-0.48]$ &    \\
\hline
 \multirow{2}{*}{$\Cc{9\mu}^{\rm NP}=-3 \Cc{9e}^{\rm NP}$} & \multirow{2}{*}{-0.93}  & $[-1.08,-0.78]$ & \multirow{2}{*}{6.2}  & \multirow{2}{*}{33.6\,\%}
  &   \multirow{2}{*}{-0.68} &    $[-0.92,-0.46]$ & \multirow{2}{*}{3.3}  & \multirow{2}{*}{60.8\,\%} \\
 & & $[-1.23,-0.63]$ &  & & & $[-1.19,-0.25]$     & \\
 \hline
\end{tabular} \scriptsize
\caption{Most prominent 1D patterns of NP in $b\to s\mu^+\mu^-$ transitions (state-of-the-art fits as of March 2020). Here, Pull$_{\rm SM}$ is quoted in units of standard deviation and the $p$-value of the SM hypothesis is 1.4\% for the fit ``All" and 12.6\% for the fit LFUV.} \scriptsize
\label{tab:Fits1D2020}
\end{table*}

\begin{table*}
\renewcommand{\arraystretch}{1.5}
\setlength{\tabcolsep}{9pt}
\begin{center}
\begin{tabular}{@{}l||c|c|c||c|c|c@{}} 
\hline
 & \multicolumn{3}{c||}{All} &  \multicolumn{3}{c}{LFUV}\\
\hline
 2D Hyp.  & Best fit  & Pull$_{\rm SM}$ & p-value & Best fit & Pull$_{\rm SM}$ & p-value\\
\hline\hline
$(\Cc{9\mu}^{\rm NP},\Cc{10\mu}^{\rm NP})$ & (-0.98,+0.19) & 6.2 & 39.8\,\% & (-0.31,+0.44) & 3.2 & 70.0\,\% \\
$(\Cc{9\mu}^{\rm NP},\Cc{7^{\prime}})$  & (-1.04,+0.01) & 6.0 & 36.5\,\% & (-0.92,-0.04) & 3.0 & 57.4\,\% \\
$(\Cc{9\mu}^{\rm NP},\Cc{9^\prime\mu})$  & (-1.14,+0.55) & 6.5 & 47.4\,\% & (-1.86,+1.20) & 3.5 & 81.2\,\% \\
$(\Cc{9\mu}^{\rm NP},\Cc{10^\prime\mu})$  & (-1.17,-0.33) & 6.6 & 50.3\,\% & (-1.87,-0.59) & 3.7 & 89.6\,\% \\
\hline
$(\Cc{9\mu}^{\rm NP}, \Cc{9e}^{\rm NP})$ & (-1.09,-0.25) & 6.0 & 36.5\,\% & (-0.72,+0.19) & 2.9 & 54.5\,\%  \\
\hline
Hyp. 1 & (-1.10,+0.28) & 6.5 & 48.9\,\% & (-1.69,+0.29) & 3.5 & 82.4\,\% \\
Hyp. 2 & (-1.01,+0.07) & 5.9 & 33.7\,\% & (-1.95,+0.22) & 3.1 & 64.3\,\% \\
Hyp. 3 & (-0.51,+0.10) & 5.4 & 24.0\,\% & (-0.39,-0.04) & 3.2 & 69.9\,\% \\
Hyp. 4  & (-0.52,+0.11) & 5.6 & 26.4\,\% & (-0.46,+0.15) & 3.4 & 77.9\,\% \\
Hyp. 5  & (-1.17,+0.23) & 6.6 & 51.1\,\% & (-2.05,+0.50) & 3.8 & 91.9\,\% \\
\hline
\end{tabular}
\caption{Most prominent 2D patterns of NP in $b\to s\mu^+\mu^-$ transitions (state-of-the-art fits as of March 2020). The last five rows correspond to Hypothesis 1: $(\Cc{9\mu}^{\rm NP}=-\Cc{9^\prime\mu} , \Cc{10\mu}^{\rm NP}=\Cc{10^\prime\mu})$,  2: $(\Cc{9\mu}^{\rm NP}=-\Cc{9^\prime\mu} , \Cc{10\mu}^{\rm NP}=-\Cc{10^\prime\mu})$, 3: $(\Cc{9\mu}^{\rm NP}=-\Cc{10\mu}^{\rm NP} , \Cc{9^\prime\mu}=\Cc{10^\prime\mu}$), 4: $(\Cc{9\mu}^{\rm NP}=-\Cc{10\mu}^{\rm NP} , \Cc{9^\prime\mu}=-\Cc{10^\prime\mu})$ and 5: $(\Cc{9\mu}^{\rm NP} , \Cc{9^\prime\mu}=-\Cc{10^\prime\mu})$.}\label{tab:Fits2D2020}\end{center}
\end{table*}

\begin{table*}
\renewcommand{\arraystretch}{1.3}
\setlength{\tabcolsep}{9pt}
\begin{tabular}{@{}c||c|c|c|c|c|c@{}}
\hline
 & $\Cc7^{\rm NP}$ & $\Cc{9\mu}^{\rm NP}$ & $\Cc{10\mu}^{\rm NP}$ & $\Cc{7^\prime}$ & $\Cc{9^\prime \mu}$ & $\Cc{10^\prime \mu}$  \\
\hline\hline
Best fit & +0.00 & -1.13 & +0.20 & +0.00 & +0.49 & -0.10 \\ \hline
1 $\sigma$ & $[-0.02,+0.02]$ & $[-1.30,-0.96]$ & $[+0.05,+0.37]$ & $[-0.01,+0.02]$ & $[+0.04,+0.95]$ &$[-0.33,+0.14]$ \\
2 $\sigma$ & $[-0.03,+0.04]$ & $[-1.46,-0.78]$ & $[-0.09,+0.57]$ & $[-0.03,+0.04]$ & $[-0.39,+1.45]$ &$[-0.55,+0.41]$\\
\hline
\end{tabular}
\caption{1 and $2\sigma$ confidence intervals for the NP contributions to Wilson coefficients in the 6D hypothesis allowing for NP in $b\to s\mu^+\mu^-$ operators dominant in the SM and their chirally-flipped counterparts, for the fit ``All'' (state-of-the-art as of March 2020). The Pull$_{\rm SM}$ is $5.8\sigma$ and the \textit{p}-value is $46.8 \%$.}
\label{tab:Fit6D2020}
\end{table*}

\subsection{Theoretical update of $B_s \to \mu\mu$}

Our analyses include ${\cal B}(B_s \to \mu^+\mu^-)$ as it constrains the space for NP contributions in ${\cal C}_{10\mu}$ and ${\cal C}_{10'\mu}$ significantly. The expression of this branching ratio can be derived from Ref.~\cite{Bobeth:2013uxa} taking into account that we use ${\cal O}_{10\ell}=\frac{e^2}{16\pi^2}(\bar{s}\gamma_{\mu}P_{L}b)(\bar{\ell}\gamma^{\mu}\gamma_5\ell)$, where ${P_{L}=\frac{1 {-}\gamma_5}{2}}$, instead of the axial operator ${\cal O}_{A}=(\bar{b}\gamma_{\mu}s)(\bar{\ell}\gamma^{\mu}\gamma_5\ell)$ used in Ref.~\cite{Bobeth:2013uxa} and therefore reads
\begin{equation}
{\cal B}_{B_s\to\mu\mu}^{\rm SM}=\frac{\lambda^2_t G^2_F m^3_{B_s} \alpha^2_{em} \tau_{B_s} f^2_{B_s}}{64\pi^3} r^2\sqrt{1-r^2}|{\cal C}_{10\mu}^{\rm SM}|^2\,,
\label{eq:BRusSM}
\end{equation}
\noindent
where $r={2m_{\mu}}/{m_{B_s}}$.
Once NP contributions in ${\cal C}_{10}$ and in the chirality-flipped Wilson coefficient ${\cal C}_{10^\prime}$ are included, the full expression in our analyses, excluding scalar and pseudoscalar operators, reads
\begin{equation}
{\cal B}_{B_s\to\mu\mu}=\frac{\lambda^2_t G^2_F m^3_{B_s} \alpha^2_{em} \tau_{B_s} f^2_{B_s}}{64\pi^3}r^2\sqrt{1-r^2}
({\cal C}_{10\mu}-{\cal C}_{10'\mu})^2
\label{eq:BRus}
\end{equation}
where ${\cal C}_{10\mu}={\cal C}^{\rm{SM}}_{10\mu}+{\cal C}^{\rm{NP}}_{10\mu}$.

As discussed in Refs.~\cite{DescotesGenon:2011pb,DeBruyn:2012wk}, the LHCb measurement of $B_s$ decays is performed after integrating the time evolution of the $B_s$ meson and its mixing with $\bar{B}_s$.
The resulting correction is an effect of
$O(\Delta\Gamma_s/\Gamma_s)$ and it is modulated by an asymmetry $A_{\Delta\Gamma}$ which depends
on the process considered. In the SM, for $B_s\to\mu\mu$, this asymmetry is known to be +1~\cite{DeBruyn:2012wk}:  the time-integrated branching ratio $\overline{{\cal B}}_{s\mu\mu}$ is then obtained from
 ${\cal B}_{B_s\to\mu\mu}$ by replacing the average of the lifetimes of the light and heavy mass eigenstates $\tau_{B_s}$ by that of the heavy mass eigenstate $\tau^s_{H}$ (see for instance the assessment performed in Ref.~\cite{Bobeth:2013uxa} within the SM). The asymmetry $A_{\Delta\Gamma}$  can be changed in the presence of NP contributions to $\C{10'\mu}$, inducing an a priori different $O(\Delta\Gamma_s/\Gamma_s)$ correction from time integration~\footnote{It can also be modified by contributions to scalar or pseudoscalar NP contributions, as well as by NP contributions with an imaginary part, but we do not consider such hypotheses among our NP scenarios.}. In principle we should thus  enlarge the error on the prediction of $\bar{\cal B}_{B_s\to\mu\mu}$
 in the case of scenarios involving NP in $\C{10'\mu}$ to take into account the uncertainty on the $O(\Delta\Gamma_s/\Gamma_s)$ correction. We checked explicitly that enlarging this uncertainty has no actual impact on the outcome of the fits and for simplicity we will thus keep the SM uncertainty on $\bar{\cal B}_{B_s\to\mu\mu}$ for all our analyses.

The most recent theoretical prediction for ${\cal B}_{B_s\to\mu\mu}$ includes a set of electromagnetic corrections at scales below $m_b$ that are dynamically enhanced by $m_b/\Lambda_{\rm{QCD}}$ and by large logarithms~\cite{Beneke:2017vpq}. The size of such corrections, found to be $1\%$, is larger than previous estimates of next-to-leading order QED effects, assessed to be $\pm 0.3\%$.
To account for these new corrections, we have rescaled our theoretical prediction Eq.~(\ref{eq:BRus}) by an overall factor $\Delta_{B_{s\mu\mu}}$
so that our own set of input parameters yields an SM result in agreement with the value presented at the Orsay workshop in 2019~\cite{MisiakOrsay}:
\begin{equation}
\overline{{\cal B}}_{s\mu\mu}^{\rm SM}=\eta_{\rm{QED}}(3.65\pm0.23)\times10^{-9}=(3.64\pm0.14)\times10^{-9}\, .
\label{eq:misiak19}
\end{equation}
where the effect of the QED corrections from Ref.~\cite{Beneke:2017vpq} is introduced as a global factor $\eta_{\rm{QED}}=0.993$.

\subsection{Updated 1D, 2D and 6D global fits to $b\to s \ell\ell$ flavour anomalies in March 2020}

Tabs.~\ref{tab:Fits1D2020},
VIII and \ref{tab:Fit6D2020} collect the updated results for the most prominent LFUV NP scenarios. These tables (updated using March 2020 data) supersede the ones presented in the main text, i.e. Tabs.~\ref{tab:results1D}, \ref{tab:results2D} and \ref{tab:Fit6D}, respectively. A discussion on the most  relevant NP scenarios can be found in the main text. Figs.~\ref{fig:FitResultAllApp} provide a graphical account of the most remarkable results.

Tab.~\ref{tab:FitsLFU2020} collects the updated NP scenarios combining LFUV and LFU, thus superseding the results presented in the main text (Tab.~\ref{Fit3Dbis}) and those presented in Ref.~\cite{Alguero:2018nvb}. Among the scenarios presented in this table, we find one of the most significant solutions in terms of sigmas (scenario 8)
as can also be seen in Figures~\ref{LFU1App} and \ref{LFU2App}.

\begin{table*}\small \begin{center}
\renewcommand{\arraystretch}{1.08}
\setlength{\tabcolsep}{9pt}
\begin{tabular}{@{}lc||c|c|c|c|c@{}}
\hline
\multicolumn{2}{c||}{Scenario} & Best-fit point & 1 $\sigma$ & 2 $\sigma$ & Pull$_{\rm SM}$ & p-value \\
\hline\hline
\multirow{ 3}{*}{Scenario 5} &$\Cc{9\mu}^{\rm V}$ & $-0.54$ & $[-1.06,-0.06]$ & $[-1.68,+0.39]$ &
\multirow{ 3}{*}{6.0} & \multirow{ 3}{*}{39.4\,\%} \\
&$\Cc{10\mu}^{\rm V}$ & $+0.58$ & $[+0.13,+0.97]$ & $[-0.48,+1.33]$ & \\
&$\Cc{9}^{\rm U}=\Cc{10}^{\rm U}$ & $-0.43$ & $[-0.85,+0.05]$ & $[-1.23,+0.67]$ &\\
\hline
\multirow{ 2}{*}{Scenario 6}&$\Cc{9\mu}^{\rm V}=-\Cc{10\mu}^{\rm V}$ & $-0.56$ & $[-0.65,-0.47]$ & $[-0.75,-0.38]$ &
\multirow{ 2}{*}{6.2} & \multirow{ 2}{*}{41.4\,\%} \\
&$\Cc{9}^{\rm U}=\Cc{10}^{\rm U}$ & $-0.41$ & $[-0.53,-0.29]$ & $[-0.64,-0.16]$ &\\
\hline
\multirow{ 2}{*}{Scenario 7}&$\Cc{9\mu}^{\rm V}$ & $-0.84$ & $[-1.15,-0.54]$ & $[-1.48,-0.26]$ &
\multirow{ 2}{*}{6.0} & \multirow{ 2}{*}{36.5\,\% }  \\
&$\Cc{9}^{\rm U}$ & $-0.25$ & $[-0.59,+0.10]$ & $[-0.92,+0.47]$  &\\
\hline
\multirow{ 2}{*}{Scenario 8}&$\Cc{9\mu}^{\rm V}=-\Cc{10\mu}^{\rm V}$ & $-0.34$ & $[-0.44,-0.25]$ & $[-0.54,-0.16]$ &
\multirow{ 2}{*}{6.5} & \multirow{ 2}{*}{48.4\,\%} \\
&$\Cc{9}^{\rm U}$ & $-0.80$ & $[-0.98,-0.60]$ & $[-1.16,-0.39]$ &\\
\hline\hline
\multirow{ 2}{*}{Scenario 9}&$\Cc{9\mu}^{\rm V}=-\Cc{10\mu}^{\rm V}$ & $-0.66$ & $[-0.79,-0.52]$ & $[-0.93,-0.40]$ &
\multirow{ 2}{*}{5.7} & \multirow{ 2}{*}{28.4\,\%} \\
&$\Cc{10}^{\rm U}$ & $-0.40$ & $[-0.63,-0.17]$ & $[-0.86,+0.07]$ &\\
\hline
\multirow{ 2}{*}{Scenario 10}&$\Cc{9\mu}^{\rm V}$ & $-1.03$ & $[-1.18,-0.87]$ & $[-1.33,-0.71]$ &
\multirow{ 2}{*}{6.2} & \multirow{ 2}{*}{41.5\,\%} \\
&$\Cc{10}^{\rm U}$ & $+0.28$ & $[+0.12,+0.45]$ & $[-0.04,+0.62]$ &\\
\hline
\multirow{ 2}{*}{Scenario 11}&$\Cc{9\mu}^{\rm V}$ & $-1.11$ & $[-1.26,-0.95]$ & $[-1.40,-0.78]$ &
\multirow{ 2}{*}{6.3} & \multirow{ 2}{*}{43.9\,\%} \\
&$\Cc{10'}^{\rm U}$ & $-0.29$ & $[-0.44,-0.15]$ & $[-0.58,-0.01]$ &\\
\hline
\multirow{ 2}{*}{Scenario 12}&$\Cc{9'\mu}^{\rm V}$ & $-0.06$ & $[-0.21,+0.10]$ & $[-0.37,+0.26]$ &
\multirow{ 2}{*}{2.1} & \multirow{ 2}{*}{2.2\,\%} \\
&$\Cc{10}^{\rm U}$ & $+0.44$ & $[+0.26,+0.62]$ & $[+0.09,+0.81]$ &\\
\hline
\multirow{ 4}{*}{Scenario 13}&$\Cc{9\mu}^{\rm V}$ & $-1.16$ & $[-1.31,-1.00]$ & $[-1.46,-0.83]$ &
\multirow{ 4}{*}{6.2} & \multirow{ 4}{*}{49.2\,\%} \\
&$\Cc{9'\mu}^{\rm V}$ & $+0.56$ & $[+0.27,+0.83]$ & $[-0.02,+1.10]$ &\\
&$\Cc{10}^{\rm U}$ & $+0.28$ & $[+0.08,+0.49]$ & $[-0.11,+0.70]$ &\\
&$\Cc{10'}^{\rm U}$ & $+0.01$ & $[-0.19,+0.22]$ & $[-0.40,+0.42]$ &\\
\hline
\end{tabular}
\caption{Most prominent patterns for LFU and LFUV NP contributions from Fit ``All'' (state-of-the-art as of March 2020). See Table~\ref{Fit3Dbis} for more detail.}
\label{tab:FitsLFU2020} \end{center}
\end{table*}

\begin{figure*} \begin{center}
\includegraphics[width=0.42\textwidth]{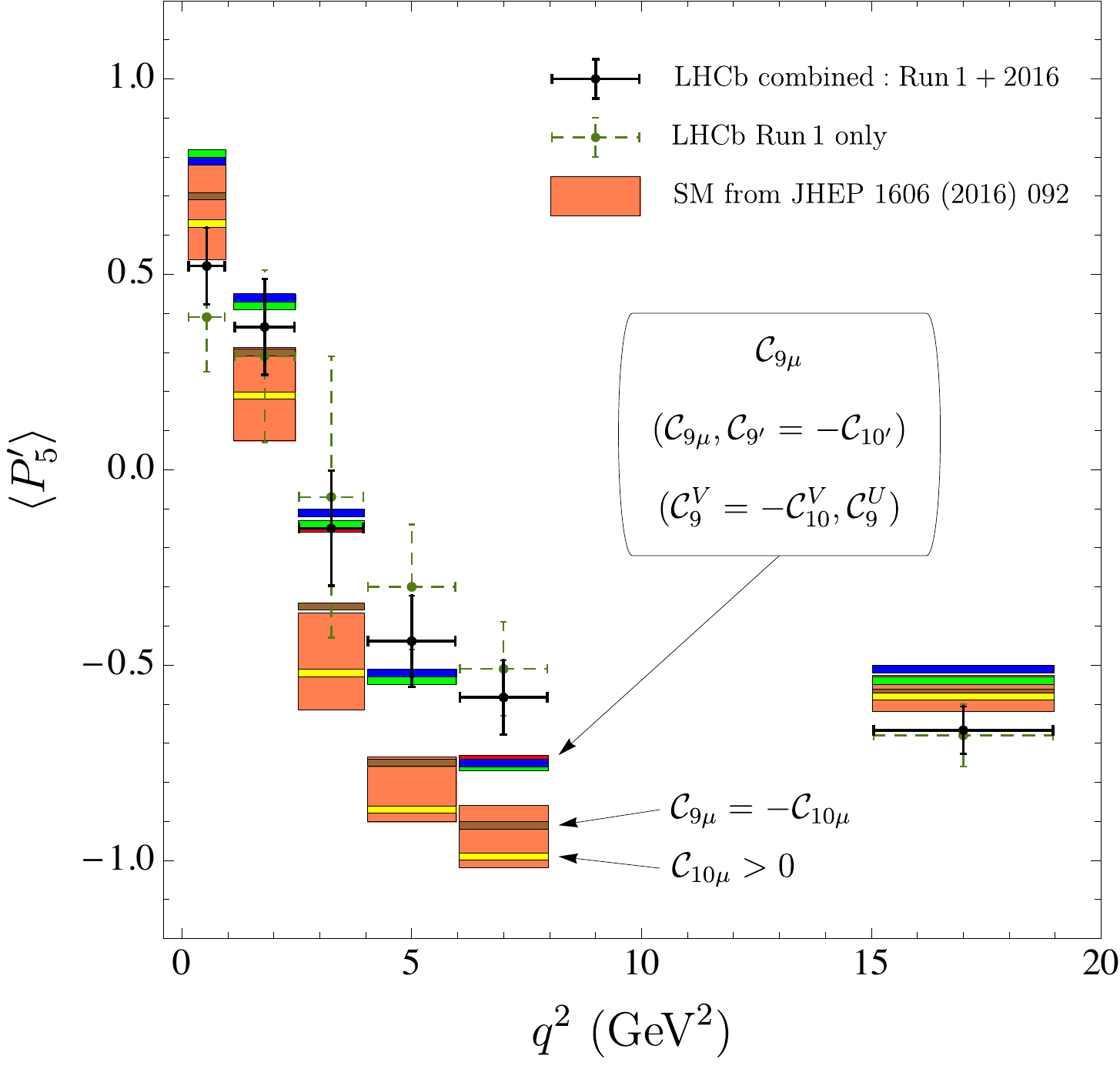}
\hspace{12mm}
\includegraphics[width=0.41\textwidth]{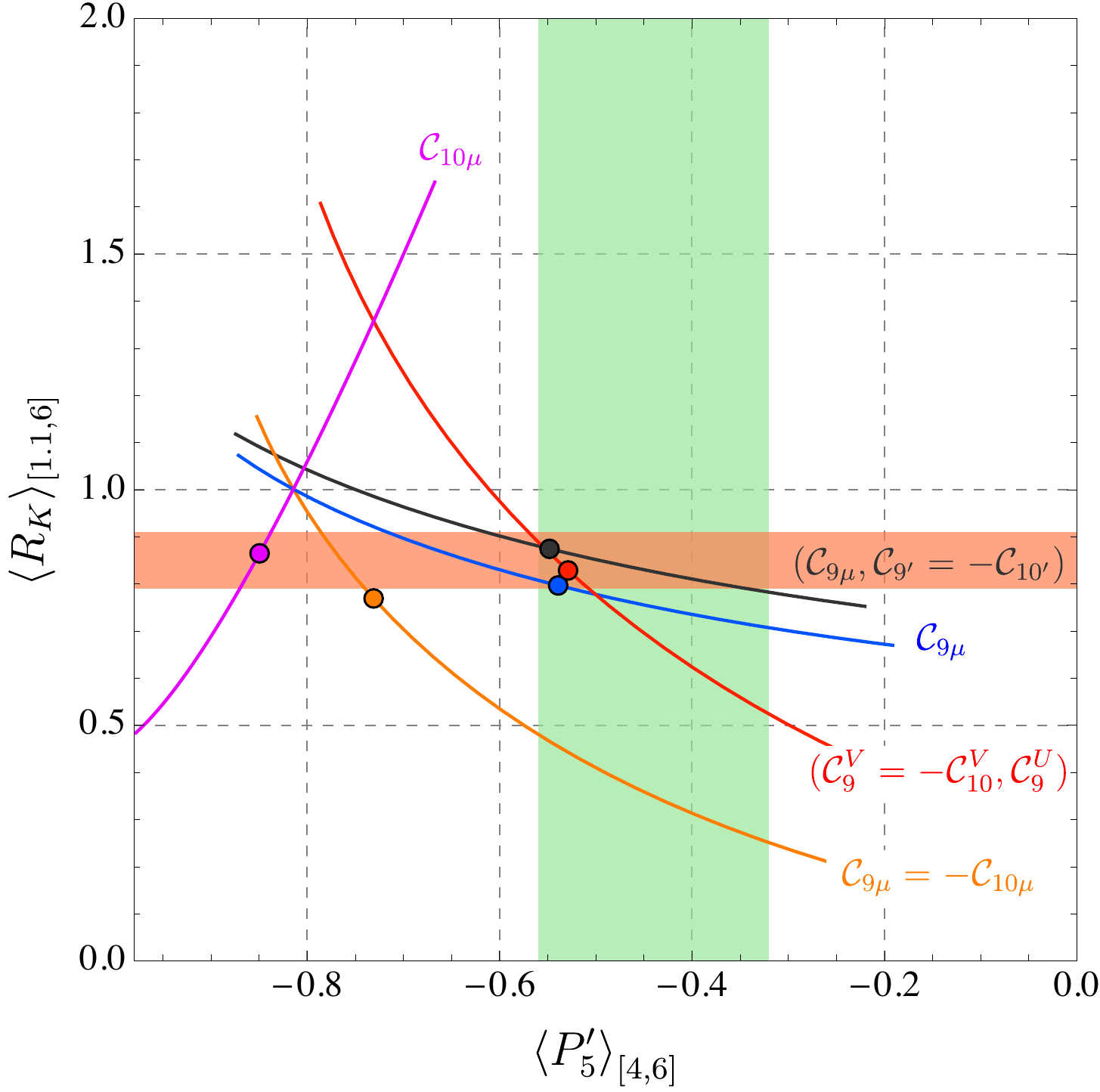}
\caption{Left: Impact of favoured NP scenarios on the observable $P_5^\prime$. This figure supersedes Fig.~\ref{fig:modelfits} of the main text. Only central values for the NP scenarios are displayed. The most interesting scenarios cluster together, ${\cal C}_{9\mu}^{\rm NP}$ in red, $({\cal C}_{9\mu}^{\rm NP},{\cal C}_{9'\mu}^{\rm NP}=-{\cal C}_{10'\mu}^{\rm NP})$ in green and $({\cal C}_{9}^{\rm V}=-{\cal C}_{10}^{\rm V},{\cal C}_9^{\rm U})$ in blue,  and they are now in better agreement with $P'_5$ data. On the other hand, ${\cal C}_{9\mu}^{\rm NP}=-{\cal C}_{10\mu}^{\rm NP}$ (brown) and ${\cal C}_{10\mu}^{\rm NP}$ (yellow, with a global significance of only 3.2$\sigma$) fail to explain the deviations observed for this observable. Right: $\langle R_K \rangle_{[1.1,6]}$ versus $\langle P_5' \rangle_{[4,6]}$ in five different scenarios: ${\cal C}_{9\mu}^{\rm NP}$ (blue), ${\cal C}_{9\mu}^{\rm NP} = - {\cal C}_{10\mu}^{\rm NP}$ (orange), and $({\cal C}^{\rm V}_{9\mu}= - {\cal C}^{\rm V}_{10\mu}, {\cal C}^{\rm U}_{9})$ (red), $({\cal C}_{9\mu}^{\rm NP}, {\cal C}_{9'\mu}^{\rm NP}= - {\cal C}_{10'\mu}^{\rm NP})$ (black), and  ${\cal C}_{10\mu}^{\rm NP}$ (pink). This figure partially supersedes Fig.~12 in Ref.~\cite{Alguero:2019pjc}. The curves correspond only to the predictions for central values. In the 2D scenarios (red and black) the Wilson coefficient not shown is set to its b.f.p. value. The current experimental values from the LHCb collaboration are also indicated (orange horizontal and green vertical bands respectively). The dots correspond to the b.f.p. values of the corresponding scenario for the fit to the "All" data set.
}\label{fig:12App}
\end{center}
\end{figure*}

\begin{figure*}
\begin{center}
\includegraphics[width=0.315\textwidth]{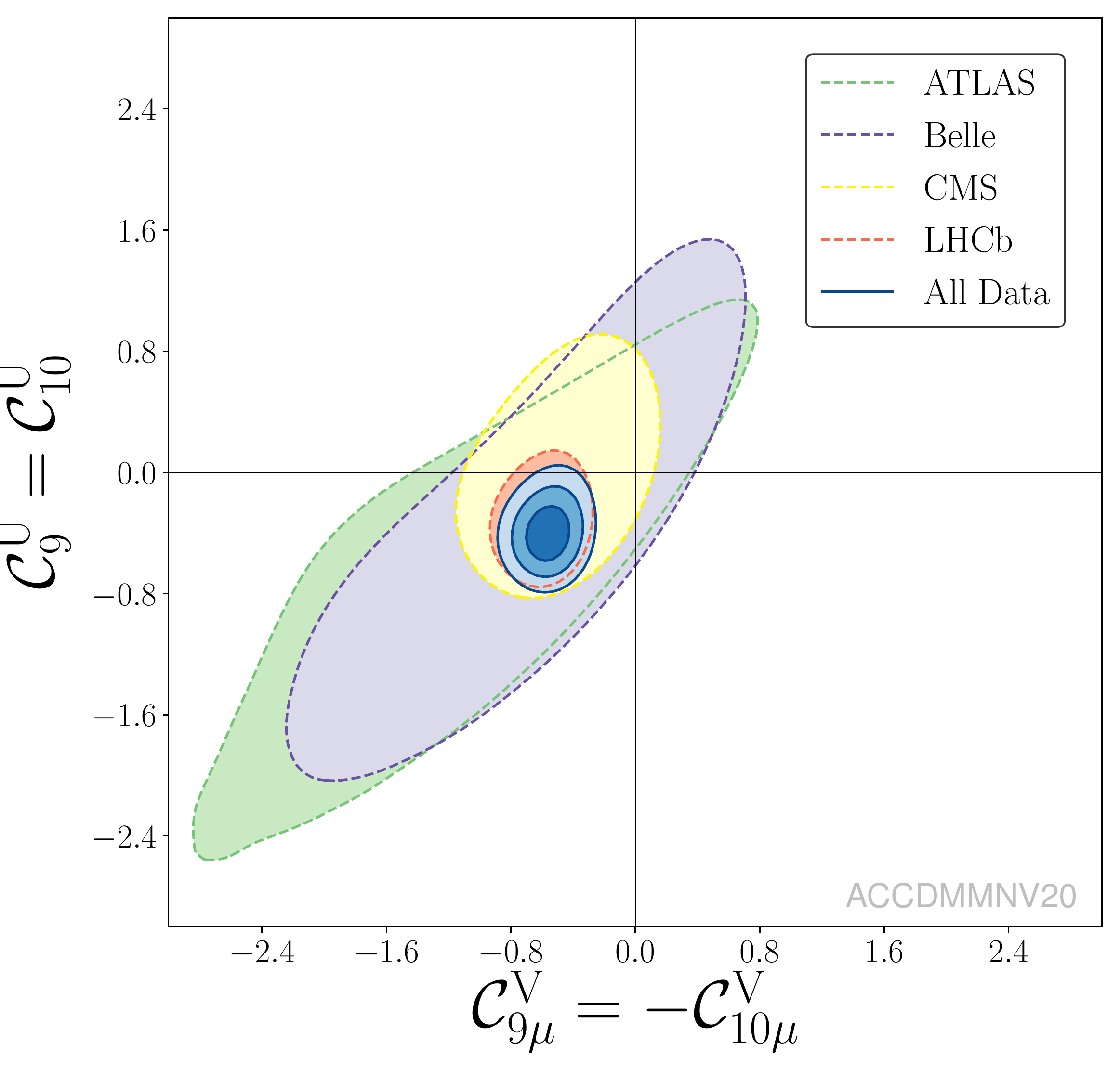}\hspace{5mm}
\includegraphics[width=0.315\textwidth]{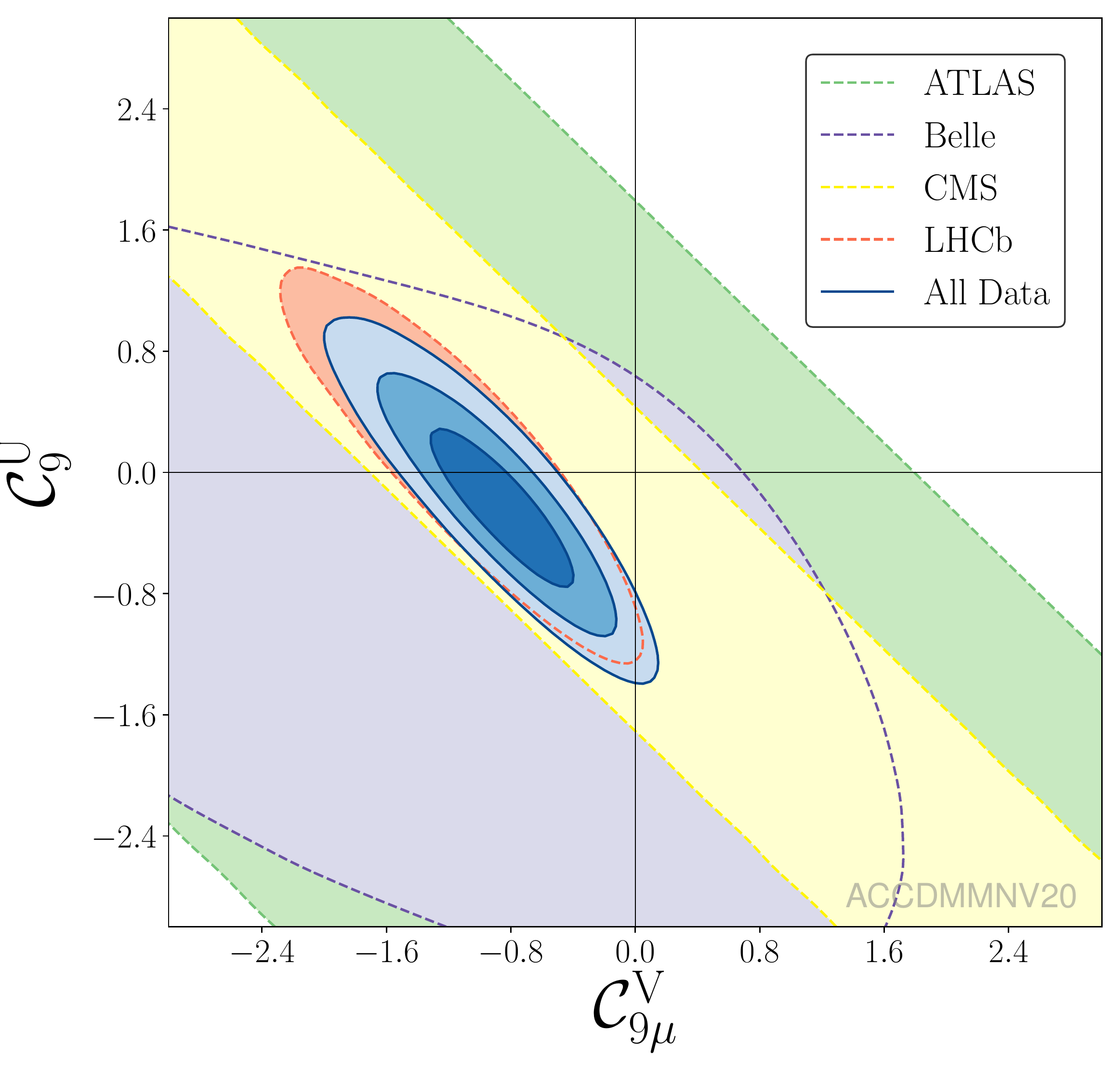}
\\
\includegraphics[width=0.315\textwidth]{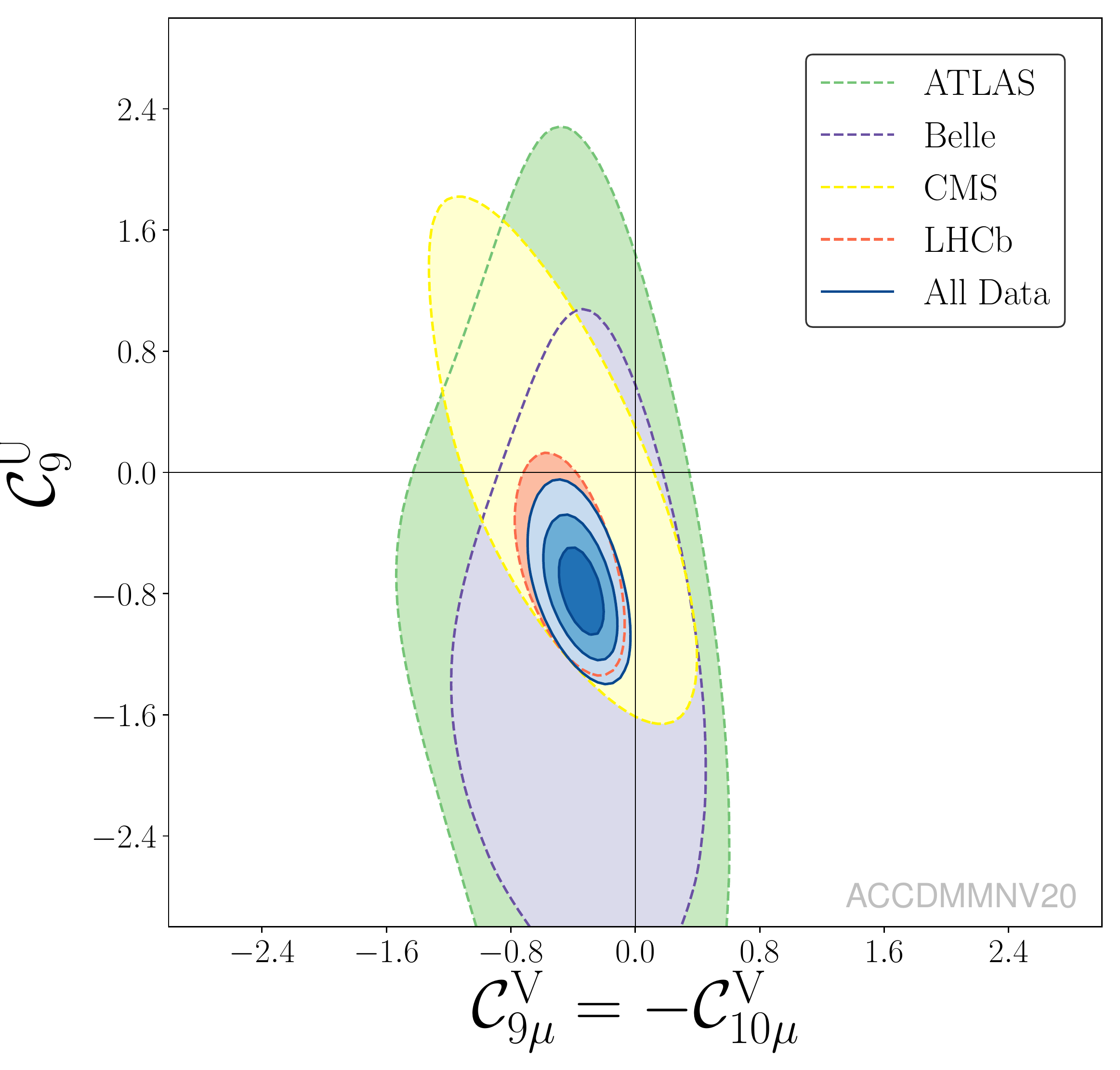}\hspace{5mm}
\includegraphics[width=0.315\textwidth]{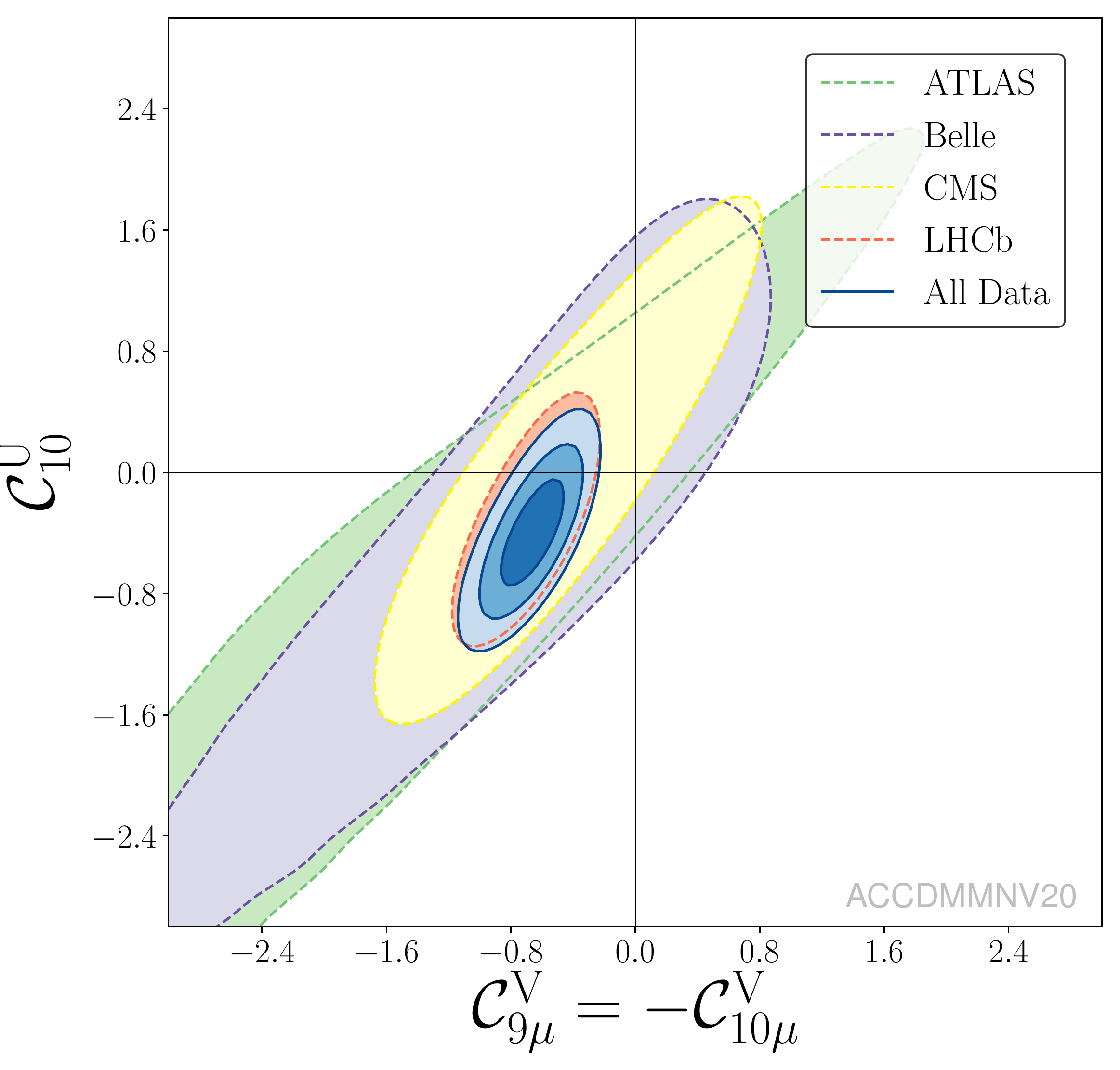}
\end{center}
\caption{Updated plots of Ref.~\cite{Alguero:2018nvb} corresponding to Scenarios 6,7,8,9.} \label{LFU1App}
\end{figure*}

\begin{figure*}
\begin{center}
\includegraphics[width=0.315\textwidth]{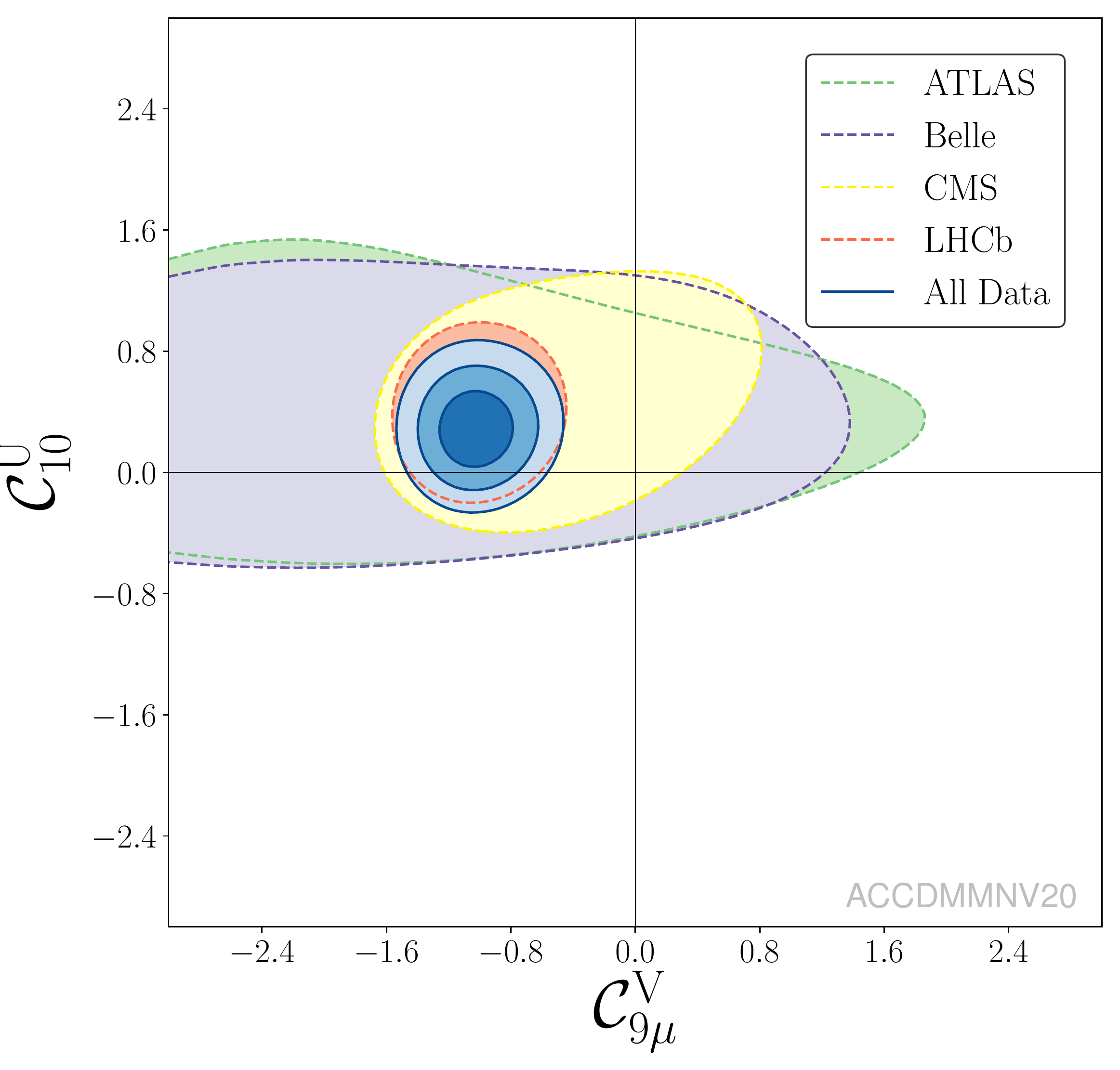} \hspace{2mm}
\includegraphics[width=0.315\textwidth]{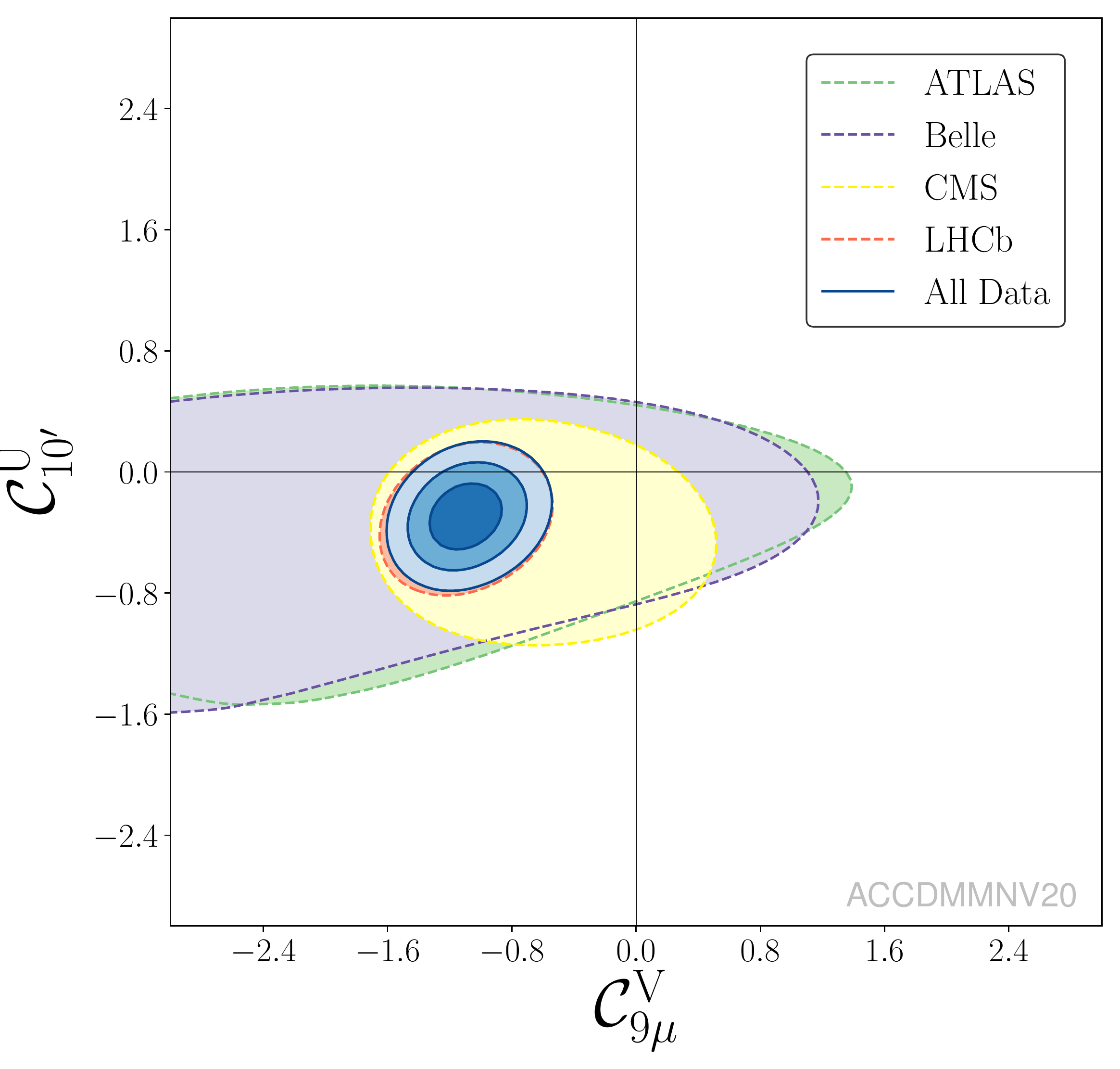}\hspace{2mm}
\includegraphics[width=0.315\textwidth]{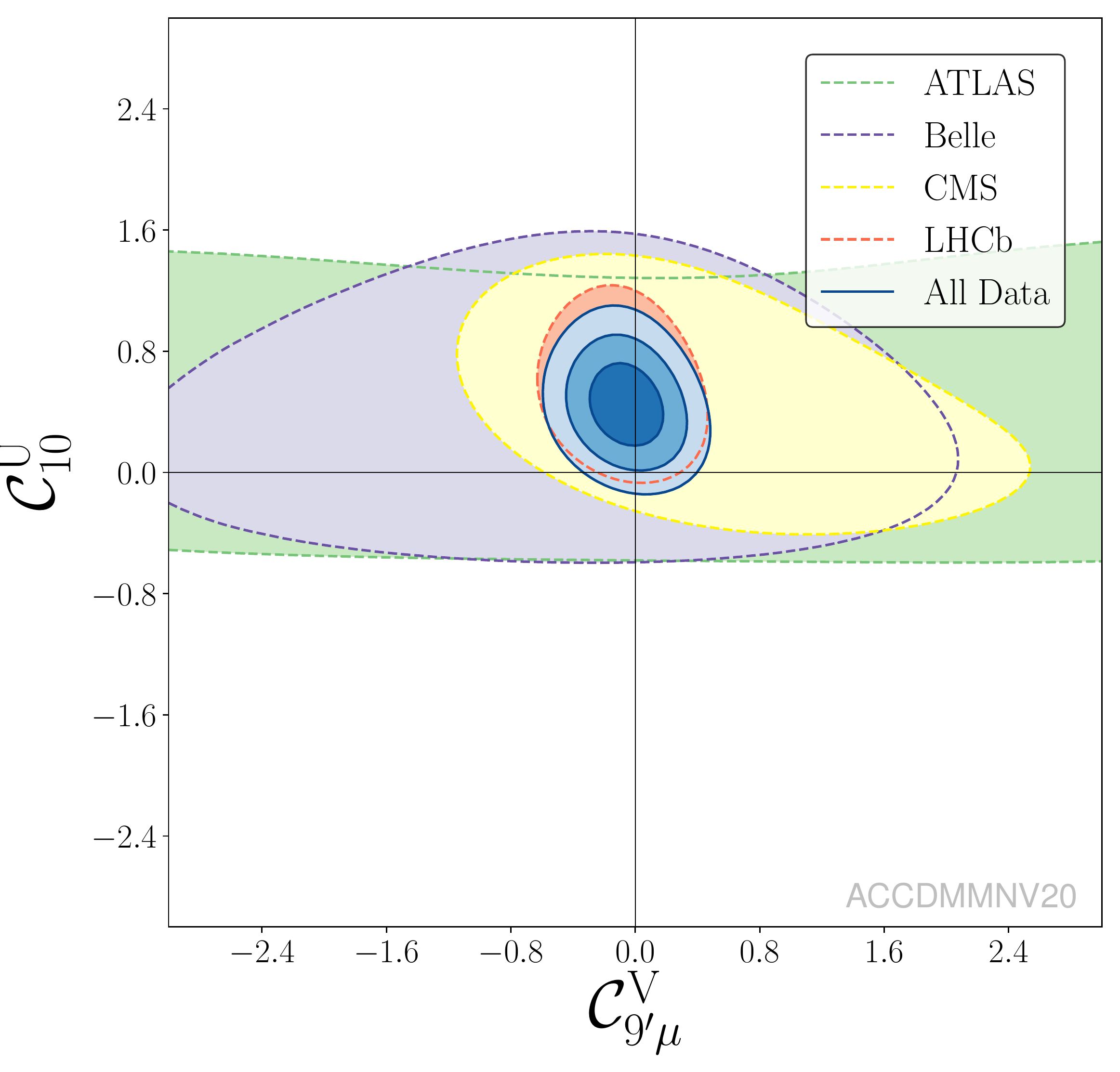}
\end{center}
\caption{Updated plots of Ref.~\cite{Alguero:2018nvb} corresponding to the Scenarios 10,11,12.} \label{LFU2App}
\end{figure*}

We observe an excellent consistency between the previous and the new data. This is a remarkable fact since almost 50 angular observables have been updated in the most recent LHCb collaboration analysis with uncertainty reductions of $30-50\%$ or more (in particular for the bins $[1.1,2.5]$ and $[2.5,4]$). The consistency between all observables previously observed is confirmed with a slightly increased tension (bin by bin) compared to the SM in basically all angular observables. New tensions with respect to the SM appear in $\langle P_3 \rangle_{[1.1,2.5]}$, $\langle P'_6 \rangle_{[6,8]}$ and $\langle P'_8 \rangle_{[1.1,2.5]}$. The tension in the first bin of $P'_5$ has decreased and it is now more similar in size with respect to other tensions~\cite{Alguero:2019pjc} (with the caveat that the experimental analysis relies on an expression of the angular distribution holding in the massless limit, which might bias the analysis in this first bin). The pull of $\langle P'_4 \rangle_{[4,6]}$ has changed sign so that $\langle P'_4 \rangle_{[4,6]}$ and
$\langle P_1 \rangle_{[4,6]}$ are not anymore in tension, favouring a contribution to
 ${\cal C}_{10'\mu}$  (see Table~VIII).

Following this increased consistency, there are two particularly positive features of the new data:
\begin{itemize}
\item[1)] On the one hand, only one of the anomalous bins in $P'_5$ ($[4,6]$) sees its individual significance marginally decreased from $2.9\sigma$ to $2.7\sigma$, while the second one ($[6,8]$) remains at $2.9\sigma$. However, the change in central value and uncertainty for $\langle P_5^\prime \rangle_{[4,6]}$   improves the agreement among the different observables, especially with $R_K$, for our most favoured NP scenarios, as illustrated in Fig. \ref{fig:12App}.
\item[2)] On the other hand, the new average value for $F_L$ in the bin $[2.5,4]$ is now more than $4\sigma$ below 1, while the previous value was at approximately $1\sigma$ from 1, which generated instability problems in some optimised observables in this bin due to a normalization. With the new data this problem is alleviated and we can use the optimised observables in all bins.
\end{itemize}

\begin{figure*}
	\begin{center}
		\includegraphics[width=7.3cm]{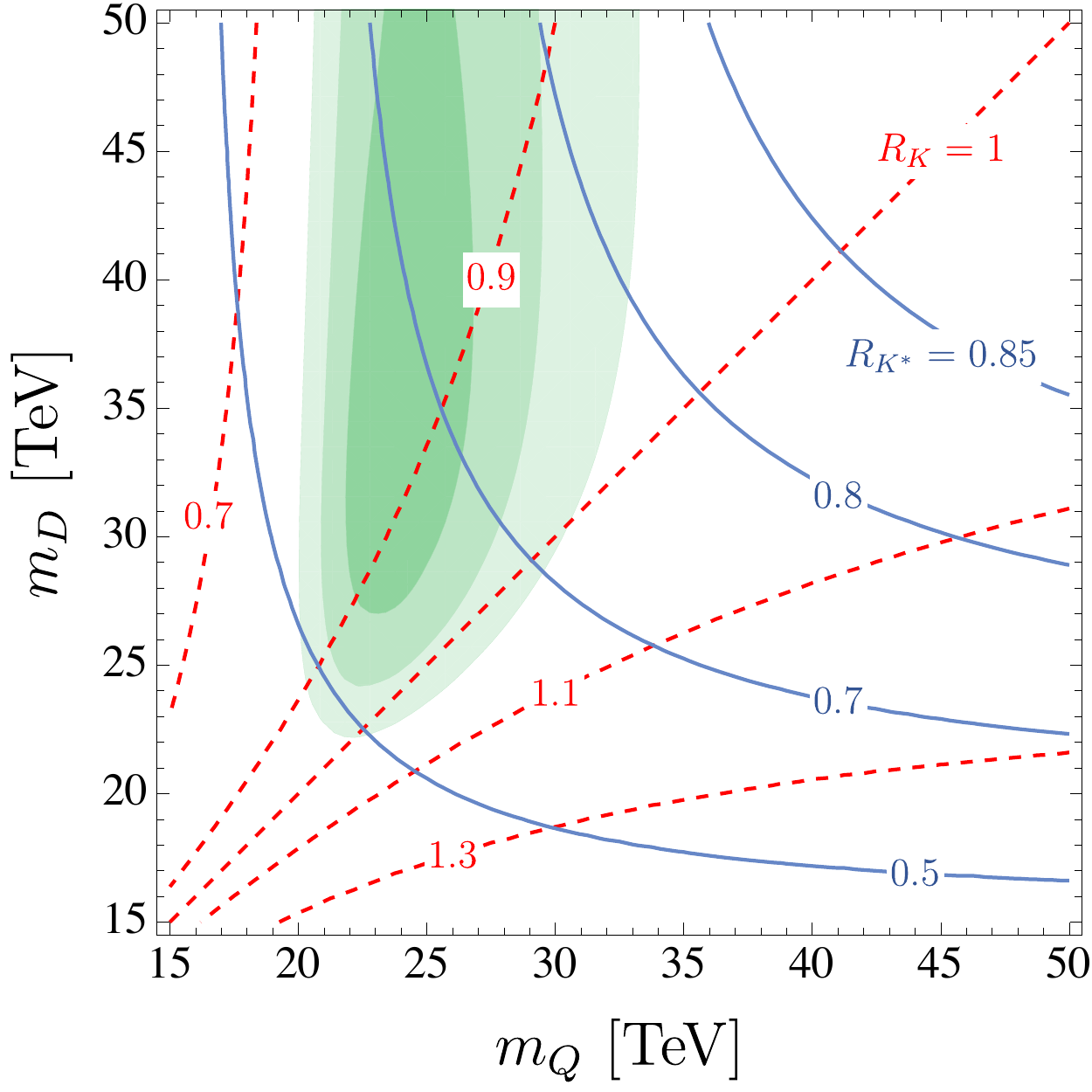}
		\hspace{10mm}
		\includegraphics[width=7.53cm]{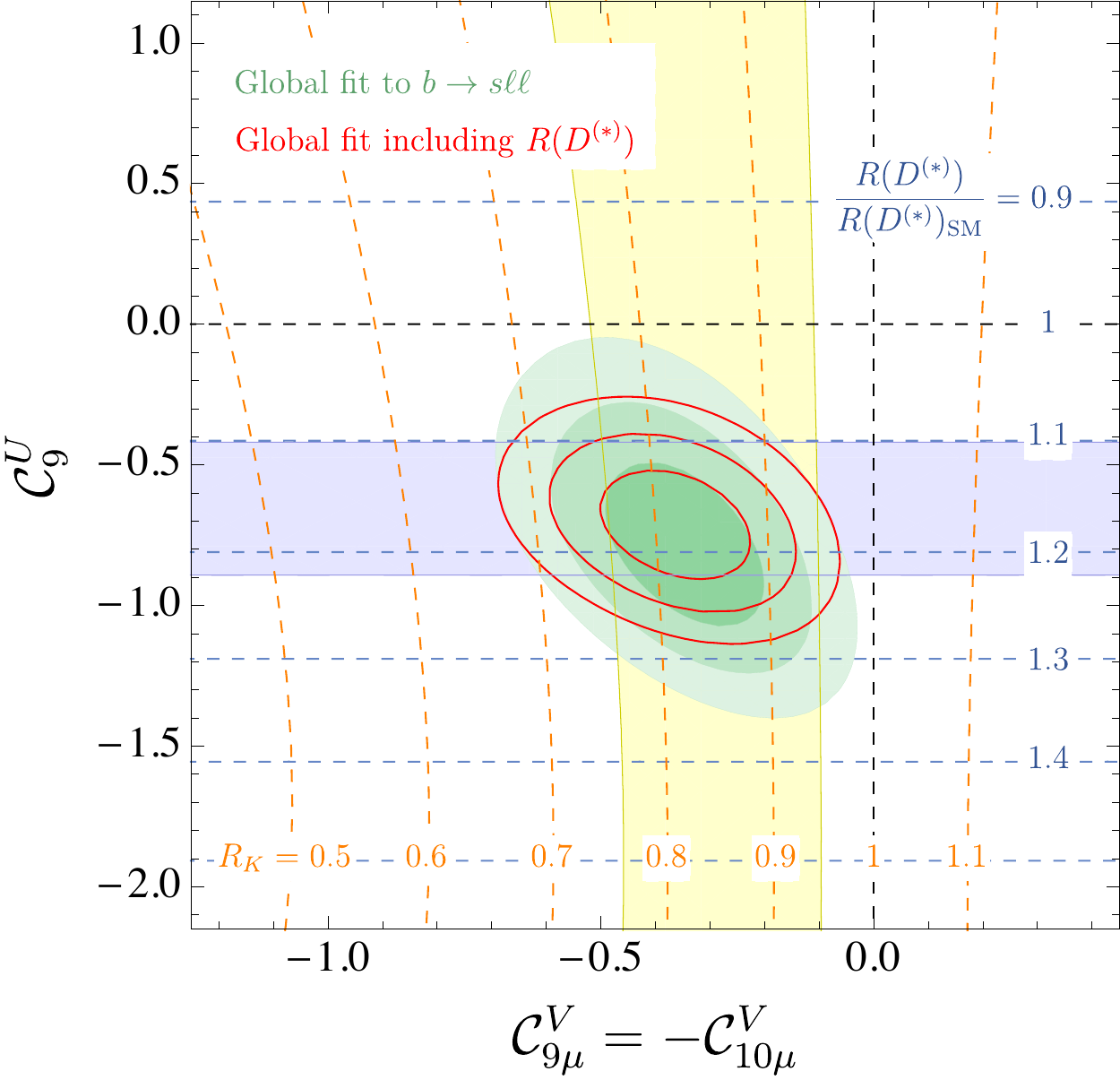}	\end{center}
	\caption{Left: This plot supersedes Fig.~\ref{fig:modelfits2} from the main text and describes the preferred regions (at the 1, 2 and 3$\,\sigma$ level) for the $L_\mu-L_\tau$ model of Ref.~\cite{Altmannshofer:2014cfa} from $b\to s\ell^+\ell^-$ data (green) in the $(m_Q,\, m_D)$ plane with $Y^{D,Q}=1$. The contour lines denote the predicted values for $R_K^{[1.1,6]}$ (red, dashed) and $R_{K^*}^{[1.1,6]}$ (blue, solid).
Right: This plot supersedes the left plot in Fig.~5 and it represents the
preferred regions at the 1, 2 and 3$\,\sigma$ level (green) in the $(\Cc{9\mu}^{\rm V}=-\Cc{10\mu}^{\rm V},\,\Cc{9}^{\rm U})$ plane from $b\to s\ell^+\ell^-$ data. The red contour lines show the corresponding regions once $R_{D^{(*)}}$ is included in the fit (for $\Lambda=2$~TeV). The horizontal blue (vertical yellow) band is consistent with $R_{D^{(*)}}$ ($R_{K}$) at the $2\,\sigma$ level and the contour lines show the predicted values for these ratios.	}
	\label{fig:modelfits2App}
\end{figure*}

In summary, all results show now the following global picture:
\begin{itemize}
 \item Besides an increase of significance of some scenarios ({up to 0.8$\sigma$}), there is no significant change, neither in the {hierarchies among scenarios}, nor in confidence intervals for the Wilson coefficients, with respect to the results presented in our earlier analysis presented in the main text. Our updated results therefore confirm the preexisting picture which calls for NP and they support the scenarios already favoured to explain the deviations.
\item There is a reduction of the internal tensions between some of the most relevant observables of the fit, in particular, between the new averages of $R_K$ and $P'_5$. This leads to an increase in consistency between the different anomalies. This is illustrated in Fig.~\ref{fig:12App} (left) showing a better agreement between the predictions for $P_5^\prime$ in the most relevant NP scenarios and its updated measurement. Furthermore, in Fig.~\ref{fig:12App} (right), the best-fit points for the three favoured NP scenarios ${\cal C}^{\rm{NP}}_{9\mu}$ (Ref.~\cite{Descotes-Genon:2013wba}), $\{{\cal C}^{\rm{NP}}_{9\mu}, {\cal C}_{9'\mu}=-{\cal C}_{10'\mu}\}$ (main text of this paper) and $\{{\cal C}^{\rm{V}}_{9\mu}=-{\cal C}^{\rm{V}}_{10\mu}, {\cal C}^{\rm{U}}_{9}\}$ (Ref.~\cite{Alguero:2018nvb}) can explain two of the most relevant anomalies, $\langle P_5^\prime\rangle_{[4,6]}$ and $R_K$, in a perfect way. On the contrary, we see that the scenarios of NP in $\C{10\mu}$ only or in $\C{9\mu}^{\rm NP}=-\C{10\mu}^{\rm NP}$ do not provide such a good agreement (this holds for any value of the NP contribution).

\item The reduced uncertainties of the $B\to K^*\mu\mu$ data and its improved internal consistency sharpen statistical statements on the hypotheses considered. There is a significant increase of the statistical exclusion of the SM hypothesis as its p-value is reduced down to $1.4\%$ (i.e. 2.5$\sigma$). The Pull$_{\rm SM}$ of the 6D fit is now higher (5.8$\sigma$).

\item Finally, we have updated the figures corresponding to specific simplified models in Fig.~\ref{fig:modelfits2App}. In particular, our scenario 8 can still be interpreted in an EFT framework explaining $b\to c\ell\nu$ and $b\to s\ell\ell$
through correlated singlet and triplet dimension-6 operators combining quark and lepton bilinears. Both $b\to s\ell\ell$ and $b\to c\ell\nu$ show a very good agreement with this interpretation (see the right-hand side of  Fig.~\ref{fig:modelfits2App}) which indicates that scenario 8 is compatible with the tensions in $R_{D^{(*)}}$ if one assumes that
the only significant contributions come from the operators ${\cal O}^{2333}$ and ${\cal O}^{2322}$ in the language of Ref.~\cite{Capdevila:2017iqn}.
The pull of  this scenario reaches {7.4$\,\sigma$} taking into account the deviations also observed in $R_{D^{(*)}}$.
\end{itemize}

\begin{figure*}
	\begin{center}
	\includegraphics[width=5.3cm]{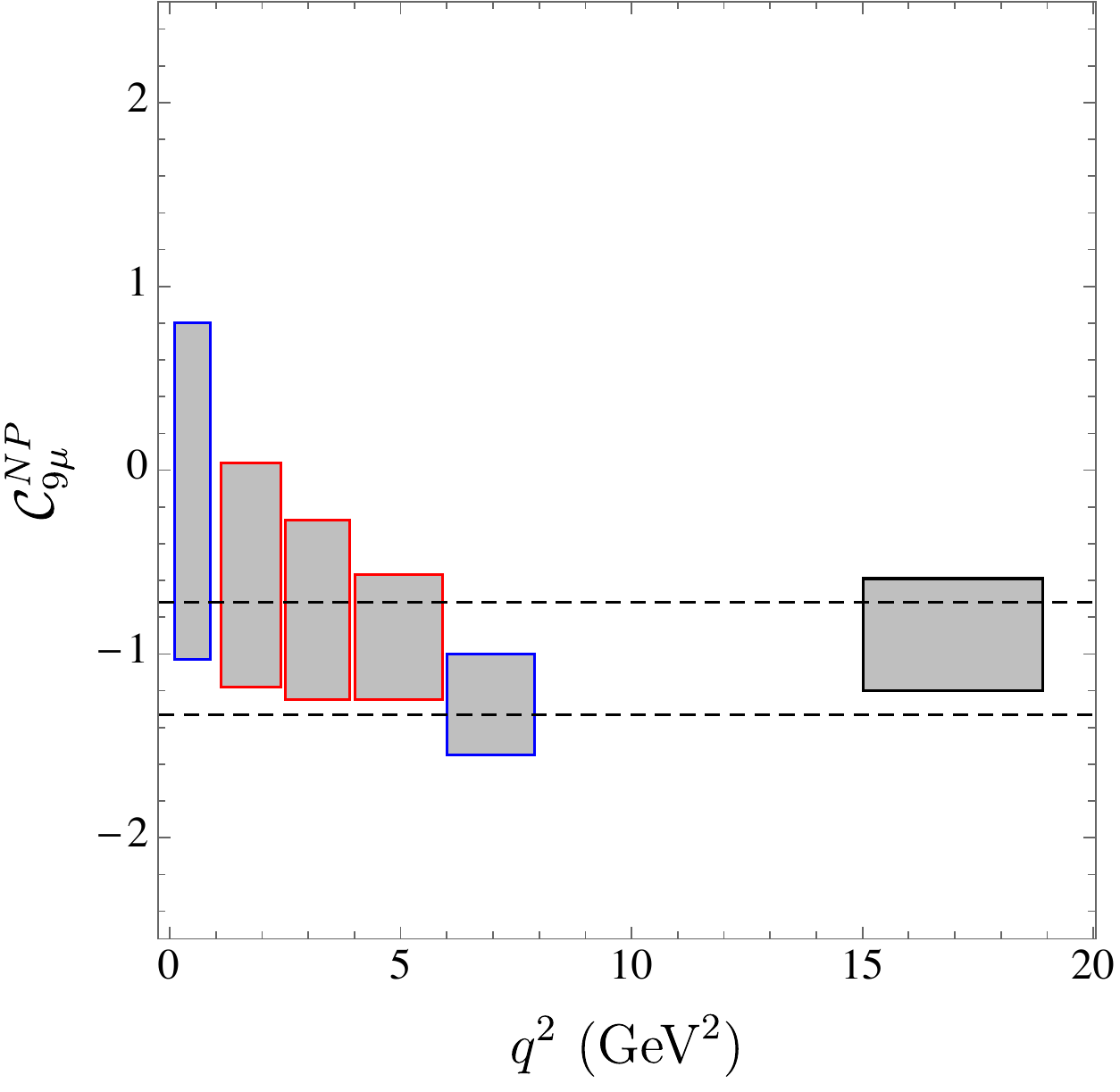}
	\hspace{2mm} \includegraphics[width=5.3cm]{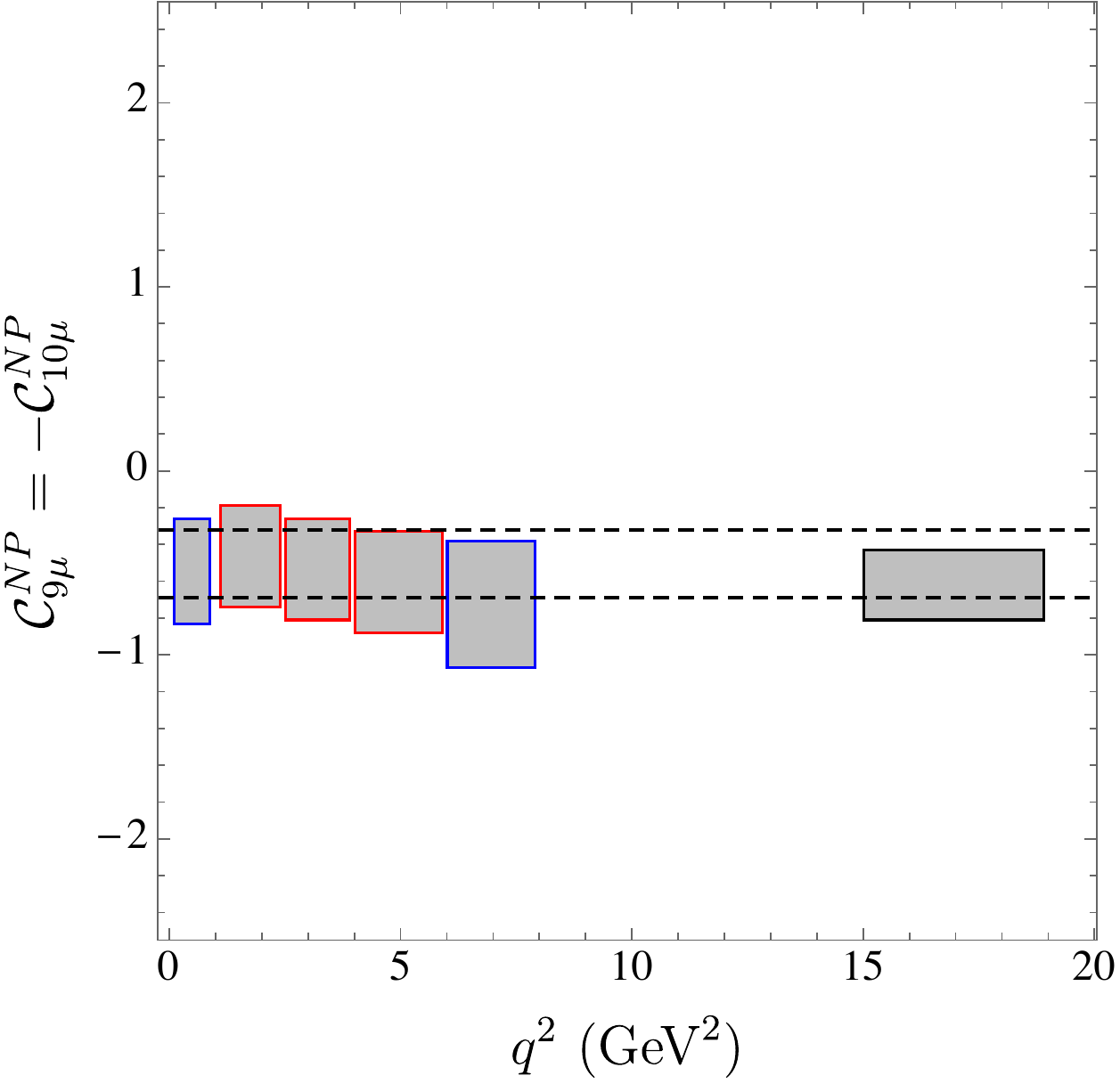}
    \hspace{2mm} 	\includegraphics[width=5.3cm]{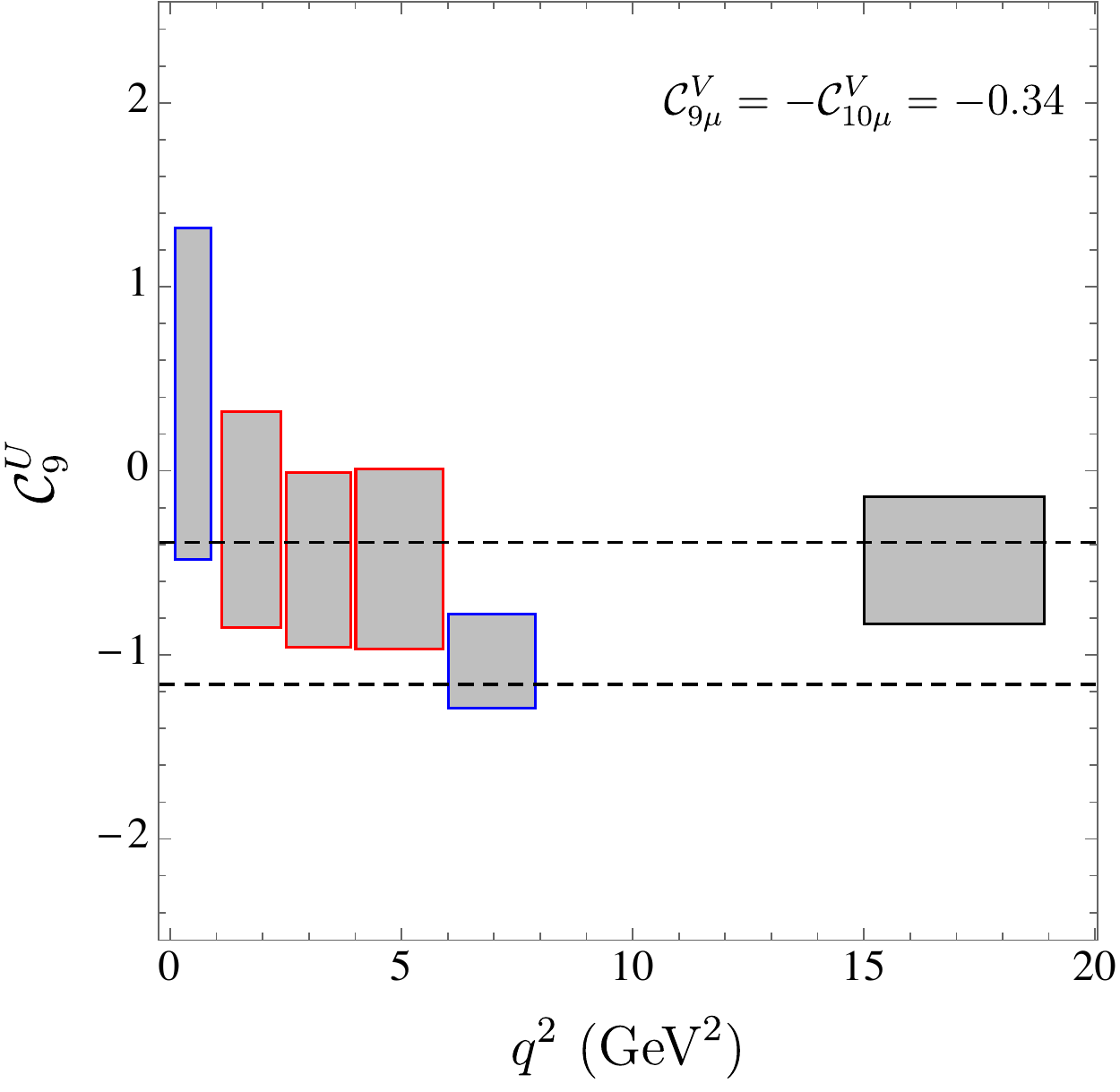}
		\end{center}
	\caption{ Determination of Wilson coefficients
	 in a bin-by-bin fit using only the new LHCb data on optimized observables, branching ratios and radiative decays. Each box correspond to the 1$\sigma$ confidence interval obtained in this bin.
	 Left: ${\cal C}_{9\mu}^{\rm NP}$ assuming NP affects only this Wilson coefficient. Middle: ${\cal C}_{9\mu}^{\rm NP}=-{\cal C}_{10\mu}^{\rm NP}$ assuming NP affects only these Wilson coefficients. Right:	${\cal C}_{9}^{\rm U}$ in scenario 8,
	setting the LFUV coefficients ${\cal C}_{9\mu}^{\rm V}=-{\cal C}_{10\mu}^{\rm V}$ to their values at the best-fit point of the "All" fit. In each case, the band corresponds to the 2$\sigma$ interval obtained from the fit of the NP hypothesis to the "All" data set.}
	\label{fig:c9}
\end{figure*}

The updated measurements of the $B\to K^*\mu\mu$ angular observables give also further possibilities to cross check the stability of our fits regarding internal inconsistencies within the data  or underestimated hadronic effects
by examining the $q^2$-dependence of our extraction (see Fig.~\ref{fig:c9}).
We perform fits testing 1D hypotheses selecting only the available LHCb data for $B\to K^*\mu\mu$ branching ratios and angular observables~\cite{Aaij:2014pli,lhcbupdated,Aaij:2016flj} in a given bin in $q^2$, together
with data on $B_s\to\mu\mu$, $B\to X_s\mu\mu$ and $b \to s\gamma$ processes. We consider 1) the
scenario with NP only in ${\cal C}_{9\mu}$, 2) the scenario with
NP in ${\cal C}_{9\mu}^{\rm NP}=-{\cal C}_{10\mu}^{\rm NP}$, 3) the scenario 8, where we fix
the LFUV part ${\cal C}_{9\mu}^{\rm V}=-{\cal C}_{10\mu}^{\rm V}$ to the b.f.p of the global fit and
determine the value of ${\cal C}_{9}^{\rm U}$ through the fit.
In all three cases, we observe an excellent agreement between the bin-by-bin determination and the outcome of the global fit, without significant $q^2$-dependence. For the scenario with NP only in ${\cal C}_{9\mu}$, a $q^2$-variation could have been the sign of underestimated hadronic effects from $c\bar{c}$-loop contributions~\cite{Capdevila:2017ert}. For the two other scenarios, a $q^2$-dependence would have been the indication of an inconsistency in the experimental data or the theoretical approaches (in particular between the low- and large-recoil bins, where very different theoretical tools are used). It is very reassuring to see that there are no hints of such problems in our analyses.

In the future, we expect more data not only to reduce the uncertainties on the $B\to K^*\mu\mu$ observables, but also to increase further the consistency between $B\to K^*\mu\mu$ data and the rest of the data. On the basis of Figs.~\ref{fig:12App} and~\ref{fig:c9}, we see that several NP scenarios currently favoured by our global fit would push the central value of $\langle P'_5 \rangle_{[6,8]}$ slightly closer to the SM value than currently measured, whereas the determination of $P_5'$ in the other bins should yield the same central values as now.

In conclusion, we see that the recent update of $B \to K^*\mu\mu$ optimised observables by the LHCb collaboration leads  to improved constraints on NP scenarios. The overall preferences for specific scenarios remain unchanged but we observe a higher consistency among the data analysed in the framework of the favoured scenarios.
We expect thus the final update of both $B \to K^*\mu\mu$ optimised observables and $R_K$ including all the remaining recorded data to be an important step forward in the clarification of the $b$-flavour anomalies and the understanding of their origin.

\subsection{Correlations among fit parameters}
 In addition to the confidence regions provided for the various scenarios in this article, we display here the correlation matrices among the Wilson coefficients for the most interesting NP scenarios including the data available in March 2020.

\subsubsection{Correlation Matrices of Fits to LFUV NP}

Following the same ordering for the correlation matrices as in App. A1, we find for the updated analysis:
\[%
\text{Corr}(\mathcal{C}_{9\mu}^\text{NP},\mathcal{C}_{10\mu}^\text{NP})= \begin{pmatrix}%
1.00&0.24\\%
0.24&1.00%
\end{pmatrix}%
\]%

\[%
\text{Corr}(\mathcal{C}_{9\mu}^\text{NP},\mathcal{C}_{9'\mu})= \begin{pmatrix}%
1.00&-0.35\\%
-0.35&1.00%
\end{pmatrix}%
\]%

\[%
\text{Corr}(\mathcal{C}_{9\mu}^\text{NP},\mathcal{C}_{10'\mu})= \begin{pmatrix}%
1.00&0.32\\%
0.32&1.00%
\end{pmatrix}%
\]%

\[%
\text{Corr}(\mathcal{C}_{9\mu}^\text{NP},\mathcal{C}_{9e}^\text{NP})= \begin{pmatrix}%
1.00&0.47\\%
0.47&1.00%
\end{pmatrix}%
\]%

\[%
\text{Corr}(\mathcal{C}_{9\mu}^\text{NP}=-\mathcal{C}_{9'\mu},\mathcal{C}_{10\mu}^\text{NP}=\mathcal{C}_{10'\mu})= \begin{pmatrix}%
1.00&-0.18\\%
-0.18&1.00%
\end{pmatrix}%
\]%

\[%
\text{Corr}(\mathcal{C}_{9\mu}^\text{NP},\mathcal{C}_{9'\mu}=-\mathcal{C}_{10'\mu})= \begin{pmatrix}%
1.00&-0.32\\%
-0.32&1.00%
\end{pmatrix}%
\]%

The last two matrices correspond to Hyp. 1 and Hyp. 5 in Tab.~VIII.

Regarding the 6D fit of Tab.~\ref{tab:Fit6D2020},

\[%
\text{Corr}_\text{6D}= \begin{pmatrix}%
1.00&-0.33&-0.06&0.04&0.04&0.01\\%
-0.33&1.00&0.21&-0.04&0.02&0.22\\%
-0.06&0.21&1.00&-0.12&0.53&0.52\\%
0.04&-0.04&-0.12&1.00&-0.14&-0.07\\%
0.04&0.02&0.53&-0.14&1.00&0.83\\%
0.01&0.22&0.52&-0.07&0.83&1.00%
\end{pmatrix}%
\]

where the columns in the matrix above are organized as in the analogous matrix in App. A1.

$\text{Corr}_\text{6D}$ shows a significant correlation in the pairs $\{\mathcal{C}_{10\mu}^\text{NP},\mathcal{C}_{9'\mu}\}$, $\{\mathcal{C}_{10\mu}^\text{NP},\mathcal{C}_{10'\mu}\}$ and $\{\mathcal{C}_{9'\mu},\mathcal{C}_{10'\mu}\}$, which is coherent with previous results in App. A1.

\vspace{0.4cm}

\subsubsection{Correlation Matrices of Fits to LFUV-LFU NP}

We also provide the correlations between fit parameters of scenarios 5 to 11 from Table~\ref{tab:FitsLFU2020}, in that order:

\[%
\text{Corr}(\mathcal{C}_{9\mu}^\text{V},\mathcal{C}_{9}^\text{U}=\mathcal{C}_{10}^\text{U},\mathcal{C}_{10\mu}^\text{V})= \begin{pmatrix}%
1.00&-0.94&0.92\\%
-0.94&1.00&-0.95\\%
0.92&-0.95&1.00%
\end{pmatrix}%
\]%

\[%
\text{Corr}(\mathcal{C}_{9\mu}^\text{V}=-\mathcal{C}_{10\mu}^\text{V},\mathcal{C}_{9}^\text{U}=\mathcal{C}_{10}^\text{U})= \begin{pmatrix}%
1.00&0.13\\%
0.13&1.00%
\end{pmatrix}%
\]%

\[%
\text{Corr}(\mathcal{C}_{9\mu}^\text{V},\mathcal{C}_{9}^\text{U})= \begin{pmatrix}%
1.00&-0.87\\%
-0.87&1.00%
\end{pmatrix}%
\]%

\[%
\text{Corr}(\mathcal{C}_{9\mu}^\text{V}=-\mathcal{C}_{10\mu}^\text{V},\mathcal{C}_{9}^\text{U})= \begin{pmatrix}%
1.00&-0.46\\%
-0.46&1.00%
\end{pmatrix}%
\]%

\[%
\text{Corr}(\mathcal{C}_{9\mu}^\text{V}=-\mathcal{C}_{10\mu}^\text{V},\mathcal{C}_{10}^\text{U})= \begin{pmatrix}%
1.00&0.71\\%
0.71&1.00%
\end{pmatrix}%
\]%

\[%
\text{Corr}(\mathcal{C}_{9\mu}^\text{V},\mathcal{C}_{10}^\text{U})= \begin{pmatrix}%
1.00&0.02\\%
0.02&1.00%
\end{pmatrix}%
\]%

\[%
\text{Corr}(\mathcal{C}_{9\mu}^\text{V},\mathcal{C}_{10'}^\text{U})= \begin{pmatrix}%
1.00&0.21\\%
0.21&1.00%
\end{pmatrix}%
\]%

The situation is very similar to the previous analysis in App. A2, showing a high anti-correlation between $\mathcal{C}_{9\mu}^\text{V}$ and $\mathcal{C}_{9}^\text{U}$, and an even more reduced correlation between the coefficients of the scenario $\{\mathcal{C}_{9\mu}^\text{V},\mathcal{C}_{10}^\text{U}\}$.

\vspace{0.4cm}


\end{document}